\documentclass{article}

\usepackage{arxiv}

\usepackage[utf8]{inputenc} 
\usepackage[T1]{fontenc}    
\usepackage{hyperref}       
\usepackage{url}            
\usepackage{booktabs}       
\usepackage{amsfonts}       
\usepackage{nicefrac}       
\usepackage{microtype}      
\usepackage{lipsum}		
\usepackage{graphicx}
\usepackage[numbers]{natbib}
\usepackage{doi}

\usepackage{enumerate}
\usepackage[centertags]{amsmath}
\usepackage{amscd}
\usepackage{amsthm}
\usepackage{amssymb}
\usepackage{bm}
\usepackage[font=footnotesize,labelfont=bf]{caption}
\usepackage{subcaption}
\usepackage{xcolor}
\usepackage{multicol}
\usepackage{multirow}

\usepackage{inconsolata}
\usepackage[sc]{mathpazo}
\newtheorem{defn0}{Definition}[section]
\newtheorem{prop0}[defn0]{Proposition}
\newtheorem{thm0}[defn0]{Theorem}
\newtheorem{lemma0}[defn0]{Lemma}
\newtheorem{corollary0}[defn0]{Corollary}
\newtheorem{example0}[defn0]{Example}
\newtheorem{remark0}[defn0]{Remark}
\newtheorem{conjecture0}[defn0]{Conjecture}
\newtheorem*{notation0}{Notation}

\newenvironment{definition}{ \begin{defn0}\rm}{\end{defn0}}

\newenvironment{theorem}{\begin{thm0}}{\end{thm0}}
\newenvironment{lemma}{\begin{lemma0}}{\end{lemma0}}

\newenvironment{example}{ \begin{example0}\rm}{\end{example0}}
\newenvironment{remark}{ \begin{remark0}\rm}{\end{remark0}}

\newenvironment{notation}{\begin{notation0}\rm}{\end{notation0}}

\newcommand{\Pa}{\operatorname{Pa}}
\newcommand{\Ch}{\operatorname{Ch}}
\newcommand{\An}{\operatorname{An}}
\newcommand{\De}{\operatorname{De}}
\newcommand{\Rt}{\operatorname{Rt}}
\newcommand{\indep}{\perp \!\!\! \perp}
\newcommand{\dashbidirectedarrow}{\ \ \dashrightarrow\!\!\!\!\!\!\!\!\!\!\!\!\dashleftarrow\ \ }

\definecolor{ForestGreen}{RGB}{10, 130, 61}

\definecolor{myOrange}{RGB}{200, 100, 0}
\definecolor{myPurple}{RGB}{100, 0, 200}
\definecolor{myDarkGreen}{RGB}{0, 128, 50}
\definecolor{myGray}{RGB}{50, 50, 90}
\usepackage{AlegreyaSans}
\newcommand{\define}[2]{\textbf{\texttt{{\color{myDarkGreen}def} {\color{myPurple}#1}(#2):}}}
\newcommand{\class}[1]{\textbf{\texttt{{\color{myOrange}#1}}}}
\newcommand{\params}[1]{\textbf{\color{myGray}\texttt{#1}}}
\newcommand{\false}{\textbf{\texttt{{\color{myDarkGreen}False}}}}
\newcommand{\true}{\textbf{\texttt{{\color{myDarkGreen}True}}}}
\newcommand{\none}{\textbf{\texttt{{\color{myDarkGreen}None}}}}

\newlength{\mylength}
\renewcommand{\footnotesize}{\scriptsize}

\counterwithout{footnote}{section}

\makeatletter
\newsavebox\myboxA
\newsavebox\myboxB
\newlength\mylenA

\newcommand*\xoverline[2][0.8]{%
    \sbox{\myboxA}{$\scriptstyle\m@th#2$}%
    \setbox\myboxB\null
    \ht\myboxB=\ht\myboxA%
    \dp\myboxB=\dp\myboxA%
    \wd\myboxB=#1\wd\myboxA
    \sbox\myboxB{$\m@th\overline{\copy\myboxB}$}
    \setlength\mylenA{\the\wd\myboxA}
    \addtolength\mylenA{-\the\wd\myboxB}%
    \ifdim\wd\myboxB<\wd\myboxA%
       \rlap{\hskip 0.5\mylenA\usebox\myboxB}{\usebox\myboxA}%
    \else
        \hskip -0.5\mylenA\rlap{\usebox\myboxA}{\hskip 0.5\mylenA\usebox\myboxB}%
    \fi}
\makeatother

\makeatletter

\newcommand*\xunderline[2][0.8]{%
    \sbox{\myboxA}{$\scriptstyle\m@th#2$}%
    \setbox\myboxB\null
    \ht\myboxB=\ht\myboxA%
    \dp\myboxB=\dp\myboxA%
    \wd\myboxB=#1\wd\myboxA
    \sbox\myboxB{$\m@th\underline{\copy\myboxB}$}
    \setlength\mylenA{\the\wd\myboxA}
    \addtolength\mylenA{-\the\wd\myboxB}%
    \ifdim\wd\myboxB<\wd\myboxA%
       \rlap{\hskip 0.5\mylenA\usebox\myboxB}{\usebox\myboxA}%
    \else
        \hskip -0.5\mylenA\rlap{\usebox\myboxA}{\hskip 0.5\mylenA\usebox\myboxB}%
    \fi}
\makeatother

\title{Algorithmic Causal Effect Identification with \texttt{causaleffect}}

\author{Martí Pedemonte \\
	Universitat de Barcelona\\
	\texttt{pedemonte96@gmail.com} \\
	\And
	Jordi Vitrià \\
	Department of Mathematics and Computer Science\\
	Universitat de Barcelona\\
	\texttt{jordi.vitria@ub.edu} \\
	\And
	Álvaro Parafita \\
	Department of Mathematics and Computer Science\\
	Universitat de Barcelona\\
	\texttt{parafita.alvaro@ub.edu} \\
}

\renewcommand{\headeright}{Technical Report}
\renewcommand{\undertitle}{Technical Report}
\renewcommand{\shorttitle}{Algorithmic Causal Effect Identification}

\hypersetup{
pdftitle={Algorithmic Causal Effect Identification},
pdfsubject={},
pdfauthor={Martí Pedemonte, Jordi Vitrià, Álvaro Parafita},
pdfkeywords={DAG, do-calculus, causality, causal model, identifiability, graph, C-component, hedge, d-separation},
}

\begin{document}
\maketitle

\begin{abstract}
Our evolution as a species made a huge step forward when we understood the relationships between causes and effects. These associations may be trivial for some events, but they are not in complex scenarios. To rigorously prove that some occurrences are caused by others, causal theory and causal inference were formalized, introducing the $do$-operator and its associated rules. The main goal of this report is to review and implement in Python some algorithms to compute conditional and non-conditional causal queries from observational data. To this end, we first present some basic background knowledge on probability and graph theory, before introducing important results on causal theory, used in the construction of the algorithms. We then thoroughly study the identification algorithms presented by Shpitser and Pearl in 2006 \cite{SP_2006a, SP_2006b}, explaining our implementation in Python alongside. The main identification algorithm can be seen as a repeated application of the rules of $do$-calculus, and it eventually either returns an expression for the causal query from experimental probabilities or fails to identify the causal effect, in which case the effect is non-identifiable. We introduce our newly developed Python library and give some usage examples.
\end{abstract}


{\let\thefootnote\relax\footnote{{This is a revised version of a thesis submitted to the Universitat de Barcelona Department of Mathematics and Compu\-ter Science in partial fulfillment of the requirements for the degree of BSc in Computer Science and Software Engineering.}}}

\section{Introduction}

What is \textit{causality}? This philosophical concept has dazzled great minds for centuries, and its definition has been debated many times \cite[Chapter 8]{pearl_why}. A possible definition is that causality is the influence by which one event (a cause) contributes to the production of another event (an effect) where the cause is partly responsible for the effect, and the effect is partly dependent on the cause. Nevertheless, the concept and definition of causation is still an ongoing debate between contemporary philosophers, but is out of the scope of this work. Instead, we are interested in how can we answer causal-effect questions, and a very helpful concept to have in mind when asking those questions is the \textit{Ladder of Causation} \cite[Chapter 1]{pearl_why}.

The Ladder of Causation is a metaphor to classify three distinct levels of cognitive ability: seeing, doing and imagining. It consists of three fundamentally different rungs: 
\begin{enumerate}[\hspace*{0.0cm} (a)]

\item[] \textbf{Rung 1: Association.} This is the first, most basic level of the Ladder of Causation, and it involves the observation of data and extraction of regularities from these observations. Examples would be how a dog figures out where a ball is going to land when its owner throws it at the park, or how IBM's Deep Blue analysed thousands of chess games to extract the moves associated with a higher percentage of wins. It is characterized by questions like ``\textit{What if I see...?}'' or ``\textit{How would seeing $X$ change my belief in $Y$?}''. For instance, what does a survey tell us about the election results? All the questions related to this level of the Ladder of Causation can be answered using standard statistical methods. Note that we cannot answer causal queries, we can only make associations (like, for example, compute the correlation of variables). Many animals and present-day Artificial Intelligence algorithms are considered to be in this rung.

\item[] \textbf{Rung 2: Intervention.} The second level of the Ladder of Causation involves intervening or doing a certain action to produce the desired outcome. Examples would be when we take paracetamol to cure a headache (we are intervening on the amount of paracetamol in our body to produce a reduction in headache pain), or when we study to pass an exam (we act on the things we learn to produce a better mark in the exam). It is characterized by questions like ``\textit{What if I do...?}'' or ``\textit{How would $Y$ be if I do $X$?}''. For instance, what would be my weight at the end of the year if I were to jog every day for thirty minutes? To answer questions in this rung of the Ladder of Causation we need to either physically perform the intervention or make use of the recently defined $do$-calculus (which will thoroughly be explained in this project). Unlike the first level, this one allows us to make causal associations between variables. Babies and also primitive humans that used intentionally-made tools are considered to be in this rung.

\item[] \textbf{Rung 3: Counterfactuals.} The highest level of the Ladder of Causation involves imagination and understanding because it compares our real world with an imaginary world. The real world is the world we live in when we do an action, and the imaginary, counterfactual world is the alternative reality in which my action would have been different. It is characterized by questions like ``\textit{What if I had done...?}'' or ``\textit{If $X$ had not occurred, would $Y$ have happened?}''. For instance, would the Theory of General Relativity had been created if Einstein had not existed? Humankind entered this rung of the Ladder when it started to imagine fictional things that they had not seen in real life before, such as divinities, religions or events that could have happened but did not. It is this counterfactual thinking that makes us different from all other intelligent life on Earth and helps us make decisions, by imagining all possible outcomes.

\end{enumerate}

What is important of this ladder is that one cannot answer queries from a level with information of lower levels alone. For instance, to be able to determine causal effects we do not have enough with only observational data, we need something else from rung two of the Ladder of Causation (or above). We will use the so-called $do$-calculus to perform interventions to our probabilities, so instead of having $P(Y|X)$, which would be read as ``\textit{the probability of $Y$ when $X$ is seen}'', we will have expressions of the form $P(Y|do(X))$, meaning ``\textit{the probability of $Y$ when $X$ is artificially imposed}''.

This tool will allow us to compute causal effects from observational data, but it will not always work. There will be cases where the mental model of the problem will not allow us to compute these causal relationships, and we will be forced to either change the model or perform a physical intervention in a real-life experiment. An example where we cannot compute a causal effect between two variables $X$ and $Y$ is when there exist some background unmeasurable variables that affect both $X$ and $Y$. In these cases where we cannot use $do$-calculus to obtain the causal effect, we say that the causal effect is \textit{not identifiable} or \textit{unidentifiable}.

The goal of this work is to address the identifiability problem (to detect in which cases we can identify a causal effect and in which cases we cannot). To do so we will study a few algorithms devised by Shpitser and Pearl \cite{SP_2006a, SP_2006b} and implement them in Python, developing a package available for everyone in the scientific community to use. In this journey, we will also study thoroughly the necessary results used in these algorithms, and we will try to explain them in the most accessible way to reach the widest audience possible. To this end we have organized this document as follows.

In the first section, we present a succinct historical background of causality, and we explain why this project is relevant and of general interest.

In section two we present important tools used in the context of causal theory. We first recall some basic probability theory definitions and theorems, before focusing on crucial aspects about graphs and more precisely about direct acyclic graphs (DAGs). We then enter the realm of probabilistic causal models and introduce $do$-calculus, an indispensable tool when querying causal effects from experimental observations. We present the identifiability problem, and we end the section by studying some criteria to identify causal effects through the so-called confounded components.

The third section is devoted to study and explain the implementation of some algorithms that can identify causal relationships from causal diagrams. We first present the \textbf{ID} algorithm, useful for unconditional causal queries, and we explain how we encode probability distributions and causal diagrams in our implementation of that algorithm. After having meticulously explored every line of the algorithm alongside its implementation, we introduce an algorithm to solve conditional causal effects, called \textbf{IDC}. We then explain how we have implemented it in our package, and give some examples of how to call these functions. We finalize this third section by concisely giving an idea of how algorithms for counterfactual queries may be constructed.

Some conclusions on this work are then presented, after which a link with the source code is provided.

\section[Why Is the Identification Problem Significant?]{Why Is the Identification Problem Significant?}

For decades causation was seen for most statisticians as a special case of correlation, and we owe this misleading association to the English statistician \footnote{Sir Francis Galton. English statistician, 1822 - 1911.}Sir Francis Galton and especially his disciple, \footnote{Karl Pearson. English mathematician and biostatistician, 1857 - 1936.}Karl Pearson. Pearson strongly believed that with data and traditional statistical methods (such as the correlation of variables) one could explain causation. \footnote{Sewall Green Wright. American geneticist, 1889 - 1988.}Sewall Wright, an American geneticist, was against that belief and thought that in causal analysis one must incorporate some understanding of the process that produces the data. He applied this technique when he constructed a path diagram to quantify the influence of developmental factors in a guinea pig's womb on the colour of the fur of its offspring. This path diagram, seen now as one of the first causal diagrams (where arrows are drawn from causes to effects), was the kind of resource Pearson was against, for he stated that different (subjective) models would lead to different conclusions, and that was not rigorous. He could not stand this idea of introducing additional, biased information into the deduction process proposed by Wright, and opposed it outright.

His influence persisted, and it was not until the late 1980's that causal theory made a significant step forward. Judea Pearl, an Israeli-American philosopher and computer scientist, was studying how to manage uncertainty in artificial intelligence systems with Bayesian networks, but this approach could not solve causal-effect queries (recall that one cannot answer questions from rung two of the Ladder of Causation with just information about rung one). With this problem in mind, he then devoted the following years of his career to the formalization of causal theory, obtaining a methodology to compute, in some cases, causal effects from causal diagrams and observational data.

Before the mathematization of causal theory, some philosophers tried to express the sentence ``\textit{$X$ causes $Y$}'' as ``\textit{$X$ raises the probability of $Y$}'' by writing $P(Y|X)>P(Y)$, but this is wrong at its core. Note that ``raises'' is a causal concept from the second level of the Ladder of Causation, while the expression $P(Y|X)>P(Y)$ uses data from observations and thus lies on the first level. This inequality really affirms that ``\textit{if I see $X$, then the probability of $Y$ increases}'', but this increase in probability could be for other reasons, like a third variable $Z$ being the cause of $X$ and $Y$.

According to Pearl, who introduced the $do$-operator, for $X$ to be the cause of $Y$ we need to state that ``\textit{doing $X$ raises the probability of $Y$}'', which would be written as $P(Y|do(X))>P(Y)$. This concept of doing or intervening is from rung two, and thus we can borrow this operator to solve causal queries. Note that doing is fundamentally different from seeing: by \textit{doing} $X$ we do not care if a third variable is causing $X$ and $Y$ because it is I who is forcing the value on $X$ and not some other background factor. If we conclude that the probability of $Y$ while I force the value of $X$ is bigger than without forcing it, then $X$ is partially responsible for $Y$.

Before the definition of the $do$-operator we could not solve causal queries because we were simply not asking the correct questions, we did not have the necessary tools to even formulate them. This operator has not only allowed us to ask the right questions, but it has also provided us with a set of rules that can help us resolve these queries. These rules constitute what is known as $do$-calculus, and under some conditions, they can be used to compute causal effects from observational data. When this is possible, this is, when we can use $do$-calculus to compute the effect of a causal relationship, we say that this effect is \textit{identifiable}.

This definition of identifiability is completely different from the one we have in statistics. In classical statistics, a statistical model $\mathcal{P} = \{P_{\theta} | \theta\in\Theta\}$ is \textit{identifiable} if the mapping $\theta\mapsto P_{\theta}$ is one-to-one, this is, if for different values of the parameter we obtain different probability distributions. In simultaneous equations models, this problem of identification arises when the value of one or more parameters of the equations in the model cannot be determined from observable variables. Note that, in this context, identification depends profoundly on the equations of the model. The concept of identifiability that Pearl introduced does not depend on the form of the equations, but only on the relationship between variables. We will study the identification problem in detail in the following section.

But why is the identification problem relevant in the framework of causal models? When trying to compute a causal effect we could perform the actual intervention in the real world, fixing the value of a variable $X$ and then measuring the other variable $Y$, and seeing if $P(Y|do(X))>P(Y)$. This is not always feasible, sometimes because it is unethical or sometimes because it is simply not viable. Therefore it is of great importance to have a way of computing these interventions without having to actually perform them in reality. This became possible with the introduction of Pearl's $do$-calculus, but lacked a systematic way of calculating causal queries. Years later, a technique to mechanize the estimation of causal effects was eventually developed. This method takes shape as algorithms, designed by Shpitser and Pearl \cite{SP_2006a, SP_2006b}, that use the rules of $do$-calculus to compute a certain causal effect, when possible, and that raise an error when the causal effect is not identifiable.

Our project will consist of studying thoroughly the theory behind these algorithms to be able to implement them in Python, and developing a package to perform these causal effect calculations. This work is relevant because it gathers a set of recent results which are unknown to many computer scientists and statisticians in general. Most of them know about $do$-calculus, but some of them are unaware of the existence of deterministic algorithms that mechanize the process of computing causal effects. There are even some scientists who still think that identifiability and the calculation of causal effects is an open problem. Through this project we want to reach more people, and to make the extraction of causal effects from observational data an effortless procedure.

There is already one implementation of these algorithms for R, by Tikka and Karavanen \cite{causaleffect_R} under the name of \texttt{causaleffect}, but we believe that implementing them in Python, a very popular programming language amongst data scientists, will make them more known worldwide. According to the TIOBE index \cite{tiobe}, at the moment of this writing Python is the second most popular programming language in the whole world just after C, and R falls back to 14\textsuperscript{th} place, so we strongly believe that developing this package for Python will boost the popularity of the results by Shpitser and Pearl.

Additionally, with this work we will also try to explain the results that support the algorithms designed by Shpitser and Pearl \cite{SP_2006a, SP_2006b} in a more clear, understandable way. The notation used to formalize causal theory is effective and very well constructed, but lacks transparency, so we believe that, in order to be accessible to a wider audience, results have to be properly organized and formulated in a friendlier way. We tried to do so in this work by first introducing a few necessary concepts of probability and graph theory, before entering the world of causal theory.

\section[Background Theory]{Background Theory}

The main purpose of this section is to lay a foundation of definitions and results needed later on in the definition and discussion of the algorithms that are the main interest of this work. Some of them can be found in a standard introductory probability book such as \cite{probability}, others in a basic graph theory book like \cite{graphs}. The more specific results on causal models and causal diagrams are available mostly in \textit{Causality} \cite{pearl_causality} by Judea Pearl and \textit{Causal Inference in Statistics} \cite{pearl_inference} by Pearl \textit{et al.}. But there are also some recent results included in this section cited from the original sources, such as \cite{pearl_back_door, pearl_do_rules, SP_2006a, Ver, verma_pearl}, and for the curious reader, \cite{pearl_why} is a great accessible book for wider audiences by the father of modern causality, Judea Pearl.

The first subsection will be focused on establishing and recalling some well-known probability definitions and basic theorems, for they will be used in the context of probabilistic causal models. Then some basic notion of graphs will be presented in the second section, given that certain types of graphs are an essential tool in causality. The section will end with the definition of causal models and causal diagrams, and with the introduction of a criterion for knowing if a causal effect is identifiable from a causal diagram.

\subsection{Probability Theory}

To be able to identify causal effects, we must first recall some basic results of probability theory. One could wonder why probability, a branch of mathematics that works with randomness and doubt, has anything to do with causality. Perhaps the most common answer would be that we live in a world surrounded by uncertainty, and in every chain of events there are observations we cannot make or factors we cannot control. For instance, the sentence ``\textit{if you don't study, you will fail the exam}'' may be true most of the time, but there are unknown and noisy factors, like chance, luck, or recalling a single memory of that only class you attended, for example, that may influence the outcome of that exam. That is why the language of probabilities is used widely in science to model not only social sciences, but natural sciences as well, and why it is also used in causal theory.

Suppose we have an event $A$. Then the probability $P(A)$ is always bounded between 0 and 1, i.e., $0\leq P(A) \leq1$, where $P(A)=0$ when that event is impossible and cannot happen, and $P(A)=1$ when $A$ always happen. If we now have another event $B$, the expression $P(A,B)$ refers to the probability of both events happening.

A basic result of probability theory is the Law of Total Probability, which will be used to simplify probability expressions in upcoming sections.

\begin{theorem}{\bf (Law of Total Probability)}
Let $A$ be an arbitrary event, and $B_1, \ldots, B_n$ mutually exclusive events such that $\sum_{i=1}^{n}P(B_{i})=1$. Then,
\begin{equation*}
P(A) = \sum_{i=1}^{n}P(A, B_{i})\ .
\end{equation*}
If $B$ is a binary event, then $P(A) = P(A, B) + P(A,\overline{B})$, where $\overline{B}$ is the complementary of $B$.
\end{theorem}

We can also wonder how an event happening influences the probability of another event. To deal with these dependent probabilities we must state some basic results on conditional probability.

\begin{definition}{\bf (Conditional Probability and Independence)}
Let $A$, $B$ two events. Then, the \textit{conditional probability} of $A$ under the condition $B$, denoted by $P(A|B)$, is the probability that the event $A$ occurs given that the event $B$ has already occurred. It can be computed from the probability of joint events,
\begin{equation*}
P(A|B) = \frac{P(A,B)}{P(B)}\ ,
\end{equation*}
which leads to a useful relation to keep in mind, $P(A,B) = P(A|B)P(B)$. We say that two events are \textit{independent} if $P(A|B) = P(A)$, meaning that knowing about either event has no effect on the likelihood of the other. Using the last derived relation, independence can also be expressed as $P(A,B) = P(A)P(B)$.
\end{definition}

\begin{notation}
Given two independent events $A$ and $B$, we will write $A\indep B$. If two events $A$ and $B$ are independent given a third event $C$, we will write $A\indep B|C$.
\end{notation}

This following result is extremely useful despite its simple formulation, and helps us change from conditioning on one variable to conditioning on another.

\begin{theorem}{\bf (Bayes' Theorem)}
Let $A$, $B$ two different events with $P(B) \neq 0$. Then,
\begin{equation*}
P(A|B) = \frac{P(B|A)P(A)}{P(B)}\ .
\end{equation*}
\end{theorem}
\begin{proof}
From the definition of conditional probability, we have
\begin{equation*}
P(A|B) = \frac{P(A,B)}{P(B)} = \frac{P(B,A)}{P(B)} = \frac{P(B|A)P(A)}{P(B)}\ .
\end{equation*}
\end{proof}

\begin{notation}
We will write $P(x|y)$ as a shorthand for $P(X=x|Y=y)$.
\end{notation}

\begin{example}\label{exp:probability_axioms}
This example illustrates how we might use the previous stated results and definitions to simplify and rewrite probability expressions. Suppose we have the following expression: $\sum_{w}P(w|z)P(x|z,w)P(y|x,z,w)$. Then, making use of the derived expression of the conditional probability definition, we can write it as
\begin{equation*}
\sum_{w}P(w|z)P(x|z,w)P(y|x,z,w) = \sum_{w}P(w|z)P(x,y|z,w) = \sum_{w}P(x,y,w|z) \ ,
\end{equation*}
and by the Law of Total Probability,
\begin{equation*}
\sum_{w}P(x,y,w|z) = P(x,y|z) \ .
\end{equation*}
\end{example}

It is sometimes useful to decompose a joint probability into individual, conditional probabilities, because if one knows information about the independence between events this procedure may lead to simpler, easier expressions. Nevertheless, the decomposition is obviously not unique. For example, $P(x,y,z) = P(x)P(y,z|x) = P(x)P(y|x)P(z|x,y)$, but also $P(x,y,z) = P(z)P(y,x|z) = P(z)P(y|z)P(x|z,y)$.

This information about dependency between events can be visualized in a graphical form, by drawing probabilistic graphical models.

\begin{definition}{\bf (Probabilistic Graphical Model)}
A \emph{Probabilistic graphical model} is a probabilistic model for which a graph shows the conditional dependence between the random variables present in the model. An example would be a \emph{Bayesian network}, which uses directed acyclic graphs to encode variable dependencies.
\end{definition}

An example of a probabilistic graphical model is shown below.

\begin{example}\label{exp:probabilistic_graphical_model}
The directed graph in Figure \ref{fig:graph_11} is an example of a probabilistic graphical model, particularly a Bayesian network. This graph encodes the dependencies between three binary variables: whether it is summer, whether it is sunny and whether I wear sunscreen.

\begin{figure}[!ht]
\captionsetup[subfigure]{labelformat=empty}
\vspace*{-0.0cm}
\centering
\includegraphics{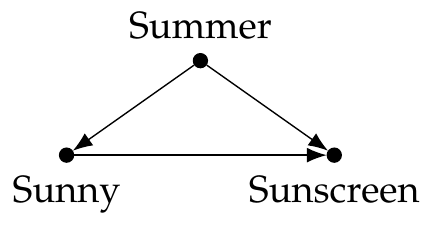}
\vspace*{-0.0cm}
\caption{Probabilistic graphical model.}
\label{fig:graph_11}
\vspace*{-0.0cm}
\end{figure}

Clearly, the fact of being summer affects both being a sunny day and my decision to wear sunscreen, but this is not true in reverse: the season will not change depending on my decision of wearing sunscreen, nor today being sunny. Additionally, my choice of wearing sunscreen will also depend on the weather, hence the directed arrow from \textit{Sunny} to \textit{Sunscreen}, but I cannot change the weather by putting on some sunscreen (thus the lack of an arrow from \textit{Sunscreen} to \textit{Sunny}).
\end{example}

From a Bayesian network one can compute many things, but all from the first rung of the Ladder of Causation. In such models we only \textit{observe} the variables of the model, and we do not \textit{intervene}, which, as we will see further down this section, is the key to solve causality, a concept from the second rung of the Ladder of Causation.

In the previous definition we have talked about directed acyclic graphs, and to understand what they are we must introduce some concepts on Graph Theory. These will help us not only to deal with Bayesian networks but also to establish the foundations of causality.

\subsection{Basics of Graph Theory}

A very useful tool when talking about causes and effects are graphs, more precisely directed graphs. Not only they are really intuitive to draw, but they are surprisingly powerful as well, as we will see throughout these following sections. The fact that even a six-year-old could sketch a causal diagram (directed graph where arrows imply causality from one node to another) and draw some basic conclusions from it makes causality often a mere puzzle. To understand how these causal diagrams are useful we first need to go through some results on graph theory.

\begin{definition}{\bf (Graph)}  \label{def:graph}  
A \emph{graph} is an ordered pair $G = (\bm{V}, \bm{E})$, where $\bm{V}$ is a finite not-empty set of \textit{vertices} or \textit{nodes} and $\bm{E}\subset \bm{V}\times \bm{V}$ a set of \textit{edges} or \textit{links} that connect some pairs of vertices. Two nodes $X$ and $Y$ connected by an edge are called \textit{adjacent}, and we say that $X$ and $Y$ are \textit{neighbours}. When the edges are ordered, represented by $X\rightarrow Y := (X, Y)$, we have a \textit{directed graph}. If $X$ and $Y$ are two nodes connected by a directed edge from $X$ to $Y$ ($X\rightarrow Y$), we say $X$ is the \textit{parent} of $Y$ and $Y$ is a \textit{child} of $X$.
\end{definition}

We can also create new, smaller graphs by selecting only some nodes and edges of a bigger graph.

\begin{definition}{\bf (Subgraph and Induced Subgraph)}
Let $G = (\bm{V}, \bm{E})$, $G' = (\bm{V'}, \bm{E'})$ be two graphs such that $\bm{V'}\subset \bm{V}$ and ${\bm{E'}\subset \bm{E}\cap (\bm{V'}\times \bm{V'})}$. Then, $G'$ is a subgraph of $G$, and we denote it by $G'\subset G$. If the equality ${\bm{E'} = \bm{E}\cap(\bm{V'}\times\bm{V'})}$ holds, then $G'$ is the \textit{node induced subgraph} by the set $\bm{V'}$, and we write $G' = G[\bm{V'}]$.
\end{definition}

It is usually convenient to study not only links between two nodes, but also the \textit{paths} between two non-adjacent nodes that are further away.

\begin{definition}{\bf (Path)}
Let $G = (\bm{V}, \bm{E})$ be a directed graph and $u, v\in \bm{V}$ two vertices. Consider a sequence of vertices $p = \{u=X_1, X_2, \ldots, X_k, X_{k+1} = v\},\ X_{i}\in \bm{V}\ \forall i$, such that
\begin{enumerate}[\hspace*{0.3cm}(a)]
\item every pair of consecutive nodes is an edge, i.e., $(X_i, X_{i+1})\in \bm{E}$ or $(X_{i+1}, X_{i})\in \bm{E}$,
\item all edges joining consecutive nodes in $p$ are different, and
\item all vertices (except $u=X_1$ and $v = X_{k+1}$) are different.
\end{enumerate}
Then, $p$ is a \textit{path} from $u$ to $v$ (or from $v$ to $u$). A path that starts and ends at the same node ($u=v$) is called a \textit{cycle}. If any node in the path does not have two incoming or outgoing edges, $u$ has an outgoing edge and $v$ and incoming edge, we say it is a \textit{directed path from $u$ to $v$}. If a path is a directed path that starts and ends at the same node ($u=v$), then it is called a \textit{directed cycle}. A path from $u$ to $v$ is called a \textit{back-door} path if it contains an arrow into $u$.
\end{definition}

A notation that will be exhaustively used all through this project is the concept of parents and children of nodes in a directed graph.

\begin{definition}{\bf (Parents, Children, Ancestors and Descendants)}
Let $G = (\bm{V}, \bm{E})$ be a directed graph and $\bm{X}\subset \bm{V}$ a subset of nodes. Then, the \textit{parents} of $\bm{X}$, denoted by $\Pa(\bm{X})_{G}$, is the set consisting of the parents of every node in $\bm{X}$ while also containing $\bm{X}$. Analogously, the \textit{children} of $\bm{X}$, denoted by $\Ch(\bm{X})_{G}$, is the set consisting of the children of every node in $\bm{X}$ while also containing $\bm{X}$. A node $Y$ is an \textit{ancestor} of a node $X_i\in\bm{X}$ if there exists a directed path $p \subset G$ from $Y$ to $X_i$, and the set of all ancestors of $\bm{X}$ while also containing $\bm{X}$ is denoted by $\An(\bm{X})_{G}$. A node $Y$ is a \textit{descendant} of a node $X_i\in\bm{X}$ if there exists a directed path $p \subset G$ from $X_i$ to $Y$, and the set of all descendants of $\bm{X}$ while also containing $\bm{X}$ is denoted by $\De(\bm{X})_{G}$. When possible we will omit the subscript $G$ to ease comprehension.
\end{definition}

\begin{remark}
Given a directed graph $G = (\bm{V}, \bm{E})$ and $\bm{X}\subset \bm{V}$, it is clear that $\bm{X}\subset\Pa(\bm{X})\subset\An(\bm{X})$ and $\bm{X}\subset\Ch(\bm{X})\subset\De(\bm{X})$.
\end{remark}

Another useful set of vertices of a graph is the so-called \textit{root set}, composed by those vertices with no descendants other than themselves.

\begin{definition}{\bf (Root Set)}
Let $G = (\bm{V}, \bm{E})$ be a directed graph. Then the \textit{root set of }$G$ is the set of nodes with no descendants, $\Rt(G) = \{X\in \bm{V}\mid \De(X)_{G}\setminus X = \varnothing\}$.
\end{definition}

As we will see down this section, performing interventions will have the effect of erasing some directed edges of a graph. To this end we present the following notation.

\begin{notation}
Let $G = (\bm{V}, \bm{E})$ be a directed graph and $\bm{X}\subset \bm{V}$. Then we will denote by $G_{\xoverline{\bm{X}}}$ (resp. $G_{\xunderline{\bm{X}}}$) the graph obtained from $G$ by removing all incoming (resp. outgoing) edges of $\bm{X}$.
\end{notation}

There is a special family of directed graphs that turns out to be very handy when dealing with causal effects, presented below.

\begin{definition}{\bf (Directed Acyclic Graph)}
A graph that contains no cycles is called \textit{acyclic}. A directed graph which has no directed cycles is called \textit{directed acyclic graph (DAG)}. These last structures will be used throughout the work, since they are a fundamental part of causal theory.
\end{definition}

\begin{remark}
Every DAG $G$ has a non empty root set, $\Rt(G) \neq \varnothing$. Note that if $\Rt(G) = \varnothing$, then every node of $G$ would have at least one child, and since the set of vertices is finite, it would contain a cycle.
\end{remark}

\begin{definition}{\bf (Topological Ordering)}
Let $G = (\bm{V}, \bm{E})$ be a DAG. Then a \textit{topological ordering} $\pi$ of $G$ is an ordering of its nodes, where for all pair of nodes $X,Y\in\bm{V}$ with $X\neq Y$ one has $X>Y$ or $Y>X$ such that if $X$ is an ancestor of $Y$ in $G$, then $X<Y$.
\end{definition}

\begin{figure}[!ht]
\newsavebox{\largestimage}
\captionsetup[subfigure]{labelformat=empty}
\vspace*{-0.0cm}
\centering
\setlength{\mylength}{\textwidth}
\savebox{\largestimage}{\includegraphics{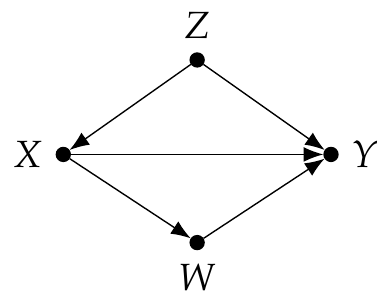}}%
\begin{subfigure}[t]{0.5\mylength}
        \centering
        \usebox{\largestimage}
        \caption{\footnotesize (a) Graph $G$.}
\end{subfigure}
\begin{subfigure}[t]{0.5\mylength}
        \centering
        \raisebox{\dimexpr.5\ht\largestimage-.5\height}{\includegraphics{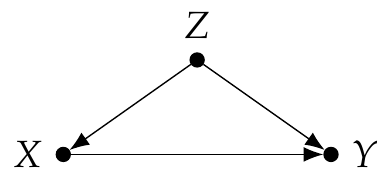}}
        \caption{\footnotesize (b) Induced subgraph $H = G[\{X, Y, Z\}]$.}
\end{subfigure}%
\vspace*{-0.0cm}
\caption{Examples of directed acyclic graphs (DAG).}
\label{fig:graph_1}
\vspace*{-0.0cm}
\end{figure}

\begin{example}
To take all these definitions in, consider the following example. In Figure \ref{fig:graph_1} (a) $G$ is a directed acyclic graph, DAG. One possible path  in $G$ could be $\{X, Y, Z\}$, which could also be represented as $X \rightarrow Y \leftarrow Z$, and a directed path could be $\{Z, X, W, Y\}$, ${Z \rightarrow X \rightarrow W \rightarrow Y}$. Since we will mostly work with directed graphs, from now on we will represent paths making use of the latter representation, i.e., specifying the directions. The parents of $X$ are $\Pa(X) = \{X, Z\}$, the ancestors of $W$ are $\An(W) = \{W, X, Z\}$, the descendants of $W$ are $\De(W) = \{W, Y\}$ and the root set of this graph is just $\Rt(G) = \{Y\}$. A topological ordering of $G$ would be $\{Z,X,W,Y\}$. Figure \ref{fig:graph_1} (b) represents graph $H$, which is the induced subgraph of $G$ by the set $\{X, Y, Z\}$.
\end{example}

\begin{definition}{\bf (Connected Graph)}
Let $G = (\bm{V}, \bm{E})$ be a directed graph. Then, $G$ is \textit{connected} if there exists a path $p\subset G$ between every pair of nodes $X, Y \in \bm{V}$.
\end{definition}

At this point we are able to consider an example of a Bayesian network and how it may answer some questions.

\begin{example}\label{exp:bayesian}
Let us retrieve the probabilistic graphical model in Example \ref{exp:probabilistic_graphical_model}. It is the same graph as $H$ in Figure \ref{fig:graph_1} (b), with renamed variables: $Z$ is the binary variable ``\textit{Summer/Not summer}'', $X$ the binary variable ``\textit{Sunny/Not sunny}'' and $Y$ the binary variable ``\textit{I wear sunscreen/I do not wear sunscreen}''. It is reasonable to think that the season of the year strongly affects the weather on a particular day ($Z\to X$), and also the probability of me wearing sunscreen ($Z\to Y$). Additionally, it is feasible to think that the decision of whether or not I must use sunscreen also depends on the weather of that particular day ($X\to Y$). The conditional probabilities between these random variables can be found in Table \ref{tab:table1}.

\begin{figure}[!ht]
\renewcommand\figurename{Table}
\captionsetup[subfigure]{labelformat=empty}
\vspace*{-0.0cm}
\centering
\setlength{\mylength}{\textwidth}
\savebox{\largestimage}{\includegraphics{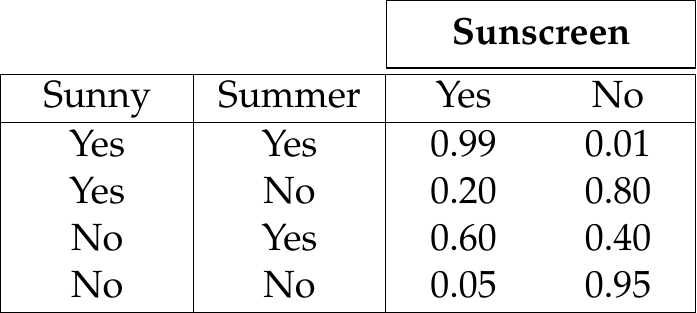}}%
\begin{subfigure}[t]{0.214\mylength}
        \centering
        \raisebox{0\height}{\includegraphics{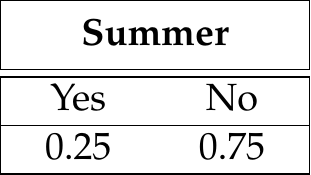}}
        \caption{\footnotesize}
\end{subfigure}
\begin{subfigure}[t]{0.32\mylength}
        \centering
        \raisebox{0\height}{\includegraphics{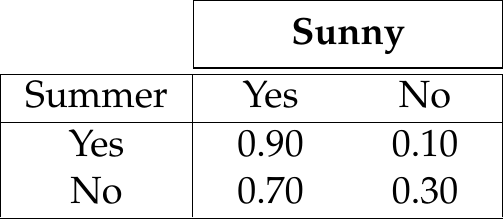}}
        \caption{\footnotesize}
\end{subfigure}%
\begin{subfigure}[t]{0.466\mylength}
        \centering
        \usebox{\largestimage}
        \caption{\footnotesize}
\end{subfigure}
\vspace*{-0.5cm}
\caption{Conditional probability tables of the different variables of the Bayesian network in example \ref{exp:bayesian}.}
\label{tab:table1}
\vspace*{-0.0cm}
\end{figure}
Now, decomposing $P(X, Y, Z)$ into $P(X, Y, Z) = P(Y|X, Z)P(X|Z)P(Z)$, we can use the given conditional probabilities to compute, for example, what is the probability of being summer knowing I do wear sunscreen:

\begin{equation*}
\begin{gathered}
P(Z=\text{T}|Y=\text{T}) = \frac{P(Y=\text{T}, Z=\text{T})}{P(Y=\text{T})} = \frac{\sum_{x\in\{\text{T}, \text{F}\}}P(X=x, Y=\text{T}, Z=\text{T})}{\sum_{x, z\in\{\text{T}, \text{F}\}}P(X=x, Y=\text{T}, Z=z)}=\\
\frac{0.99\cdot0.90\cdot0.25 + 0.60\cdot0.10\cdot0.25}{0.99\cdot0.90\cdot0.25 + 0.60\cdot0.10\cdot0.25 + 0.20\cdot0.70\cdot0.75 + 0.05\cdot0.30\cdot0.75}=\frac{0.23775}{0.354} \approx 0.67.
\end{gathered}
\vspace*{0.3cm}
\end{equation*}

Therefore, the probability that it is summer knowing I wear sunscreen is around 67\%, according to the conditional probabilities shown in Table \ref{tab:table1} and the dependencies encoded in the Bayesian network in Figure \ref{fig:graph_1} (b).
\end{example}

Bayesian networks are a powerful tool to compute probabilities from a dependency graph of variables, but they do not shed light on the problem of causality. To understand how we can tackle such a puzzle we must introduce more results, starting with causal models.

\subsection{Causal Models}

Causal models are a mathematical representation that will help us solve how a first action causes a second, and are the central construction of the algorithms that will be analysed in the following section. But before formally defining what a causal model is, consider this plausible made-up example.

\begin{example}\label{exp:gender}
Suppose we want to know how the salary of an employee in a given company depends on the gender of the worker. It is reasonable to suppose that salary ($Y$) could depend on the level of education ($E$) of the employee (better academic records usually translate to higher payroll), the field the employee is working in ($F$) and the amount of time he has been working for the company ($S$ for seniority). Clearly, the age ($A$) of the worker influences both the level of education and the seniority, and gender ($X$) may have an impact on the seniority as well as the field of the employee. In addition, we might think that both age and gender may be related through a third unobserved variable ($U$), which could be that perhaps the company used to hire only men in the past, and thus the average age of men is higher than that of women in the company.

The so-called \textit{exogenous variables} would be the unobserved, unmeasurable variables, $U$ in this example, and the variables $Y, E, F, S, A, X$ are called \textit{endogenous variables}. These variables, together with the relations of dependence described above, form what is known as \textit{causal model}.
\end{example}
\pagebreak
\begin{definition}{\bf (Causal Model)}
A \textit{causal model} is a triple $M = (\bm{U}, \bm{V}, \bm{F})$, where:
\begin{enumerate}[\hspace*{0.3cm}(a)]
\item $\bm{U}$ is a set of background random variables, called exogenous variables, determined by factors from outside the model.
\item $\bm{V} = \{V_1,\ldots, V_n\}$ is a set of random variables, called endogenous variables, that are determined by variables in the model, i.e., by variables in $\bm{U}\cup\bm{V}$.
\item $\bm{F} = \{f_{i},\ldots, f_{n}\}$ is a set of functions such that for every ${V_i\in\bm{V}}$, there is a mapping ${f_{i}:\bm{S_i}\cup\{U_{V_i}\}\to V_i}$, and such that the whole set $\bm{F}$ forms a mapping from $\bm{U}$ to $\bm{V}$. That is, for every ${V_i\in\bm{V}}$ there is a mapping (named structural equation) $f_i\in\bm{F}$ such that
\begin{equation*}
V_i = f_i(\bm{S_i}, U_{V_i}),\ i\in\{1,\ldots,n\}\ ,
\end{equation*}
where $U_{V_i}\in\bm{U}$ is the error term linked to $V_i$, and $\bm{S_i}\subset (\bm{U}\cup\bm{V})\setminus\{V_i, U_{V_i}\}$, known as the \textit{parent set}.
\end{enumerate}
\end{definition}

It is important to emphasise the concept of exogenous variables. Those variables can affect endogenous variables, which are measurable in our model, but we cannot see nor measure said exogenous variables. They encompass all kinds of unmeasurable perturbations, including the small deviations due to error terms or noise.

Every causal model $M$ has its corresponding directed acyclic graph $G = (\bm{W}, \bm{E})$, where the node set $\bm{W} = \bm{U}\cup\bm{V}$ contains a node for each observed (endogenous, $\bm{V}$) and unobserved (exogenous, $\bm{U}$) variable of the model. We usually ignore the unobservable vertices $U_{V_i}$ correspondent to the error term of measurable variables, for we know that they are always there and are implicitly taken into account in the model. Then, the set of edges $\bm{E}$ is determined by the functional relationships between the variables in the model, meaning that $\bm{E}$ contains an edge coming into $V_i$ from every node required to uniquely define $f_i$. This graph $G$ is known as the \textit{causal diagram} of $M$.

\begin{example}
The DAG induced by the the causal model in Example \ref{exp:gender} can be seen in Figure \ref{fig:graph_9}.
\begin{figure}[!ht]
\captionsetup[subfigure]{labelformat=empty}
\vspace*{-0.0cm}
\centering
\includegraphics{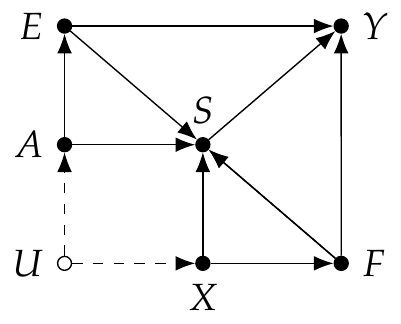}
\vspace*{-0.0cm}
\caption{DAG $G$ encodes the relations between variables of Example \ref{exp:gender}.}
\label{fig:graph_9}
\vspace*{-0.0cm}
\end{figure}
\end{example}

\begin{definition}{\bf (Probabilistic Causal Model)}
A \textit{probabilistic causal model} is a tuple $M = (\bm{U}, \bm{V}, \bm{F}, P(\bm{U}))$, where $(\bm{U}, \bm{V}, \bm{F})$ is a causal model and $P(\bm{U})$ is the joint distribution of the variables in $\bm{U}$. The distribution on $\bm{V}$ induced by $P(\bm{U})$ and $\bm{F}$ will be denoted $P(\bm{V})$.
\end{definition}

\begin{definition}{\bf (Semi-Markovian Causal Model)}
Given a causal model $M$, if every unobserved node is a parent of exactly two observed nodes, then $M$ is called a \textit{semi-Markovian causal model}.
\end{definition}

We are going to focus only on semi-Markovian causal models, since by the result proved by Verma \cite{Ver}, any causal model with unobserved variables can be redesigned into a semi-Markovian causal model while conserving all dependencies between variables.

\begin{figure}[!ht]
\captionsetup[subfigure]{labelformat=empty}
\vspace*{-0.0cm}
\centering
\setlength{\mylength}{\textwidth}
\savebox{\largestimage}{\includegraphics{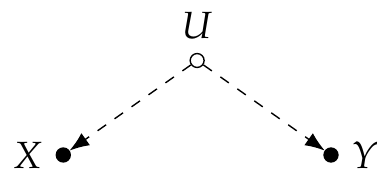}}%
\begin{subfigure}[t]{0.5\mylength}
        \centering
        \usebox{\largestimage}
        \caption{\footnotesize (a)}
\end{subfigure}
\begin{subfigure}[t]{0.5\mylength}
        \centering
        \raisebox{\dimexpr.5\ht\largestimage-.5\height}{\includegraphics{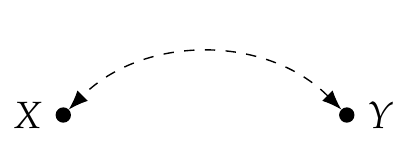}}
        \caption{\footnotesize (b)}
\end{subfigure}%
\vspace*{-0.0cm}
\caption{(a) $U$ is a confounder of nodes $X$ and $Y$. (b) We will usually denote confounded nodes with a dashed bidirected edge.}
\label{fig:graph_3}
\vspace*{-0.0cm}
\end{figure}
Figure \ref{fig:graph_3} (a) shows how unobserved nodes in semi-Markovian causal models are often represented. In this example, node $U$ would be an exogenous variable with two children, $X$ and $Y$. Despite exogenous variables can be explicitly drawn in the causal, we will usually omit unobserved variables, since we cannot measure nor control them. We will represent their effect with a dashed bidirected edge between nodes $X$ and $Y$, which corresponds to the effect of an unobserved confounding variable, this is, a hidden common cause. Note that this bidirected edge is not equivalent to two directed edges $X\to Y$ and $X \leftarrow Y$, as this would form a cycle in the graph which is not allowed in DAGs.

\begin{definition}{\bf ($d$-separation)}{\label{def:d_separation}}
Let $G = (\bm{V}, \bm{E})$ be a DAG, $p$ a path in $G$ and $\bm{Z}$ a set of nodes $\bm{Z}\subset \bm{V}$. Then, the path $p$ is \textit{d-separated} by the set $\bm{Z}$ in $G$ if and only if either
\begin{enumerate}[\hspace*{0.3cm}(a)]
\item $p$ contains a chain $I\to M \to J$ or a fork $I\leftarrow M \to J$, such that $M\in\bm{Z}$ and $I,J\in\bm{V}$, or
\item $p$ contains an inverted fork or collider $I \to M \leftarrow J$, such that $\De(M)_{G}\cap\bm{Z}=\varnothing$.
\end{enumerate}
Two disjoint sets $\bm{X}$ and $\bm{Y}$ are \textit{d-separated} by $\bm{Z}$ in $G$ if all paths from $\bm{X}$ to $\bm{Y}$ are $d$-separated by $\bm{Z}$ in $G$. A path that is not $d$-separated is said to be \textit{d-connected}.
\end{definition}

The following is an important result proved by Verma and Pearl \cite{verma_pearl}, although we present the clearer and more succinct version proposed by Shpitser and Pearl in \cite{SP_2006a}.
\begin{theorem}{\bf ({\rm Theorem 1 in  \cite{SP_2006a}})}\label{thm:independence_d_separation}
Let $M$ be a causal model with the corresponding DAG $G = (\bm{V}, \bm{E})$, and $\bm{X}, \bm{Y}, \bm{Z}\subset \bm{V}$ be sets of variables or nodes in $G$. If $\bm{X}$ and $\bm{Y}$ are $d$-separated by $\bm{Z}$, then $\bm{X}$ is independent of $\bm{Y}$ given $\bm{Z}$ in $G$, i.e., $(\bm{X}\indep\bm{Y}|\bm{Z})_{G}$.
\end{theorem}

\begin{example}{\label{exp_d_separation}}
Consider the DAG $G$ shown in Figure \ref{fig:graph_4}, from \cite[p.p. 17-18]{pearl_causality}. There are two different paths between $X$ and $Y$ in $G$, the one that uses the bidirected arc, $X\to Z_1 \dashbidirectedarrow Z_3 \leftarrow Y$, and $X\to Z_1 \leftarrow Z_2\leftarrow Z_3 \leftarrow Y$.
\begin{figure}[!ht]
\vspace*{-0.0cm}
\centering
\includegraphics{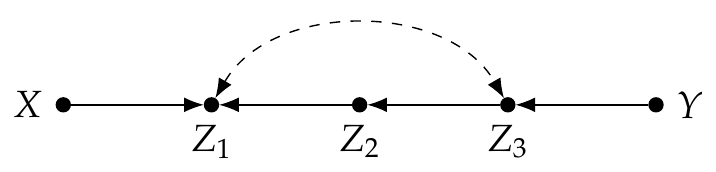}
\vspace*{-0.0cm}
\caption{DAG $G$ showing $d$-separation in Example \ref{exp_d_separation}.}
\label{fig:graph_4}
\vspace*{-0.0cm}
\end{figure}
Note that, without measuring $\{Z_1, Z_2, Z_3\}$, $X$ and $Y$ are $d$-separated, since both paths have a collider ($X\to Z_1 \leftarrow U$, where $U$ is a confounder of $Z_1$ and $Z_3$, and $X\to Z_1 \leftarrow Z_2$). Nevertheless, when $Z_1$ is measured the path $X\to Z_1 \dashbidirectedarrow Z_3 \leftarrow Y$ becomes unblocked. This is so because measuring $\bm{Z} = \{Z_1\}$ unblocks both colliders at $Z_1$ and $Z_3$: at $X\to Z_1 \leftarrow U$ we have $\De(Z_1)\cap\{Z_1\} = \{Z_1\} \neq\varnothing$, and at $U\to Z_3 \leftarrow Y$ we have $\De(Z_3)\cap\{Z_1\} = \{Z_1, Z_2, Z_3\}\cap\{Z_1\} \neq\varnothing$, and thus none of the conditions in Definition \ref{def:d_separation} are met, making $X$ and $Y$ $d$-connected given $Z_1$.
\end{example}

\subsubsection{Causal Effects, \textit{do}-Calculus and Identifiability}

\begin{figure}[!ht]
\captionsetup[subfigure]{labelformat=empty}
\vspace*{-0.0cm}
\centering
\setlength{\mylength}{\textwidth}
\savebox{\largestimage}{\includegraphics{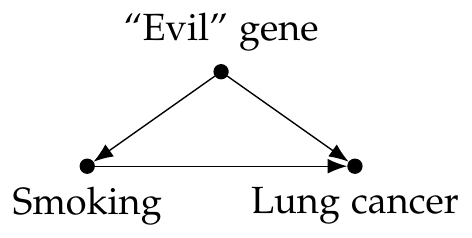}}%
\begin{subfigure}[t]{0.5\mylength}
        \centering
        \usebox{\largestimage}
        \caption{\footnotesize (a)}
\end{subfigure}
\begin{subfigure}[t]{0.5\mylength}
        \centering
        \raisebox{\dimexpr.5\ht\largestimage-.5\height}{\includegraphics{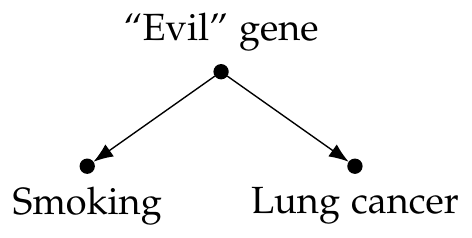}}
        \caption{\footnotesize (b)}
\end{subfigure}%
\vspace*{-0.0cm}
\caption{(a) Causal diagram where Smoking is causally related to Lung cancer, while confounded by an ``Evil'' gene. (b) Causal diagram where Lung cancer is not a causal effect of Smoking.}
\label{fig:graph_5}
\vspace*{-0.0cm}
\end{figure}

Does smoking cigarettes increase the likelihood of developing lung cancer? This apparently obvious question was not that obvious sixty years ago before the foundations of causality were established. A typical argument against the claim that smoking caused lung cancer was that it could be an unknown ``evil'' gene confounding the variables ``Smoking'' and ``Lung cancer'', fully accounting for the correlation between both variables. Such a gene would make an individual crave nicotine (and thus consume more tobacco) while at the same time would make them prone to developing lung cancer (see Figure \ref{fig:graph_5} (b)).

A randomized controlled trial (RTC) was (and it still is) usually performed in similar problems (for example, in drug testing) to get rid of confounders or other sources of bias. This experiment consists of separating the subjects into two or more groups \textit{randomly} and treating them differently, so one can be sure that the differences between them after the experiment are only a product of the different treatments given. The key point in these trials is the random selection process, which eliminates (or at least reduces) the biases of known and unknown factors. In our situation performing an RTC would not be possible: if both groups were separated by mere observation and not randomness it would not be an RTC because the ``evil'' gene would be present on most of the smokers, thus inducing a higher probability of cancer in that group. On the other hand, creating the groups at random and forcing people to smoke for twenty years to see their evolution and harming their health in the process is greatly unethical.

We have already stated that to solve problems from the second level of the Ladder of Causation (such as causal effects queries) we must fix, act on some variables to remove confounder effects. So the point now is, how can we intervene in an experiment without actually physically doing so? The so-called \textit{do-operator} can solve this issue.

\begin{definition}{\bf (\textit{do}-operator)}
Let $M = (\bm{U}, \bm{V}, \bm{F}, P(\bm{U}))$ be a probabilistic causal model, and $\bm{X}$ a set of variables of the model. Then the \textit{do-operator} is the intervention that sets the values of $\bm{X}$ to $\bm{x}$, and it is denoted by $do(\bm{X}=\bm{x})$. This is, every action $do(\bm{X}=\bm{x})$ on $M$ produces a new model $M_{\bm{x}} = (\bm{U}, \bm{V}, \bm{F_{\bm{x}}}, P(\bm{U}))$, where $\bm{F_{\bm{x}}}$ is obtained by, for every $X\in\bm{X}$, replacing $f_{X}\in\bm{F}$ with a new constant function of value $x$ given by $do(\bm{X}=\bm{x})$.
\end{definition}

\begin{remark}
Despite being apparently similar, we must not misinterpret intervening for conditioning over a variable. $P(Y=y|X=x)$ is the probability that $Y=y$ conditional on finding $X=x$, in other words, we are considering the distribution of $Y$ among individuals whose $X$ value is $x$. $P(Y=y|do(X=x))$, however, is the probability that $Y=y$ when we intervene to make $X=x$, namely, we are now considering the distribution of $Y$ if every individual in the population had their $X$ value fixed at $x$. 
\end{remark}

\begin{definition}{\bf (Causal Effect)}
Let $M = (\bm{U}, \bm{V}, \bm{F}, P(\bm{U}))$ be a probabilistic causal model and $\bm{X}, \bm{Y}\subset\bm{V}$. Then the \textit{causal effect} of $do(\bm{X}=\bm{x})$ on $\bm{Y}$ in $M$ is the marginal distribution of $\bm{Y}$ in $M_{\bm{x}}$, noted by $P(\bm{Y}|do(\bm{X}=\bm{x})) = P_{\bm{x}}(\bm{Y})$.
\end{definition}

\begin{remark}
For every intervention $do(\bm{X}=\bm{x})$, to ensure that $P_{\bm{x}}(\bm{V})$ and its marginals are well defined, is required that $P(\bm{x}|\Pa(\bm{X})_{G}\setminus\bm{X})>0$. It is not possible to force $\bm{X}$ to have values not observed in the data.
\end{remark}

A special case of causal effects are direct effects, where the intervened variables are the parents of the studied variable.

\begin{definition}{\bf (Direct Effect)}
Given a probabilistic causal model $M$ with variables $\bm{V}$ and $\bm{Y}\subset\bm{V}$, a \textit{direct effect} is a causal effect of the form ${P(\bm{Y}|do(\Pa(\bm{Y})\setminus\bm{Y}=\bm{y'}))}$, this is, when the parents of the variables are the ones intervened.
\end{definition}

The ultimate goal of solving a causal effect problem of some variables $\bm{X}$ over $\bm{Y}$ given a causal model $M$ has therefore been reduced to finding the probability $P_{\bm{x}}(\bm{Y})$. Nevertheless, how can we obtain a value for this probability given that, most of the time, we will only have access to data from observational studies (information of the first rung of the Ladder of Causation)? In 1993 the computer scientist and philosopher Judea Pearl \cite{pearl_back_door} showed that, when some conditions are fulfilled, the intervening probability can be computed from just observational data, making use of the so-called \textit{back-door criterion}.

\begin{definition}{\bf (Back-Door Criterion)}
Let $M = (\bm{U}, \bm{V}, \bm{F}, P(\bm{U}))$ be a probabilistic causal model with DAG $G$ and $\bm{Z}\subset \bm{V}$, $X_i, X_j\in\bm{V}$ with $X_i \neq X_j$. Then, we say that $\bm{Z}$ satisfies the \textit{back-door criterion} relative to $(X_i, X_j)$ if:
\begin{enumerate}[\hspace*{0.3cm} (a)]
\item $\bm{Z}\cap\De(X_i) = \varnothing$, and
\item $\bm{Z}$ blocks every path between $X_i$ and $X_j$ that contains an arrow into $X_i$ (back-door path).
\end{enumerate}
More generally, if $\bm{X}, \bm{Y}\subset\bm{V}$ with $\bm{X} \cap \bm{Y} = \varnothing$, then $\bm{Z}$ satisfies the \textit{back-door criterion} relative to $(\bm{X}, \bm{Y})$ if it satisfies the criterion for every pair $(X_i, X_j)\in\bm{X} \times \bm{Y}$.
\end{definition}

If such a condition is fulfilled, then the next result follows.

\begin{theorem}{\bf (Back-Door Adjustment {\rm\cite{pearl_back_door}})}
If a set of variables $\bm{Z}$ satisfies the back-door criterion relative to $(\bm{X}, \bm{Y})$, then the causal effect of $X$ on $Y$ is given by
\begin{equation*}
P_{\bm{x}}(\bm{y}) = \sum_{z}P(\bm{y}|\bm{x}, z)P(z).
\end{equation*}
\end{theorem}

Although this criterion was a big step in the right direction, it couldn't be applied to every scenario, so more adjustments like the previous were required to account for all possible causal diagrams. This dream of his of obtaining causal information such as $P_{\bm{x}}(\bm{Y})$ from observational data was finally a reality thanks to the development of \textit{do-calculus} by Pearl and some of his colleagues in 1995 \cite{pearl_do_rules}. His theory is constructed on three simple, yet powerful rules that allow the removal of the $do-$operator under some specific scenarios, thus letting one travel from unmeasurable probabilities involving $do$ expressions to observational, standard probabilities. These rules were first proven by Pearl himself in \cite{pearl_do_rules}.

\begin{theorem}{\bf (Rules of $do$-Calculus {\rm\cite[Theorem 3.4.1]{pearl_causality}})}
Let $M = (\bm{U}, \bm{V}, \bm{F}, P(\bm{U}))$ be a probabilistic causal model and $G$ its associated DAG. For any pairwise disjoint subsets of nodes $\bm{X}, \bm{Y}, \bm{Z}\subset\bm{V}$, the following rules apply:
\begin{enumerate}[\hspace*{0.3cm} (a)]

\item[] \textbf{Rule 1} (Insertion and deletion of observations)
\begin{equation*}
P_{\bm{x}}(\bm{y}|\bm{z}, \bm{w}) = P_{\bm{x}}(\bm{y}|\bm{w}),\quad \text{if}\quad (\bm{Y}\indep\bm{Z}|\bm{X}, \bm{W})_{G_{\mbox{\tiny\xoverline{\bm{X}}}}}
\end{equation*}

\item[] \textbf{Rule 2} (Exchanging actions and observations)
\begin{equation*}
P_{\bm{x}, \bm{z}}(\bm{y}|\bm{w}) = P_{\bm{x}}(\bm{y}|\bm{z}, \bm{w}),\quad \text{if}\quad (\bm{Y}\indep\bm{Z}|\bm{X}, \bm{W})_{G_{\mbox{\tiny\xoverline{\bm{X}}, \xunderline{\bm{Z}}}}}
\end{equation*}

\item[] \textbf{Rule 3} (Insertion and deletion of actions)
\begin{equation*}
P_{\bm{x}, \bm{z}}(\bm{y}|\bm{w}) = P_{\bm{x}}(\bm{y}|\bm{w}),\quad \text{if}\quad (\bm{Y}\indep\bm{Z}|\bm{X}, \bm{W})_{G_{\mbox{\tiny\xoverline{\bm{X}}, \xoverline{Z(\bm{W})}}}}
\end{equation*}
where $Z(\bm{W}) = \bm{Z}\setminus\An{(\bm{W})}_{G_{\mbox{\tiny\xoverline{\bm{X}}}}}$, i.e., the set of nodes in $\bm{Z}$ that are not ancestors of any node in $\bm{W}$ in $G_{\xoverline{\bm{X}}}$.
\end{enumerate}
\end{theorem}

\begin{remark}
Rule 1 of insertion and deletion of observations is a generalization of $d$-separa\-tion (Theorem \ref{thm:independence_d_separation}) in a graph with interventions (that is why independence in $G_{\xoverline{\bm{X}}}$ is required). Rule 2 of exchanging actions and observations is a generalization of the back-door criterion: the only paths in DAG $G_{\xoverline{\bm{X}}, \xunderline{\bm{Z}}}$ between $\bm{Z}$ and $\bm{Y}$ are back-door paths, and if we block those paths conditioning over $\bm{X}$ and $\bm{W}$ we can swap the intervention for the observation. Rule 3 provides conditions for introducing or deleting other interventions.
\end{remark}

These rules have proven to be enough to compute interventional probabilities whenever these probabilities can be identified. Informally speaking, when it is possible to compute the interventional term of a distribution from just observational data, we say that the effect is \textit{identifiable}.

\begin{theorem}{\bf ($do$-Calculus Completeness {\rm\cite[Theorem 7]{SP_2006a}})}
The three rules of $do$-calculus, together with standard probability manipulations, are complete for determining identifiability of all effects of the form $P_{\bm{x}}(\bm{Y})$.
\end{theorem}

So we know that, if the effect is identifiable, using only the three rules of $do$-calculus we can obtain an algebraic expression for $P_{\bm{x}}(\bm{Y})$ which does not involve the $do$-operator, meaning that we can use only observable data to infer causal conclusions. But what does it really mean for an effect to be \textit{identifiable}?

\begin{definition}{\bf (Causal Effect Identifiability)}
Let $M = (\bm{U}, \bm{V}, \bm{F}, P(\bm{U}))$ be a probabilistic causal model with DAG $G$, and $\bm{X}, \bm{Y}\subset\bm{V}$. The causal effect of an action $do(\bm{x})$ on $\bm{Y}$ such that $\bm{X}\cap\bm{Y} = \varnothing$ is said to be \textit{identifiable} from $P$ in $G$ if $P_{\bm{x}}(\bm{Y})$ is uniquely computable from $P(\bm{V})$ in any causal model which induces $G$. This is, if for every pair of causal models $M^{1}$ and $M^{2}$ such that $P^{1}(\bm{V}) = P^{2}(\bm{V})$, the causal effect coincides, $P^{1}_{\bm{x}}(\bm{Y}) = P^{2}_{\bm{x}}(\bm{Y})$.
\end{definition}

In the following example we will use $do$-calculus to determine the causal effect of a very similar causal diagram to Figure \ref{fig:graph_5} (a).

\begin{example}\label{exp:do_calculus}
Consider the causal diagram $G$ shown in Figure \ref{fig:graph_6} (a), where the variables $S$, $T$ and $C$ stand for Smoking, Tar and Cancer. This causal diagram is a slight variation of the one shown in Figure \ref{fig:graph_5} (a). Here, the main modification of the model is that we are supposing that lung cancer only develops through tar deposited in the lungs (and not from smoking directly), which in turn is \textit{only} produced by smoking. In this scenario, Tar is called a \textit{mediator} between $S$ and $C$, because it is the variable that explains the causal effect of $S$ on $C$.

\begin{figure}[!ht]
\captionsetup[subfigure]{labelformat=empty}
\vspace*{-0.0cm}
\centering
\setlength{\mylength}{\textwidth}
\savebox{\largestimage}{\includegraphics{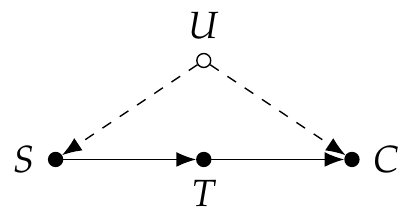}}%
\begin{subfigure}[t]{0.5\mylength}
        \centering
        \raisebox{\dimexpr.5\ht\largestimage-.5\height}{\includegraphics{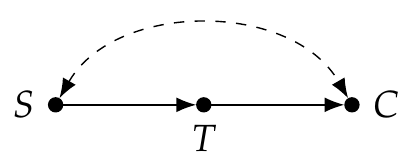}}
        \caption{\footnotesize (a)}
\end{subfigure}
\begin{subfigure}[t]{0.5\mylength}
        \centering
        \usebox{\largestimage}
        \caption{\footnotesize (b)}
\end{subfigure}%
\vspace*{-0.0cm}
\caption{(a) Causal diagram where Smoking is causally related to Lung cancer through Tar, and Smoking and Cancer confounded. (b) Same causal diagram but we explicit the unobserved confounder for ease when applying $do$-calculus.}
\label{fig:graph_6}
\vspace*{-0.0cm}
\end{figure}

Our goal is to determine the causal effect $P_{\bm{s}}(c)$ from Figure \ref{fig:graph_6}. First of all, we use the Law of Total Probability, followed by some conditional dependence manipulations (similar to Example \ref{exp:probability_axioms}):

\begin{equation*}
P_{\bm{s}}(c) = P(c|do(s)) = \sum_{t}P(c, t|do(s)) = \sum_{t}P(c|t, do(s))P(t|do(s))
\end{equation*}

Then we apply rule 2 of do calculus, exchanging the $t$ for $do(t)$ in the first term, since $(C\indep T|S)_{G_{\mbox{\tiny\xoverline{S}, \xunderline{T}}}}$ (see Figure \ref{fig:graph_7} (a)):

\begin{equation*}
\sum_{t}P(c|t, do(s))P(t|do(s)) = \sum_{t}P(c|do(t), do(s))P(t|do(s))
\end{equation*}

Applying rule 2 again we can change $do(s)$ for $s$ in the second term, because $(T\indep S)_{G_{\mbox{\tiny\xunderline{S}}}}$ (see Figure \ref{fig:graph_7} (b)). The fact that $(T\indep S)_{G_{\mbox{\tiny\xunderline{S}}}}$ follows from Theorem \ref{thm:independence_d_separation}, since $T$ and $S$ are $d$-separated in $G_{\xunderline{S}}$ by the collider $C$:

\begin{equation*}
\sum_{t}P(c|do(t), do(s))P(t|do(s)) = \sum_{t}P(c|do(t), do(s))P(t|s)
\end{equation*}

We now see that $(C\indep S|T)_{G_{\mbox{\tiny\xoverline{T}, \xoverline{S}}}}$ (see Figure \ref{fig:graph_7} (c)), so we can apply rule 3 by deleting the intervention $do(s)$ from the first term:

\begin{equation*}
\sum_{t}P(c|do(t), do(s))P(t|s) = \sum_{t}P(c|do(t))P(t|s)
\end{equation*}

Using some probability axioms as in the first step, we can write:

\begin{equation*}
\sum_{t}P(c|do(t))P(t|s) = \sum_{s'}\sum_{t}P(c, s'|do(t))P(t|s) = \sum_{s'}\sum_{t}P(c|do(t), s')P(s'|do(t))P(t|s)
\end{equation*}

Using Theorem \ref{thm:independence_d_separation} once again we see that $(C\indep T|S)_{G_{\mbox{\tiny\xunderline{T}}}}$ (see Figure \ref{fig:graph_7} (d)), because the chain $U \rightarrow S \rightarrow T$ while conditioning on $S$ $d$-separates the only path between $C$ and $T$. So using rule 2 once more we can replace $do(t)$ for $t$ in the first term:

\begin{equation*}
\sum_{s'}\sum_{t}P(c|do(t), s')P(s'|do(t))P(t|s) = \sum_{s'}\sum_{t}P(c|t, s')P(s'|do(t))P(t|s)
\end{equation*}

Finally we are able to delete the only $do$-expression in the second term, $do(t)$, by using rule 3, because $G_{\xunderline{S}} = G_{\xoverline{T}}$ and we have seen previously that $(S\indep T)_{G_{\mbox{\tiny\xoverline{T}}}}$ (see Figure \ref{fig:graph_7} (b)):

\begin{equation*}
\sum_{s'}\sum_{t}P(c|t, s')P(s'|do(t))P(t|s) = \sum_{s'}\sum_{t}P(c|t, s')P(s')P(t|s) = \sum_{t}P(t|s)\left(\sum_{s'} P(c|t, s')P(s')\right).
\end{equation*}

\begin{figure}[!ht]
\captionsetup[subfigure]{labelformat=empty}
\vspace*{-0.0cm}
\centering
\setlength{\mylength}{\textwidth}
\begin{subfigure}[t]{0.25\mylength}
        \centering
        \includegraphics{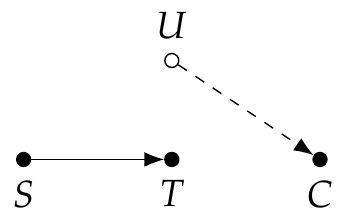}
        \caption{\footnotesize (a) $G_{\xoverline{S}, \xunderline{T}}$.}
\end{subfigure}
\begin{subfigure}[t]{0.25\mylength}
        \centering
        \includegraphics{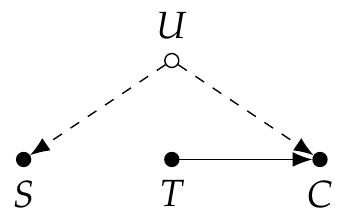}
        \caption{\footnotesize (b) $G_{\xunderline{S}} = G_{\xoverline{T}}$.}
\end{subfigure}%
\begin{subfigure}[t]{0.25\mylength}
        \centering
        \includegraphics{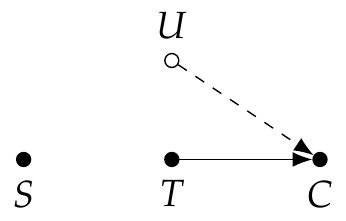}
        \caption{\footnotesize (c) $G_{\xoverline{T}, \xoverline{S}}$.}
\end{subfigure}%
\begin{subfigure}[t]{0.25\mylength}
        \centering
        \includegraphics{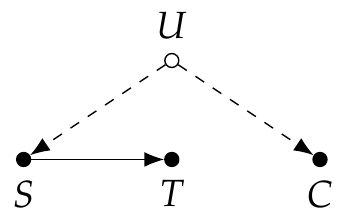}
        \caption{\footnotesize (d) $G_{\xunderline{T}}$.}
\end{subfigure}%

\vspace*{-0.0cm}
\caption{Different subgraphs used throughout Example \ref{exp:do_calculus}.}
\label{fig:graph_7}
\vspace*{-0.0cm}
\end{figure}

So we conclude that, in the causal diagram shown in Figure \ref{fig:graph_6}, the causal effect $P_{\bm{s}}(c)$ is identifiable and can be computed from observational data as

\begin{equation}\label{eq:example_do_calculus}
P_{\bm{s}}(c) = \sum_{t}P(t|s)\left(\sum_{s} P(c|t, s)P(s)\right).
\end{equation}

The causal diagram in Figure \ref{fig:graph_6} appears often in bigger causal diagrams, and the formula obtained to identify the causal effect in Equation \ref{eq:example_do_calculus} is known as the \textit{Front Door Adjustment}. The diagrams shown in Figure \ref{fig:graph_6} and the resolution applying $do$-calculus are from Pearl \cite[pp. 236]{pearl_why}.
\end{example}

Even though with the three rules of $do$-calculus we can solve any identifiable causal effect, we do not have a specific policy on the order we should use those rules, and more importantly, we do not know a priori if a certain causal effect is identifiable or not. If this is the case, we might be using the rules uselessly without being able to reach a successful $do$-free algebraic expression. To overcome this problem, Shpitser and Pearl devised an algorithm \cite{SP_2006a} that checks if a causal effect is identifiable, and if so, returns a $do$-free algebraic expression. This will be the main topic in the following section, and its implementation in Python the central goal of this project.

\subsubsection{Identifiability Criteria}

Before introducing the algorithm we should understand some criteria on identifiability, since they are the basis of said algorithm. The question we want to answer now is simple: how can we check if the causal effect of one variable on another is identifiable in a certain causal model? If this causal effect were not identifiable, we would not even bother to try to use $do$-calculus to compute it.

A useful but partial solution was found by Pearl for Markovian models, those which do not have bidirected edges (i.e., confounders).

\begin{theorem}{\bf (Identifiability of Markovian Models {\rm\cite[Corollary 3.2.6]{pearl_causality}})}
Given the causal diagram $G$ of any Markovian model (i.e., that do not contain bidirected edges) in which all variables are measured, all effects are identifiable.
\end{theorem}

But we do have confounders, so we need another approach for semi-Markovian causal models. What is essentially different between Markovian and semi-Markovian causal models is the appearance of bidirected edges, so it seems reasonable to study them in detail. To do so, we first need some definitions regarding properties of directed acyclic graphs, in particular, we will look at sets of nodes interconnected by bidirected paths, for they play an important role in identifiability.

\begin{definition}{\bf (C-component)}
Let $G = (\bm{V}, \bm{E})$ be a graph. If there exists a set $\bm{F}\subset \bm{E}$ which contains only bidirected edges and the graph $(\bm{V}, \bm{F})$ is connected, then $G$ is a \textit{C-component} (\textit{confounded component}).
\end{definition}

\begin{definition}{\bf (Maximal C-component)}
Let $G$ be a graph and $S = (\bm{V}, \bm{E})$ a C-component with $S\subset G$. Then $S$ is a \textit{maximal C-component} (with respect to $G$) if, for every bidirected path in $G$ containing at least one node of $\bm{V}$, that path is also a path in $S$.
\end{definition}

If $G$ is not a C-component, it can be uniquely partitioned into a set of graphs, each a maximal C-component with respect to $G$.

\begin{lemma}\label{lem:decomposition}
Every directed graph $G = (\bm{V}, \bm{E})$ can be decomposed into a unique set $C(G) = \{G[S_1],\ldots,G[S_k]\}$ of subgraphs such that every $G[S_i]$ $\forall i\in\{1, \ldots, k\}$ is a maximal C-component of $G$. 
\end{lemma}
\begin{proof}
Given two nodes $X, Y\in\bm{V}$, they belong to the same maximal C-component if and only if there exists a bidirected path between $X$ and $Y$, from the definition of maximal C-component. Therefore every maximal C-component is unique, hence the bidirected paths of $G$ define its maximal C-components.
\end{proof}

This decomposition will ultimately help us to reduce the identification problem into several smaller identification subproblems. A useful special case of C-components are C-trees, 
which are closely related to direct effects.

\begin{definition}{\bf (C-tree)}
Let $G = (\bm{V}, \bm{E})$ be a C-component such that every node has at most one child. If there exists a node $X\in\bm{V}$ such that $\An(X)_{G} = \bm{V}$, then $G$ is a $X$\textit{-rooted C-tree}.
\end{definition}

The following is just a generalization of a C-tree with multiple roots.

\begin{definition}{\bf (C-forest)}
Let $G = (\bm{V}, \bm{E})$ be a C-component such that every node has at most one child. Then, if $\bm{X} = \Rt(G)$, i.e., if $\bm{X}$ have no descendants, we say $G$ is a $\bm{X}$\textit{-rooted C-forest}.
\end{definition}

If a DAG contains a pair of different C-forests, under some conditions this pair of C-forests are called a \textit{hedge}, structures that play a fundamental part in identifiability.

\begin{definition}{\bf (Hedge)}
Let $G = (\bm{V}, \bm{E})$ be a directed graph, and $\bm{X}, \bm{Y} \subset \bm{V}$ disjoint sets of nodes, i.e., ${\bm{X}\cap\bm{Y}=\varnothing}$. Suppose that there exist two $\bm{R}$-rooted C-forests $F = (\bm{V}_{F}, \bm{E}_{F})$, $\widetilde{F} = (\bm{V}_{\widetilde{F}}, \bm{E}_{\widetilde{F}})$ such that ${\bm{X}\cap\bm{V}_{F}\neq\varnothing}$, ${\bm{X}\cap\bm{V}_{\widetilde{F}}=\varnothing}$, $\widetilde{F}\subseteq F$ and $\bm{R}\subset\An(\bm{Y})_{G_{\xoverline{\bm{X}}}}$. Then, $F$ and $\widetilde{F}$ form a \textit{hedge} for $P_{\bm{x}}(\bm{y})$ in $G$.
\end{definition}

\begin{figure}[!ht]
\captionsetup[subfigure]{labelformat=empty}
\vspace*{-0.0cm}
\centering
\setlength{\mylength}{\textwidth}
\savebox{\largestimage}{\includegraphics{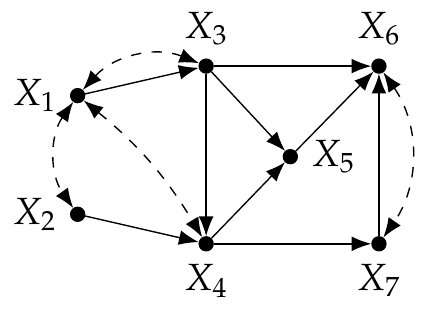}}%
\begin{subfigure}[t]{0.333\mylength}
        \centering
        \usebox{\largestimage}
        \caption{\footnotesize (a) Graph $G$.}
\end{subfigure}
\begin{subfigure}[t]{0.333\mylength}
        \centering
        \raisebox{\dimexpr.5\ht\largestimage-.5\height}{\includegraphics{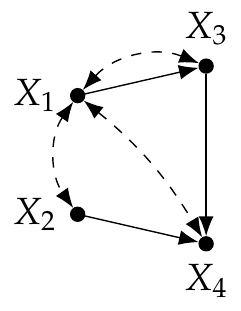}}
        \caption{\footnotesize (b) Maximal C-component $G[\bm{S}_1]\subset G$.}
\end{subfigure}%
\begin{subfigure}[t]{0.333\mylength}
        \centering
        \raisebox{\dimexpr.5\ht\largestimage-.5\height}{\includegraphics{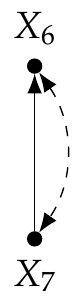}}
        \caption{\footnotesize (c) Maximal C-component $G[\bm{S}_2]\subset G$.}
\end{subfigure}%
\vspace*{-0.0cm}
\caption{Causal diagram $G$ contains three maximal C-components, $G[\bm{S}_1]$, $G[\bm{S}_2]$ and $G[\bm{S}_3]$, that are also C-trees.}
\label{fig:graph_8}
\vspace*{-0.0cm}
\end{figure}

\begin{example}\label{exp:c-components}
Figure \ref{fig:graph_8} (a) shows a causal diagram $G$ of a probabilistic causal model $M$. It can easily be seen that $G$ is not a C-component, because its bidirected edges do not connect all vertices in $G$. Nevertheless, $G$ can be decomposed into three maximal C-components, $G[\bm{S}_1]$ and $G[\bm{S}_2]$, as seen in Figures \ref{fig:graph_8} (b) and (c), and also the trivial component $G[\bm{S}_3] = \{X_5\}$, which is the graph containing $X_5$ with no edges. These three C-components are, in turn, also C-trees, because every node in $G[\bm{S}_1]$, $G[\bm{S}_2]$ and $G[\bm{S}_3]$ has at most one child: $G[\bm{S}_1]$ is a $X_4$-rooted C-tree, since $\An(X_4)_{G[\bm{S}_1]} = \{X_1, X_2, X_3, X_4\}$, $G[\bm{S}_2]$ is a $X_6$-rooted C-tree, for $\An(X_6)_{G[\bm{S}_2]} = \{X_6, X_7\}$, and finally $G[\bm{S}_3]$ is a $X_5$-rooted C-tree.

We can also detect a hedge for $P_{x_1}(x_6)$ in $G$. We can easily see that $G[\bm{S}_1]$ is a $\{X_4\}$-rooted C-forest with $\{X_1\}\cap\bm{S}_1\neq\varnothing$ and $\{X_4\}\subset \An(X_6)_{G_{\mbox{\tiny\xoverline{X_1}}}}$. Now consider $F$, the $\{X_4\}$-rooted C-forest formed only by the vertex $X_4$ and no edges. It is clear that $\{X_1\}\cap\bm{V}_{F} = \{X_1\}\cap\{X_4\}=\varnothing$, and that $F\subseteq \bm{S}_1$. Therefore, $G[\bm{S}_1]$ and $F$ form a hedge for $P_{x_1}(x_6)$ in $G$.
\end{example}

The following result is finally what we were searching for in this section, a criterion to check if a causal effect is identifiable.

\begin{theorem}{\bf (Hedge Identifiability Criterion {\rm\cite[Theorem 4]{SP_2006a}})}
Let $G$ be the causal diagram of a model $M = (\bm{U}, \bm{V}, \bm{F}, P(\bm{U}))$, and $\bm{X}, \bm{Y}\subset\bm{V}$. Then the causal effect $P_{\bm{x}}(\bm{y})$ is not identifiable in $G$ if and only if there exist two $\bm{R}$-rooted C-forests $F$ and $\widetilde{F}$ that form a hedge for $P_{\bm{x}}(\bm{y})$ in $G$.
\end{theorem}

Knowing this we can state that the causal effect $P_{x_1}(x_6)$ is not identifiable in Figure \ref{fig:graph_8} (a), since we have seen in Example \ref{exp:c-components} that there is a hedge for $P_{x_1}(x_6)$.

\begin{remark}
In Example \ref{exp:c-components} we have introduced a well-known non-identifiable graph, $G[\bm{S}_2]$, which is the so-called \textit{bow arc} graph. It is the simplest non-identifiable graph, easily provable with the hedge criterion. Consider the $\{X_6\}$-rooted C-forests $G[\bm{S}_2]$ and $\{X_6\}$, which clearly form a hedge for $P_{x_7}(x_6)$ in $G[\bm{S}_2]$.
\end{remark}

A complete characterization of identifiability involves hedges, and will be the discussed in the coming section.

\section[Identification Algorithms and Their Implementation]{Identification Algorithms and Their Implementation}

$do$-calculus was the first step into solving causal effects from causal diagrams, but it has a major problem: even if the effect is identifiable, no one tells us in what order should we apply the three rules, and this gets even more complex when the number of variables in the model grows.

In this section, we will introduce two algorithms that help us compute causal queries, when those are identifiable, called \textbf{ID} and \textbf{IDC}. We will examine meticulously each and every line of both algorithms, justifying why the operations involved are correct, while also explaining our implementation in Python. This implementation is the central objective of this project, and all objects and functions involved form a newly developed package for Python named \texttt{causaleffect}.

We will first introduce an algorithm to solve non-conditional causal effects and its implementation, then we will do the same for an algorithm to compute conditional causal effects. To conclude this section we will shortly explain the existence of some algorithms to compute counterfactual queries from interventional distributions, which have not been developed in our package.

\subsection{Identification of Interventional Distributions}

We have seen that if the causal diagram $G$ of a causal model $M$ is not a C-component it can be decomposed into a unique set of maximal C-components (Lemma \ref{lem:decomposition}). This in turn will help to reduce the identification problem in more manageable subproblems, making use of the following result by Jin Tian \cite{tian_2002}.

\begin{lemma}{\bf ({\rm\cite[Corollary 1 pp. 56]{tian_2002}})}\label{lem:tian}
Let $G = (\bm{V}, \bm{E})$ be the induced DAG from the causal model $M = (\bm{U}, \bm{V}, \bm{F}, P(\bm{U}))$, and $C(G) = \{G[\bm{S}_1],\ldots,G[\bm{S}_k]\}$ a decomposition of $G$ in C-components, where $\bm{S}_i$ are the vertices in $G[\bm{S}_i]$. Then, we have

\begin{enumerate}[\hspace*{0.3cm} (a)]

\item $P(\bm{v})$ factorizes as

\begin{equation*}
P(\bm{v}) = \prod_{i=1}^{k}P(\bm{s_i}|do(\bm{v}\setminus \bm{s_i})) = \prod_{i=1}^{k}P_{\bm{v}\setminus \bm{s_i}}(\bm{s_i})\ .
\end{equation*}

\item Let a topological order over $\bm{V}$ be $V_1<\cdots<V_n$, and let $\bm{V}^{(i)}_{G} = \{V_1, \ldots, V_i\}$ for $1\leq i\leq n$, and $\bm{V}^{(0)}_{G} = \varnothing$. Then every factor from the previous product is identifiable in $G$ as

\begin{equation*}
P_{\bm{v}\setminus \bm{s_j}}(\bm{s_j}) = \prod_{\{i|V_i\in \bm{S}_j\}}P(v_i|\bm{v}_{G}^{(i-1)})
\end{equation*}

\end{enumerate}
\end{lemma}

\begin{figure}[!ht]
\vspace*{-0.0cm}
\centering
\includegraphics{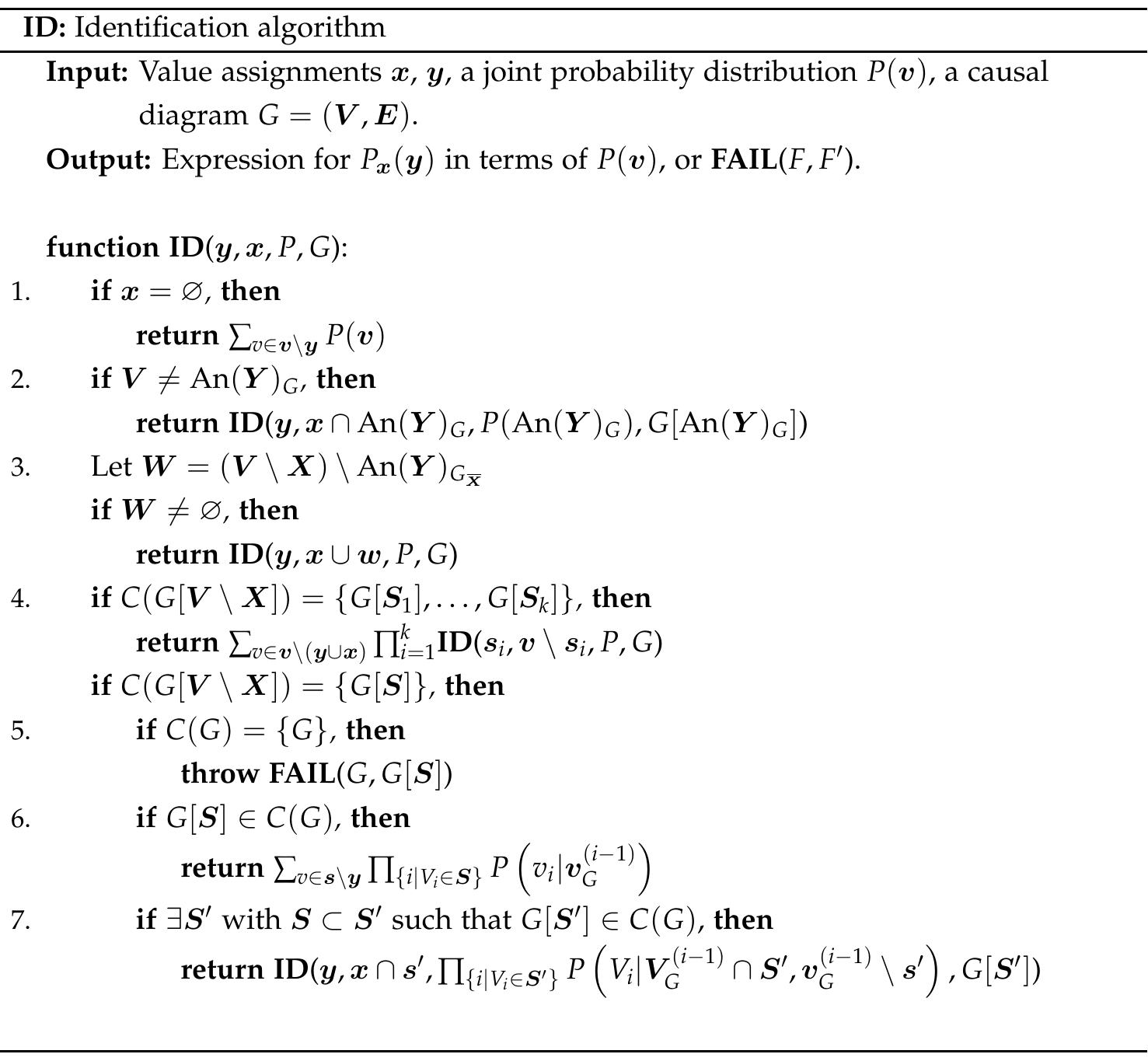}
\vspace*{-0.0cm}
\caption{Algorithm proposed by Shpitser and Pearl \cite{SP_2006a} to compute $P_{\bm{x}}(\bm{y})$.}
\label{fig:ID}
\vspace*{-0.0cm}
\end{figure}

We are now able to define the identification algorithm presented in Figure \ref{fig:ID}.

This algorithm systematically takes advantage of the properties of C-components to decompose, recursively, the identification problem into smaller subproblems until either we find an expression for $P_{\bm{x}}(\bm{y})$ or a hedge that indicates the causal effect is not identifiable.

In the same paper where Shpitser and Pearl defined this algorithm \cite{SP_2006a}, they also proved that it is sound, that is, that when \textbf{ID} returns an expression for $P_{\bm{x}}(\bm{y})$ it is correct. They also proved that it is complete.

\begin{theorem}{\bf (Soundness and Completeness of ID {\rm\cite[Lemma 3 and Theorem 5]{SP_2006a}})}
\textbf{ID} always terminates, and whenever it returns an expression for $P_{\bm{x}}(\bm{y})$, it is correct.
\end{theorem}

The algorithm in Figure \ref{fig:ID} makes use of topological ordering, which is encoded in the graph structure and can be computed beforehand. This is practical, since a topological ordering of a graph $G$ is conserved for any subgraph, and it does not need to be computed again. It is also easy to see that one and only one line of the algorithm will be invoked at every step of the recursion because after checking if a certain condition is fulfilled, it either returns a probability expression, calls the \textbf{ID} again with other parameters or throws an error revealing a hedge in the DAG.

Before seeing in more detail every line of the algorithm and how it has been implemented we will show how the algorithm operates given a causal diagram. To do so, consider again the causal diagram introduced in Example \ref{exp:do_calculus}.

\begin{example}\label{exp:algorithm}
In the example where we showed the Front Door Adjustment we had the causal diagram shown again in Figure \ref{fig:graph_10}, with renamed variables to ease comprehension. We will also write expressions like $G[X, Y]$ instead of $G[\{X, Y\}]$ to avoid notation overload.

\begin{figure}[!ht]
\vspace*{-0.0cm}
\centering
\includegraphics{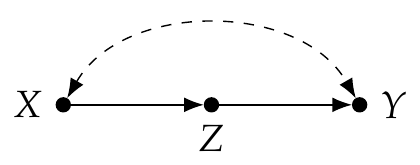}
\vspace*{-0.0cm}
\caption{Causal diagram first introduced in Example \ref{exp:do_calculus}.}
\label{fig:graph_10}
\vspace*{-0.0cm}
\end{figure}
We want to obtain the causal effect $P_{x}(y)$ from the probability distribution $P(X, Y, Z)$ and the causal diagram $G$ shown in Figure \ref{fig:graph_10}. First we have to compute the topological ordering of the vertices, which will be encoded in the graph structure, but there is only a single possibility: $X<Z<Y$. We are now ready to apply the algorithm. The line triggered in this first step is line 4, since clearly $\bm{x}=\{x\}\neq\varnothing$, $\An(\bm{Y})_{G} = \An(Y)_{G} = \bm{V}$ and $\bm{W} = \varnothing$, but $C(G[\bm{V}\setminus\bm{X}]) = C(G[Y, Z]) = \{G[Y], G[Z]\}$. Hence we have

\begin{equation}\label{eq:exp_algorithm}
P_{x}(y) = \sum_{z}P_{x, z}(y)P_{x, y}(z)\ .
\end{equation}
Now, for the first term the conditions in first three lines are not fulfilled by the same reason as before, but neither line 4, since $C(G[\bm{V}\setminus\bm{X}]) = C(G[Y]) = \{G[Y]\}$, so $\bm{S} = \{Y\}$. Additionally, $C(G) = \{G[X, Y], G[Z]\}$, so we have that $\bm{S} = \{Y\}\subset \{X, Y\} = \bm{S'}$ with ${G[\bm{S}']\in C(G)}$. Therefore line 7 is triggered,
\begin{equation*}
P_{x, z}(y) = P'_{x}(y)\ ,
\end{equation*}
where $G' = G[X, Y]$ (see Figure \ref{fig:graph_10_bis} (a)) and $P'(X, Y) = P(X)P(Y|X, z)$. Now line 2 is triggered, because $\An(Y)_{G'} = \{Y\}\neq\{X, Y\}$, so 
\begin{equation*}
P'_{x}(y) = P''_{\varnothing}(y)\ ,
\end{equation*}
where $G'' = G'[Y] = \{Y\}$ and $P''(Y) = \sum_{x}P'(Y, x)$. Finally line 1 is triggered, and so we obtain $P''_{\varnothing}(y) = P''(y)$. Going backwards to write this probability in terms of $P$, 
\begin{equation*}
P_{x, z}(y) = P''(y) = \sum_{x}P'(x, y) = \sum_{x}P(x)P(y|x, z)\ .
\end{equation*}

\begin{figure}[!ht]
\captionsetup[subfigure]{labelformat=empty}
\vspace*{-0.0cm}
\centering
\setlength{\mylength}{\textwidth}
\begin{subfigure}[t]{0.5\mylength}
        \centering
        \includegraphics{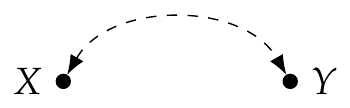}
        \caption{\footnotesize (a) $G' = G[X, Y]$.}
\end{subfigure}
\begin{subfigure}[t]{0.5\mylength}
        \centering
        \includegraphics{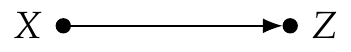}
        \caption{\footnotesize (b) $G''' = G[X, Z]$.}
\end{subfigure}%
\vspace*{-0.0cm}
\caption{Induced subgraphs used throughout Example \ref{exp:algorithm}.}
\label{fig:graph_10_bis}
\vspace*{-0.0cm}
\end{figure}
Now we focus on the second term in Equation \ref{eq:exp_algorithm}, and we see that line 2 is triggered, for $\An(\bm{Y})_{G} = \An(Z)_{G} = \{X, Z\}\neq\bm{V}$. Hence 
\begin{equation*}
P_{x, y}(z) = P'''_{x}(z)\ ,
\end{equation*}
where $G''' = G[X, Z]$ (see Figure \ref{fig:graph_10_bis} (b)) and $P'''(X, Z) = \sum_{y}P(X, Z, y)$. The next step is to see that $\An(\bm{Y})_{G'''} = \An(Z)_{G'''} = \{X, Z\}=\bm{V}$, and also $\bm{W} = \varnothing$, so we compute the confounded components of $G'''[\bm{V}\setminus\bm{X}]$ and $G'''$: $C(G'''[\bm{V}\setminus\bm{X}]) = \{G'''[Z]\}$ and $C(G''') = \{G'''[X], G'''[Z]\}$, so line 6 is invoked:
\begin{equation*}
P'''_{x}(z) = P'''(z|x)\ .
\end{equation*}
This means that
\begin{equation*}
P_{x, y}(z) = P'''(z|x) = \sum_{y}P(z, y|x) = P(z|x)\ .
\end{equation*}
Putting together the two terms of Equation \ref{eq:exp_algorithm} we finally obtain the desired causal effect:
\begin{equation*}
P_{x}(y) = \sum_{z}\left(\sum_{x}P(x)P(y|x, z)\right)P(z|x) = \sum_{z}P(z|x)\sum_{x}P(x)P(y|x, z)\ .
\end{equation*}
Indeed, we obtain the same result as in Example \ref{exp:do_calculus}, as we see that Equation \ref{eq:example_do_calculus} is identical to the one obtained using the identification algorithm with renamed variables $\{X = S, Z = T, Y = C\}$.
\end{example}

\textbf{ID} algorithm in Figure \ref{fig:ID} can also be used to detect unidentifiability. We are going to see another example of the algorithm that raises an error due to the existence of a hedge.

\begin{example}\label{exp:algorithm_hedge}
Consider the DAG $G$ in Figure \ref{fig:graph_2} (a), and we will to compute $P_{x}(y)$. The only possible topological order would be $X<Z<W<Y$, and we see that the ancestors of $Y$ in $G$ are all the nodes in $G$.

\begin{figure}[!ht]
\captionsetup[subfigure]{labelformat=empty}
\vspace*{-0.0cm}
\centering
\setlength{\mylength}{\textwidth}
\savebox{\largestimage}{\includegraphics{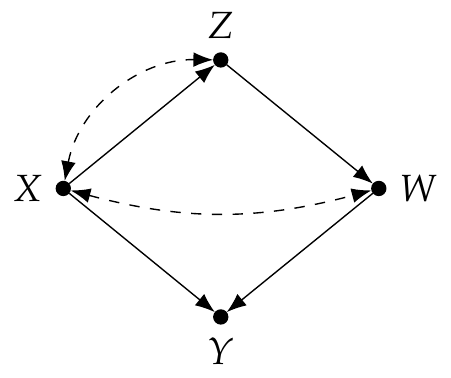}}%
\begin{subfigure}[t]{0.5\mylength}
        \centering
        \usebox{\largestimage}
        \caption{\footnotesize (a) Graph $G$.}
\end{subfigure}
\begin{subfigure}[t]{0.5\mylength}
        \centering
        \raisebox{\dimexpr.5\ht\largestimage-.5\height}{\includegraphics{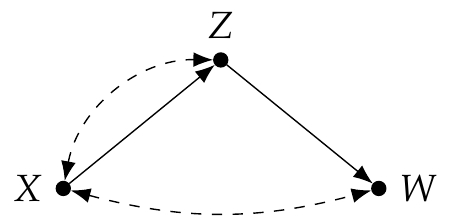}}
        \caption{\footnotesize (b) $G' = G[X, W, Z]$.}
\end{subfigure}%
\vspace*{-0.0cm}
\caption{Causal diagrams in Example \ref{exp:algorithm_hedge}.}
\label{fig:graph_2}
\vspace*{-0.0cm}
\end{figure}
First of all, line 4 will be executed, because the ancestors of $Y$ (in $G$ and in $G_{\overline{X}}$) is the set containing all vertices of $G$ and also $C(G[\bm{V}\setminus X]) = \{G[Y], G[W], G[Z]\}$. So we will have
\begin{equation}\label{eq:exp_algorithm_hedge}
P_{x}(y) = \sum_{w, z}P_{x, w, z}(y)P_{x, y, z}(w)P_{x, y, w}(z)\ .
\end{equation}
For the first term, line 6 is triggered, because $G[Y]\in C(G) = \{G[Y], G[X, W, Z]\}$, so
\begin{equation*}
P_{x, w, z}(y) = P(y|x, w, z)\ .
\end{equation*}
The second term invokes line 2, because $Y\notin\An(W)_G$, thus we obtain
\begin{equation*}
P_{x, y, z}(w) = P_{x, z}(w)\ ,
\end{equation*}
where $G' = G[X, W, Z]$ (see Figure \ref{fig:graph_2} (b)) and $P'(X, W, Z) = \sum_{y}P(X, W, Z, y)$. Now we have $C(G'[W]) = \{G'[W]\}$ and $C(G') = \{G'\}$, so line 5 is triggered, throwing the hedge $(G', G'[W])$ for $P_{x, z}(w)$ in $G'$. Clearly both are C-components of $G'$, they are $W$-rooted (with $W$ being an ancestor of $Y$ in $G'_{\overline{X}}$), $G'[W]\subset G'$, $G'\cap \{X\} \neq\varnothing$ and $G'[W]\cap \{X\} =\varnothing$. So we conclude that $(G', G'[W])$ form a hedge for $P_{x, z}(w)$ in $G'$, and thus the original causal effect $P_{x}(y)$ is not identifiable in $G$.
\end{example}

We are now ready to explore in detail every line of the algorithm and to show our implementation of \textbf{ID} in Python.

\subsubsection{Python Implementation of Graphs and Distributions}

First of all, to handle graphs and probabilities we need some classes. In our implementation to deal with graphs we have used the \texttt{igraph} library for Python since it provides some useful methods (although, as we will see, we will also need to implement our own).

\begin{table}[!ht]
\renewcommand\tablename{Function}
\centering
\begin{tabular}{p{0.13\textwidth}p{0.10\textwidth}p{0.675\textwidth}}
\hline
\multicolumn{3}{p{0.8\textwidth}}{\define{createGraph}{edges, verbose=\false}}\\ \hline
\textbf{Description}                                        & \multicolumn{2}{p{0.8\textwidth}}{Creates a Graph object from a list of edges in string format.}\\ \hline
\multirow{2}{*}[0.60em]{\textbf{Parameters}}    & \params{edges}            & List of edges of the graph. Each edge must either be directed (\texttt{'X->Y'}) or bidirected (\texttt{'X<->Y'}).            \\ 
                                                                                            & \params{verbose}      & Boolean. If enabled, some useful debugging information will be printed.                  \\ \hline
\textbf{Returns}                                                & \multicolumn{2}{p{0.8\textwidth}}{A Graph object from the \texttt{igraph} library, with the directed and bidirected edges given as the \texttt{edges} parameter. Each bidirected edge will be encoded as two directed edges. It will contain exactly all vertices appearing in the edges list. Each edge will have a property called \texttt{confounding}, which will be 0 for directed edges and $\pm1$ for bidirected edges (edges of a bidirected pair will have opposite signs for the \texttt{confounding} property).}\\ \hline
\end{tabular}
\vspace*{0.2cm}
\caption{Implemented \texttt{createGraph} function.}
\label{fun:createGraph}
\end{table}
The first thing we need to be able to do is to create a DAG in a simple, intuitive way. For this purpose, we devised the function \texttt{createGraph} (see Function \ref{fun:createGraph}).

Another useful function that helps visualize causal diagrams that we have implemented is \texttt{plotGraph} (see Function \ref{fun:plotGraph}).

\begin{table}[!ht]
\renewcommand\tablename{Function}
\centering
\begin{tabular}{p{0.13\textwidth}p{0.10\textwidth}p{0.675\textwidth}}
\hline
\multicolumn{3}{p{0.8\textwidth}}{\define{plotGraph}{graph, name=\none}}\\ \hline
\textbf{Description}                                        & \multicolumn{2}{p{0.8\textwidth}}{Plots a causal diagram. Needs \texttt{pycairo} library.}\\ \hline
\multirow{2}{*}[0.60em]{\textbf{Parameters}}    & \params{graph}            & Graph object with an edge property named \texttt{confounding}.\\
                                                                                            & \params{name}             & Name of the png file with the plotted graph. If not introduced it does not produce a png image. \\ \hline
\textbf{Returns}                                                & \multicolumn{2}{p{0.8\textwidth}}{Nothing. It makes use of the function \texttt{plot} of the \texttt{igraph} library to plot the causal diagram.}\\ \hline
\end{tabular}
\vspace*{0.2cm}
\caption{Implemented \texttt{plotGraph} function.}
\label{fun:plotGraph}
\end{table}
To ease the comparison of causal effects of different causal models between our implementation the one made with R by Tikka and Karvanen \cite{causaleffect_R}, we have created a function that ``translates'' our edge notation into theirs (see Function \ref{fun:to_R_notation}).

\begin{figure}[!ht]
\captionsetup[subfigure]{labelformat=empty}
\vspace*{-0.0cm}
\centering
\setlength{\mylength}{\textwidth}
\savebox{\largestimage}{\includegraphics[width=0.3\mylength]{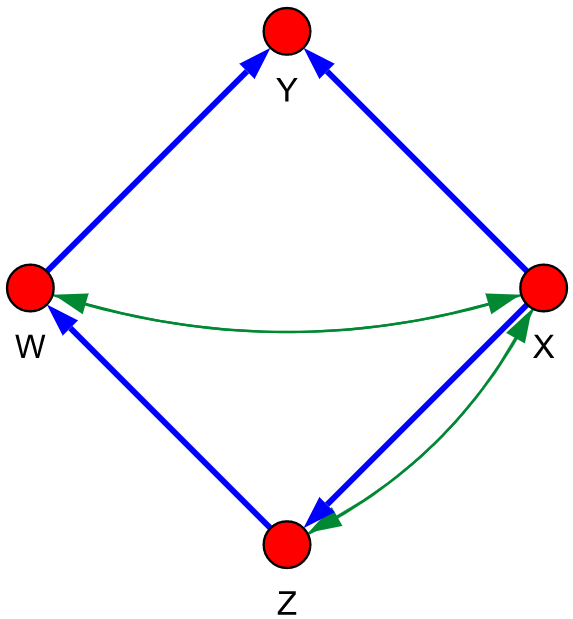}}%
\begin{subfigure}[t]{\mylength}
        \centering
        \includegraphics{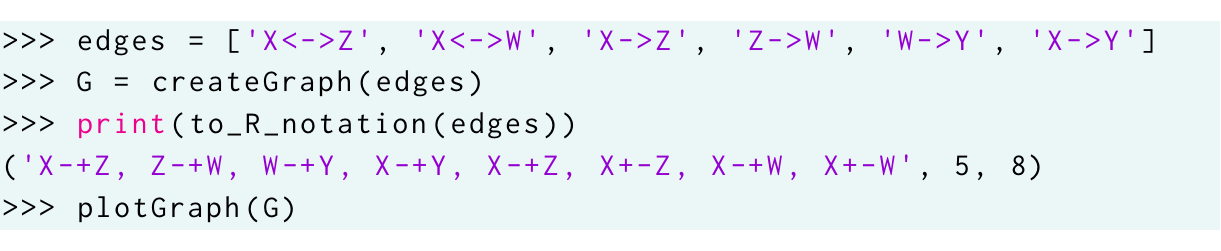}
        \caption{\footnotesize (a) Code executed.}
\end{subfigure}
\vspace*{0.1cm}
\begin{subfigure}[t]{0.3\mylength}
        \centering
        \usebox{\largestimage}
        \caption{\footnotesize (b) Plot of graph $G$.}
\end{subfigure}%
\vspace*{-0.0cm}
\caption{Executing \texttt{createGraph}, \texttt{to\_R\_notation} and \texttt{plotGraph}.}
\label{fig:code_1}
\vspace*{-0.0cm}
\end{figure}

\begin{table}[!ht]
\renewcommand\tablename{Function}
\centering
\begin{tabular}{p{0.13\textwidth}p{0.10\textwidth}p{0.675\textwidth}}
\hline
\multicolumn{3}{p{0.8\textwidth}}{\define{to\_R\_notation}{edges}}\\ \hline
\textbf{Description}                                        & \multicolumn{2}{p{0.8\textwidth}}{Converts a list of edges from our notation to the notation used in the R package \texttt{causaleffect}.}\\ \hline
\multirow{1}{*}{\textbf{Parameters}}    & \params{edges}            & List of strings containing the edges of a graph.            \\ \hline
\textbf{Returns}                                                & \multicolumn{2}{p{0.8\textwidth}}{Tuple of three elements. The first is a string encoding the edges of a graph. The next two elements are integers, and are the indexes used by the \texttt{causaleffect} package to set confounding properties to the edges of the graph.}\\ \hline
\end{tabular}
\vspace*{0.2cm}
\caption{Implemented \texttt{to\_R\_notation} function.}
\label{fun:to_R_notation}
\end{table}
To see how these three functions work we can create the causal diagram in Figure \ref{fig:graph_2} (a) by executing the code in Figure \ref{fig:code_1} (a).

The output of the plot of graph $G$ can be seen in Figure \ref{fig:code_1} (b), where bidirected edges are coloured in green and are slightly thinner than directed edges.

We have developed many more functions to obtain properties of causal diagrams needed in the implementation of \textbf{ID}, but they will be explained in due course when required by the algorithm. The next step is to manage distribution probabilities. For that purpose, we have created a Python class named \class{Probability}, which can be constructed recursively to embrace the nature of the algorithm, also recursive.

\begin{figure}[!ht]
\captionsetup[subfigure]{labelformat=empty}
\vspace*{-0.0cm}
\centering
\setlength{\mylength}{\textwidth}
\begin{subfigure}[t]{\mylength}
        \centering
        \includegraphics{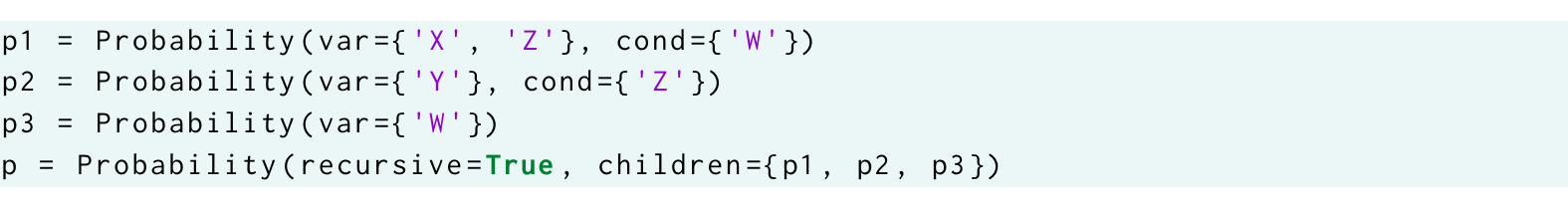}
        \caption{\footnotesize (a) $P^{*}(X, Y, Z, W)$.}
\end{subfigure}
\vspace*{0.1cm}
\begin{subfigure}[t]{\mylength}
        \centering
        \includegraphics{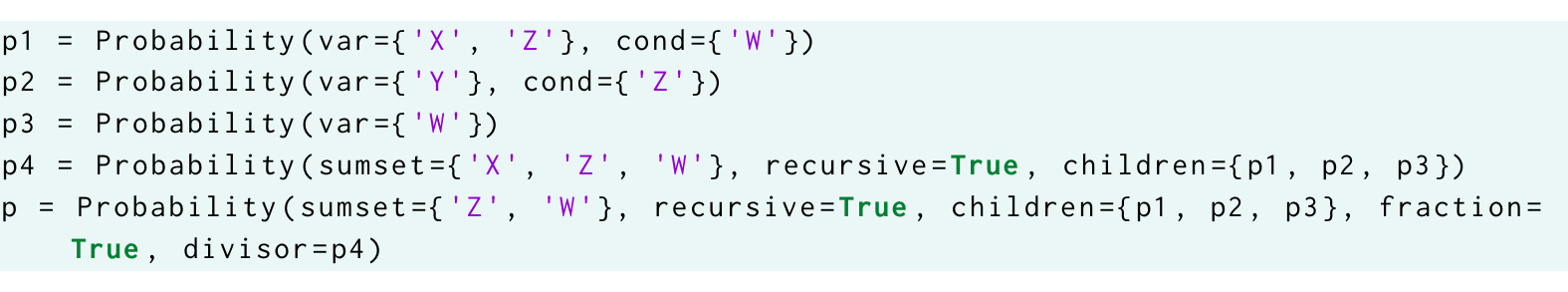}
        \caption{\footnotesize (b) $P^{*}(X | Y)$.}
\end{subfigure}%
\vspace*{-0.0cm}
\caption{Representing probability distributions with the \class{Probability} class.}
\label{fig:code_2}
\vspace*{-0.0cm}
\end{figure}

The \class{Probability} class has several attributes. The string sets \params{var} and \params{cond} allow simple conditional distributions (distributions of \params{var} conditioned on \params{cond}). For instance, the object \verb+Probability(var={'Y'}, cond={'X'})+ represents the distribution $P(Y|X)$. To model products of distributions, and mimicking \cite{causaleffect_R}, the boolean attribute \params{recursive} is defined, allowing multiple \class{Probability} objects to be nested inside one another when \true. When this is the case, the attribute \params{children}, which is a set, is filled with \class{Probability} objects, and the variables \params{var} and \params{cond} are ignored. We can now have more complex probability distributions such as $P^{*}(X, Y, Z, W) = P(Z,X|W)P(Y|Z)P(W)$ by creating the object in Figure \ref{fig:code_2} (a).

We also need to consider how to manage marginal distributions, and thus a set containing variables to be summed over (in the discrete case, integrated in the continuous case) is defined. The attribute \params{sumset} is the set of variables that makes this possible. So if we wanted to represent $P^{*}(X, Y)$, we would need to set the attribute \verb+sumset={'Z', 'W'}+ for the object \verb+p+ previously defined.

Another level of complexity is needed when computing conditional probabilities. Sometimes to form conditional distribution expressions one has to introduce fractions, and that is why the \class{Probability} class has two last attributes, \params{fraction} and \params{divisor}. When the boolean attribute \texttt{fraction} is set to \true, the \params{divisor} containing a probability distribution object is enabled. This final step allows us to represent complex conditional distributions such as
\begin{equation*}
P^{*}(X | Y) = \frac{P^{*}(X, Y)}{P^{*}(Y)} = \frac{\sum_{W, Z}P(Z,X|W)P(Y|Z)P(W)}{\sum_{X, W, Z}P(Z,X|W)P(Y|Z)P(W)}\ ,
\end{equation*}
which can be stored as seen in Figure \ref{fig:code_2} (b).

To understand and read the \class{Probability} objects easily, a method to get the \LaTeX\ string representation has been added, named \texttt{printLatex} (see Function \ref{fun:printLatex}).

\begin{table}[!ht]
\renewcommand\tablename{Function}
\centering
\begin{tabular}{p{0.13\textwidth}p{0.10\textwidth}p{0.675\textwidth}}
\hline
\multicolumn{3}{p{0.9\textwidth}}{\define{printLatex}{self, tab=0, simplify=\true, complete=\true, verbose=\false}}\\ \hline
\textbf{Description}                                        & \multicolumn{2}{p{0.8\textwidth}}{Constructs the \LaTeX\ string representation of a \class{Probability} object.}\\ \hline
\multirow{4}{*}[1.80em]{\textbf{Parameters}}    & \params{tab}          & Integer, keeps track of the depth of the recursion.\\
                                                                                            & \params{simplify} & Boolean, simplifies the expression if enabled. \\
                                                                                            & \params{complete} & Boolean, performs additional simplifications if enabled.\\
                                                                                            & \params{verbose}  & Boolean, prints some useful debugging information if enabled. \\ \hline
\textbf{Returns}                                                & \multicolumn{2}{p{0.8\textwidth}}{A string containing the \LaTeX\ representation of the probability distribution.}\\ \hline
\end{tabular}
\vspace*{0.2cm}
\caption{Implemented \texttt{printLatex} function.}
\label{fun:printLatex}
\end{table}
In the latter function defined we have seen that, in some cases, simplifications are made to the distribution. Those are handled by the \texttt{simplify} method in Function \ref{fun:simplify}.

\begin{table}[!ht]
\renewcommand\tablename{Function}
\centering
\begin{tabular}{p{0.13\textwidth}p{0.10\textwidth}p{0.675\textwidth}}
\hline
\multicolumn{3}{p{0.9\textwidth}}{\define{simplify}{self, complete=\true, decouple=\true, verbose=\false}}\\ \hline
\textbf{Description}                                        & \multicolumn{2}{p{0.8\textwidth}}{Simplifies the probability distribution object.}\\ \hline
\multirow{2}{*}[0.60em]{\textbf{Parameters}}    & \params{complete} & Boolean, performs additional simplifications if enabled.\\
                                                                                            & \params{verbose}  & Boolean, prints some useful debugging information if enabled. \\ \hline
\textbf{Returns}                                                & \multicolumn{2}{p{0.8\textwidth}}{Nothing.}\\ \hline
\end{tabular}
\vspace*{0.2cm}
\caption{Implemented \texttt{simplify} function.}
\label{fun:simplify}
\end{table}
There are some steps to perform basic simplifications. The first step is to decouple the \class{Probability} objects when those have children that, in turn, have children. If possible, we will delete these children of children by moving them up one level in the recursion definition, to have just a single level of children. Then the basic simplifications are made for those distributions that are not recursive, which are $\params{newsumset} = \params{sumset}\setminus(\params{sumset}\cap\params{var})$ and $\params{newvar} = \params{var}\setminus(\params{sumset}\cap\params{var})$. This is an application of the Law of Total Probability, 
\begin{equation*}
\sum_{X, Y} P(Y, W|X) = \sum_{X} P(W|X)\ .
\end{equation*}

If the probability to simplify is a fraction it will do the same in the denominator, and then it will check if the variable set is empty. Additionally, if the conditioning set of the denominator is empty and the variable set of the denominator is a subset of the variables of the numerator, it will delete the fraction and change the numerator accordingly to the conditional probability. An example of this simplification is shown below:
\begin{equation*}
\frac{\sum_{X} P(Y,W|X)}{P(Y)} = \sum_{X} P(W|X, Y)\ .
\end{equation*}

If the \params{complete} boolean is enabled and the probability distribution is recursive it will try to perform additional simplifications. For every different pair of non-recursive children {(\texttt{p1, p2})} it will check if \texttt{p1.cond == (p2.var)}$\cup$\texttt{(p2.cond)}, and if this is the case, it will delete \texttt{p2} from the children and change \texttt{p1.var = (p1.var)}$\cup$\texttt{(p2.var)} and \texttt{p1.cond = (p1.cond)}$\setminus$\texttt{(p2.var)}. This simplification is the same made in Example \ref{exp:probability_axioms}, and an example can be seen below:
\begin{equation*}
P(Y, W|X, Z)P(X|Z) = P(X, Y, W|Z)\ .
\end{equation*}

With all these preparations discussed we find ourselves ready to explore the \textbf{ID} algorithm line by line and see how it has been implemented.

\subsubsection{Python Implementation of the \textbf{ID} Algorithm}

We have already reviewed how graphs and probability distribution objects will be encoded in our implementation, and later on, when used, some useful functions regarding these objects will be explained. The function implementing the \textbf{ID} algorithm in Figure \ref{fig:ID} in our package is called \texttt{ID\_rec}, referring to its recursive nature (see Function \ref{fun:ID_rec}).

\begin{table}[!ht]
\renewcommand\tablename{Function}
\centering
\begin{tabular}{p{0.13\textwidth}p{0.10\textwidth}p{0.675\textwidth}}
\hline
\multicolumn{3}{p{0.9\textwidth}}{\define{ID\_rec}{Y, X, P, G, ordering, verbose=\false, tab=0}}\\ \hline
\textbf{Description}                                        & \multicolumn{2}{p{0.8\textwidth}}{Recursive function that implements the identification algorithm \textbf{ID}, computing the causal effect $P_{\bm{x}}(\bm{y})$ of a DAG $G$.}\\ \hline
\multirow{7}{*}[3.60em]{\textbf{Parameters}} & \params{Y}    & Set of strings containing the variables in $\bm{y}$.\\
                                                                                            & \params{X}    & Set of strings containing the intervened variables in $\bm{x}$.\\
                                                                                            & \params{P}    & \class{Probability} object with the probability distribution $P$.\\
                                                                                            & \params{G}    & Graph object, encoding the DAG of the causal model $G$.\\
                                                                                            & \params{ordering} & List of strings containing a topological ordering of the nodes of $G$.\\
                                                                                            & \params{verbose}  & Boolean, prints some useful debugging information if enabled. \\
                                                                                            & \params{tab}          & Integer, keeps track of the depth of the recursion.\\ \hline
\textbf{Returns}                                                & \multicolumn{2}{p{0.8\textwidth}}{If the effect is identifiable it returns a \class{Probability} object with de computed causal effect $P_{\bm{x}}(\bm{y})$. If the algorithm encounters a hedge it raises an error, providing the two forests that form the hedge for $P_{\bm{x}}(\bm{y})$ in $G$.}\\ \hline
\end{tabular}
\vspace*{0.2cm}
\caption{Implemented \texttt{ID\_rec} function.}
\label{fun:ID_rec}
\end{table}
When the function is called it first retrieves all vertices of $G$ and stores them as a set in a variable \texttt{V}. It then separates the graph $G$ in the directed (only containing directed edges) and the bidirected (only containing bidirected edged) parts, calling the developed function \texttt{get\_directed\_bidirected\_graphs} in Function \ref{fun:get_directed_bidirected_graphs}.

\begin{table}[!ht]
\renewcommand\tablename{Function}
\centering
\begin{tabular}{p{0.13\textwidth}p{0.10\textwidth}p{0.675\textwidth}}
\hline
\multicolumn{3}{p{0.9\textwidth}}{\define{get\_directed\_bidirected\_graphs}{G}}\\ \hline
\textbf{Description}                                        & \multicolumn{2}{p{0.8\textwidth}}{Decouples the causal diagram $G$ into two separate graphs: one containing only bidirected edges, and the other containing directed edges.}\\ \hline
\multirow{1}{*}[0.0em]{\textbf{Parameters}}     & \params{G}    & Graph object $G$.\\ \hline
\textbf{Returns}                                                & \multicolumn{2}{p{0.8\textwidth}}{A tuple with two Graph objects (one with bidirected edges and the other with directed edges).}\\ \hline
\end{tabular}
\vspace*{0.2cm}
\caption{Implemented \texttt{get\_directed\_bidirected\_graphs} function.}
\label{fun:get_directed_bidirected_graphs}
\end{table}
The observed portion of $G$ will be saved as \texttt{G\_dir}. We now proceed to explain every line of the algorithm in detail, revealing our implementation.

\noindent\textbf{\large Line 1}

\begin{figure}[!ht]
\vspace*{0.2cm}
\centering
\includegraphics{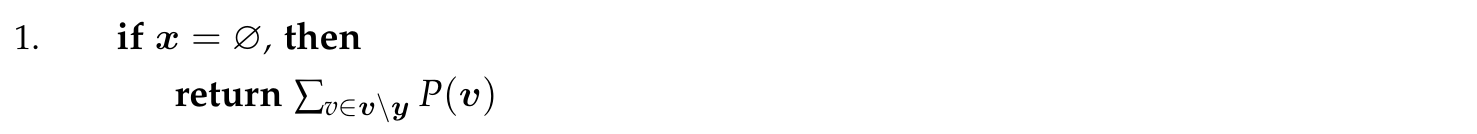}
\vspace*{-0.5cm}
\end{figure}
This line is rather straightforward, since we do not have the $do$-operator any more ($\bm{x}=\varnothing$). Therefore, we are computing the marginal distribution $P(\bm{y})$ instead of a causal effect. This action of marginalizing considers two cases: if the probability distribution is simple, this is, if it is not a product of probabilities, changes the variables of $P$ to the variables in $\bm{y}$. If, on the contrary, the probability is a product of probabilities, adds the pertinent variables to the sum set, $\params{sumset} = \params{sumset}\cup(\texttt{V}\setminus\texttt{Y})$.

\noindent\textbf{\large Line 2}

\begin{figure}[!ht]
\vspace*{0.2cm}
\centering
\includegraphics{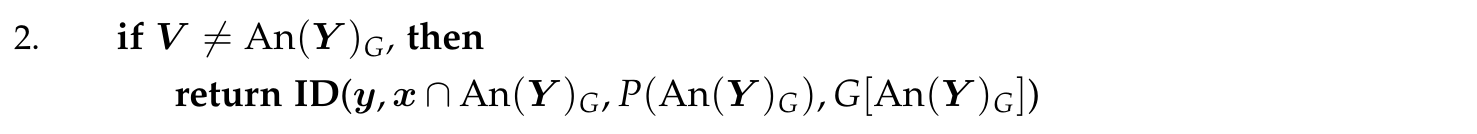}
\vspace*{-0.5cm}
\end{figure}
This line eliminates all non-ancestors of $Y$ in $G$. To see clearly how this is possible, consider the following result.

\begin{lemma}{\bf ({\rm\cite[Lemma 5]{SP_2006a}})}
Let $\bm{X}' = \bm{X}\cap\An(\bm{Y})_{G}$. Then, $P_{\bm{x}}(\bm{y})$ obtained from $P$ in $G$ is equal to $P'_{\bm{x}'}(\bm{y})$ obtained from $P'=P(\An(\bm{Y})_G)$ in $G[\An(\bm{Y})_G]$.
\end{lemma}
\begin{proof}
Let $\bm{W} = \bm{V}\setminus\An(\bm{Y})_{G}$, and consider the submodel $M_{\bm{w}}$, i.e., where the variables in $\bm{W}$ are intervened. The induced causal diagram is $G[\bm{V}\setminus\bm{W}] = G[\An(\bm{Y})_{G}]$, and the induced distribution is $P' = P_{\bm{w}}(\An(\bm{Y})_{G})$.

Now let $\bm{X}'' = \bm{X}\setminus\An(\bm{Y})_{G}$, and clearly $\bm{X} = \bm{X}' \sqcup \bm{X}''$. Therefore, we have
\begin{equation*}
P_{\bm{x}}(\bm{y}) = P_{\bm{x}', \bm{x}''}(\bm{y}) = P_{\bm{x}'}(\bm{y})\ ,
\end{equation*}
where in the last equality we have used rule 3 of $do$-calculus. Indeed, note that we have ${(\bm{Y}\indep\bm{X}''|\bm{X}')_{G_{\mbox{\tiny\xoverline{\bm{X}'}, \xoverline{\bm{X}''}}}}} = {(\bm{Y}\indep\bm{X}''|\bm{X}')_{G_{\mbox{\tiny\xoverline{\bm{X}}}}}}$, because there are no paths from $\bm{Y}$ to $\bm{X}''$: directed paths from $\bm{X}''$ to $\bm{Y}$ are non-existing because, by definition, $\bm{X}''$ has no ancestors of $\bm{Y}$, and back-door paths are deleted by removing all incoming edges to $\bm{X}''\subset\bm{X}$ in $G_{\xoverline{\bm{X}}}$. We apply rule 3 of $do$-calculus again, finally obtaining
\begin{equation*}
P_{\bm{x}'}(\bm{y}) = P_{\bm{x}', \bm{w}}(\bm{y}) = P_{\bm{x}'}'(\bm{y})\ ,
\end{equation*}
because ${(\bm{Y}\indep\bm{W}|\bm{X}')_{G_{\mbox{\tiny\xoverline{\bm{X}'}, \xoverline{\bm{W}}}}}}$. By the same reasoning as before, there are no directed paths from $\bm{W} = \bm{V}\setminus\An(\bm{Y})_{G}$ to $\bm{Y}$, neither back-door paths (deleted by looking at $G_{\xoverline{\bm{X}'}, \xoverline{\bm{W}}}$).
\end{proof}

To implement this line, we first have developed a function to compute the ancestors of a node or a set of nodes (see Function \ref{fun:get_ancestors}).

\begin{table}[!ht]
\renewcommand\tablename{Function}
\centering
\begin{tabular}{p{0.13\textwidth}p{0.10\textwidth}p{0.675\textwidth}}
\hline
\multicolumn{3}{p{0.9\textwidth}}{\define{get\_ancestors}{G, V}}\\ \hline
\textbf{Description}                                        & \multicolumn{2}{p{0.8\textwidth}}{Computes the set containing all ancestors of a vertex or a set of vertices, including itself.}\\ \hline
\multirow{2}{*}[0.60em]{\textbf{Parameters}}    & \params{G}    & Graph object, encoding the direct portion of the causal diagram $G$ (which is a DAG).\\
                                                                                            & \params{V}    & String with the name of a vertex, or a set of strings containing the names of the vertices.\\ \hline
\textbf{Returns}                                                & \multicolumn{2}{p{0.8\textwidth}}{A set of strings containing the names of vertices in $G$ which are ancestors of \params{V}.}\\ \hline
\end{tabular}
\vspace*{0.2cm}
\caption{Implemented \texttt{get\_ancestors} function.}
\label{fun:get_ancestors}
\end{table}
If a single vertex is inputted, Function \ref{fun:get_ancestors} computes its ancestors by exploring the directed acyclic graph using BFS (breadth-first search). If, on the other hand, a set of vertices is given, it calls itself for every single vertex in the set and performs the union of the results.

To check if the condition in line 2 is fulfilled we compute the ancestors of \texttt{Y} in $G$ and save it as \texttt{anc}, before querying the length of the set $\texttt{V}\setminus\texttt{anc}$. If this length is not zero, we create a \class{Probability} object which is the marginalized distribution of the one given (considering the two cases discussed in line 1), and we call \texttt{ID\_rec} again with the new parameters specified. The only parameter worth mentioning is the induced subgraph, $G[\An(\bm{Y})_{G}]$, which we have computed using the function \texttt{induced\_subgraph} from the \texttt{igraph} library for efficiency purposes. This function takes a graph and a set of nodes and constructs a subgraph with all given nodes and edges between them as in the original graph.

\noindent\textbf{\large Line 3}

\begin{figure}[!ht]
\vspace*{0.2cm}
\centering
\includegraphics{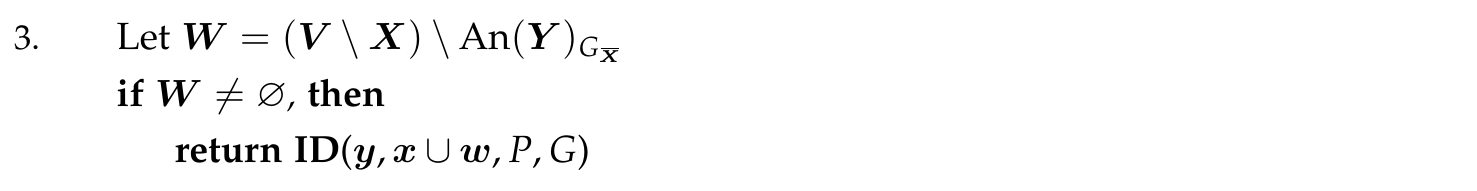}
\vspace*{-0.5cm}
\end{figure}
This line adds interventions to the original causal effect, which is possible due to rule 3 of $do$-calculus.

\begin{lemma}{\bf ({\rm\cite[Lemma 6]{SP_2006a}})}\label{lem:line_3}
Let $\bm{W} = (\bm{V}\setminus\bm{X})\setminus\An(\bm{Y})_{G_{\mbox{\tiny\xoverline{\bm{X}}}}}$. Then, $P_{\bm{x}}(\bm{y}) = P_{\bm{x}, \bm{w}}(\bm{y})$, where $\bm{x}$ are arbitrary values of $\bm{W}$ within its domain.
\end{lemma}
\begin{proof}
By assumption, we have that ${(\bm{Y}\indep\bm{W}|\bm{X})_{G_{\mbox{\tiny\xoverline{\bm{X}}, \xoverline{\bm{W}}}}}}$, because there are no paths from $\bm{Y}$ to $\bm{W}$: directed paths from $\bm{W}$ to $\bm{Y}$ are non-existing because, by definition, $\bm{W}$ has no ancestors of $\bm{Y}$, and back-door paths are deleted by removing all incoming edges to $\bm{W}$ in $G_{\xoverline{\bm{X}}, \xoverline{\bm{W}}}$. Hence the result holds by applying rule 3 of $do$-calculus.
\end{proof}

In our code, to construct $\bm{W}$ we first have created a copy of the graph and deleted all incoming edges into the set $\bm{X}$, thus constructing $G_{\xoverline{\bm{X}}}$. This has been achieved using the \texttt{igraph} function \texttt{delete\_edges} and inputting the edges to remove using the method \texttt{select}, also provided by the package. We then have computed the ancestors of $\bm{Y}$ in $G_{\xoverline{\bm{X}}}$ using the previously discussed function, and we have successfully constructed $\bm{W}$ with set differences. By checking the length of the set $\bm{W}$ we have determined whether to execute line 3, and if the condition is fulfilled, the \texttt{ID\_rec} function is called again with the appropriate parameters.

Note that Lemma \ref{lem:line_3} does not fix the values of added interventions $do(\bm{W}=\bm{w})$. This means that the resulting expression does not depend on the value assigned to $\bm{w}$, even though it appears in the expression $P_{\bm{x}, \bm{w}}(\bm{y})$.

\noindent\textbf{\large Line 4}

\begin{figure}[!ht]
\vspace*{0.2cm}
\centering
\includegraphics{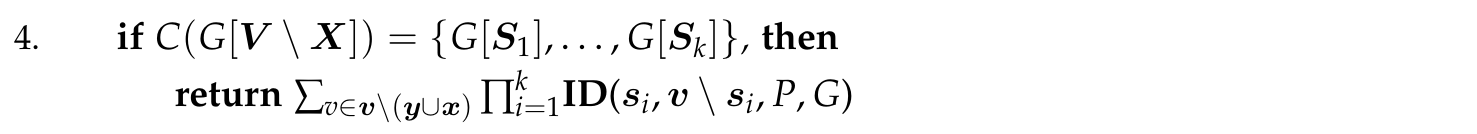}
\vspace*{-0.5cm}
\end{figure}
This follows directly from Lemma \ref{lem:tian} (a), considering the submodel model $M_{\bm{x}}$. This model induces the causal diagram $G[\bm{V}\setminus\bm{X}]$ and probability distribution ${P_{\bm{x}}(\bm{V}\setminus\bm{X})}$. Then, if we marginalize the distribution we get
\begin{equation*}
P_{\bm{x}}(\bm{y}) = \sum_{\bm{v}\setminus(\bm{y}\cup\bm{x})}P_{\bm{x}}(\bm{v}) = \sum_{\bm{v}\setminus(\bm{y}\cup\bm{x})}P_{\bm{x}}(\bm{v}\setminus\bm{x}) = \sum_{\bm{v}\setminus(\bm{y}\cup\bm{x})}\prod_{i=1}^{k}P_{\bm{v}\setminus \bm{s_i}, \bm{x}}(\bm{s_i}) = \sum_{\bm{v}\setminus(\bm{y}\cup\bm{x})}\prod_{i=1}^{k}P_{\bm{v}\setminus \bm{s_i}}(\bm{s_i})\ ,
\end{equation*}
where $do(\bm{X}=\bm{x})$ is in each term of the product, because $\bm{V}\setminus\bm{X} = \bigcup_{i}\bm{S}_i$ and thus $\bm{X}\subset\bm{V}\setminus\bm{S}_i$ for all $i\in\{1, \ldots, k\}$. A full alternative proof can be found in \cite[Lemma 4]{SP_2006a}.

We have created a method (refer to Function \ref{fun:get_C_components}) that computes the C-components of a causal diagram.

\begin{table}[!ht]
\renewcommand\tablename{Function}
\centering
\begin{tabular}{p{0.13\textwidth}p{0.10\textwidth}p{0.675\textwidth}}
\hline
\multicolumn{3}{p{0.9\textwidth}}{\define{get\_C\_components}{G}}\\ \hline
\textbf{Description}                                        & \multicolumn{2}{p{0.8\textwidth}}{Computes the set of maximal C-components of a given graph.}\\ \hline
\multirow{1}{*}[0.00em]{\textbf{Parameters}}    & \params{G}    & Graph object, the causal diagram $G$.\\ \hline
\textbf{Returns}                                                & \multicolumn{2}{p{0.8\textwidth}}{A list of subgraphs containing all maximal C-components of $G$.}\\ \hline
\end{tabular}
\vspace*{0.2cm}
\caption{Implemented \texttt{get\_C\_components} function.}
\label{fun:get_C_components}
\end{table}
Function \ref{fun:get_C_components} has been created by decomposing $G$ into its directed and bidirected part. Then, for every directed edge in $G$, we have checked if both source and target vertices of said edge were in the same connected component in the bidirected part of $G$ and if so, we have added this edge into the bidirected part. By performing this algorithm for all directed vertices we will obtain a modified graph of the old bidirected part, which may be no longer only bidirected. This new graph might be disconnected, and if so every connected subcomponent will be a maximal C-component of the original graph $G$. Finally, we have created a list of connected subcomponents of this potentially disconnected graph, which was the initial goal. In the making of this function, we have used the methods \texttt{subcomponent} and \texttt{decompose} of \texttt{igraph}, for retrieving a connected subcomponent of a graph and for making a list of connected subcomponents of a graph, respectively.

The length of the set containing the maximal C-components of $G[\bm{V}\setminus\bm{X}]$ will determine if line 4 is triggered or not. If invoked, a set of probabilities will be filled by calling the \texttt{ID\_rec} function again, with the proper parameters, one time for each C-component. If identifiable, each \texttt{ID\_rec} call will return a \class{Probability} object, stored in the set of probabilities aforementioned. Finally, a recursive \class{Probability} object will be returned, with the set of probabilities as children and the pertinent \params{sumset}.

If line 4 has not been triggered, then there is only one C-component, $C(G[\bm{V}\setminus\bm{X}]) = \{G[\bm{S}]\}$, so lines 5, 6 or 7 will be executed.

\noindent\textbf{\large Line 5}

\begin{figure}[!ht]
\vspace*{0.2cm}
\centering
\includegraphics{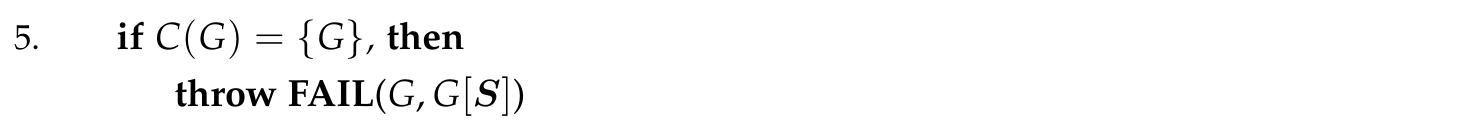}
\vspace*{-0.5cm}
\end{figure}

The discovery of a hedge in $G$ for $P_{\bm{x}}(\bm{y})$ is the only source of unidentifiability in the algorithm. The conditions of line 5 imply the existence of a hedge for the current recursion stage, as it states the following Theorem.

\begin{theorem}{\bf ({\rm\cite[Theorem 6]{SP_2006a}})}
Suppose that line 5 in \textbf{ID} is executed. Then there exist $\bm{X}'\subseteq\bm{X}$ and $\bm{Y}'\subseteq\bm{Y}$ such that the graph pair $G$, $G[\bm{S}]$ returned by the fail condition of \textbf{ID} contain as edge subgraphs two C-forests $F$, $F'$ that form a hedge for $P_{\bm{x}'}(\bm{y}')$.
\end{theorem}
\begin{proof}
Let $\bm{R}$ the root set of $G$ (which in this step of the recursion will be a subgraph of the original input). Since $G$ is a single C-component, it is possible to remove a set of directed arrows from $G$ while preserving the root set $\bm{R}$ such that the resulting graph $F$ is an $\bm{R}$-rooted C-forest (i.e., such that each node has at most one child). Additionally, by lines 2 and 3 of the algorithm, we know that $\bm{R}\subset\An(\bm{Y})_{G_{\mbox{\tiny\xoverline{\bm{X}}}}}$.

Now consider the graph $F' = F\cap G[\bm{S}]$, which is also an $\bm{R}$-rooted C-forest because only single directed arrows were removed from $G[\bm{S}]$ to obtain $F'$. Moreover we know that $F'\subset F$, $\bm{V}_{F}\cap\bm{X}\neq\varnothing$ and $\bm{V}_{F'}\cap\bm{X}=\varnothing$ by construction, so we have a hedge for $\bm{X}$ and $\bm{Y}$, a subset of the original input.
\end{proof}

In our source code we have called once again the function \texttt{get\_C\_components} for the original graph. If it only produces a single maximal C-component it must be the graph itself, so by checking the length of the set of maximal C-components of $G$ we decide whether to invoke line 5. If so, it raises a \class{HedgeFound} exception, which returns information about the pair $(G, G[\bm{S}])$.

\noindent\textbf{\large Line 6}

\begin{figure}[!ht]
\vspace*{0.2cm}
\centering
\includegraphics{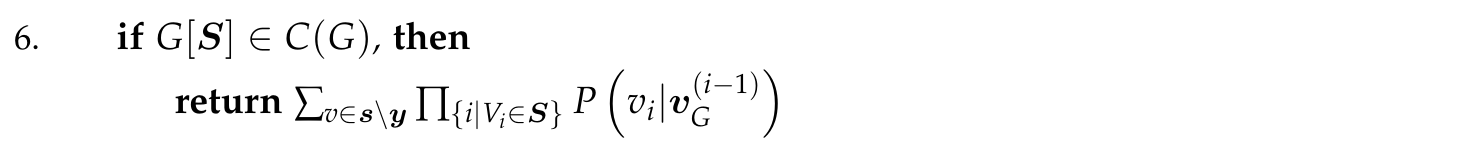}
\vspace*{-0.5cm}
\end{figure}
This line also follows from Lemma \ref{lem:tian}, but this time from part (b). If line preconditions are met, this is, if ${C(G[\bm{V}\setminus\bm{X}]) = \{G[\bm{S}]\}}$, then $\bm{V}\setminus\bm{X} = \bm{S}$ follows. This implies that $G$ local to that recursive call is partitioned into $G[\bm{S}]\in C(G)$ and $G[\bm{X}]$, with no bidirected arcs between these two partitions, and that $\bm{X} = \bm{V}\setminus\bm{S}$. Therefore,
\begin{equation*}
P_{\bm{x}}(\bm{y}) = P_{\bm{v}\setminus\bm{s}}(\bm{y}) = \sum_{\bm{s}\setminus\bm{y}}P_{\bm{v}\setminus\bm{s}}(\bm{s}) = \sum_{\bm{s}\setminus\bm{y}}\prod_{\{i|V_i\in \bm{S}\}}P(v_i|\bm{v}_{G}^{(i-1)})\ ,
\end{equation*}
where the last equality follows from Lemma \ref{lem:tian} (b) given that $G[\bm{S}]\in C(G)$.

Note that in line 4 we could not use Lemma \ref{lem:tian} (b) for each product term $P_{\bm{v}\setminus \bm{s_i}}(\bm{s_i})$ as we have done here because, unlike this case, we did not know if $G[\bm{S}_i]\in C(G)$.

To implement this line of the algorithm we must first check if $G[\bm{S}]\in C(G)$, and we have constructed a function called \texttt{check\_subcomponent} that does just that: it iterates through a list of graphs, called \texttt{components}, and returns \true\ if a given graph is in that list. The complicated part is to create a function to compare graphs, this is, to analyse if two graphs are equal. This function, named \texttt{graphs\_are\_equal} in our package, has been implemented relying on another function, \texttt{check\_subgraph} (see Function \ref{fun:check_subgraph}), that checks if a given graph $G^1$ is a subgraph of another graph $G^2$, $G^1\subseteq G^2$.
\begin{table}[!ht]
\renewcommand\tablename{Function}
\centering
\begin{tabular}{p{0.13\textwidth}p{0.10\textwidth}p{0.675\textwidth}}
\hline
\multicolumn{3}{p{0.9\textwidth}}{\define{check\_subgraph}{G1, G2}}\\ \hline
\textbf{Description}                                        & \multicolumn{2}{p{0.8\textwidth}}{Given two graphs $G^1$ and $G^2$ this function verifies if $G^1\subseteq G^2$.}\\ \hline
\multirow{2}{*}[0.60em]{\textbf{Parameters}}    & \params{G1}   & Graph object, with both directed and bidirected parts.\\
                                                                                            & \params{G2}   & Graph object, with both directed and bidirected parts.\\ \hline
\textbf{Returns}                                                & \multicolumn{2}{p{0.8\textwidth}}{A boolean whose value is \true\ if \params{G1}$\subseteq$\params{G1}, \false\ otherwise.}\\ \hline
\end{tabular}
\vspace*{0.2cm}
\caption{Implemented \texttt{check\_subgraph} function.}
\label{fun:check_subgraph}
\end{table}
Provided with this function, to see if two graphs are equal we just have to see if $G^1\subseteq G^2$ and $G^2\subseteq G^1$, and that is precisely what the \texttt{graphs\_are\_equal} function does.

Function \ref{fun:check_subgraph} first checks if the set of nodes of $G^1$ is a subset of the set of nodes of $G^2$, and then looks if the number of edges of $G^1$ is larger than the number of edges in $G^2$. If the former condition is not satisfied or the latter is fulfilled, the function would immediately return \false. Then it checks if every edge in $G^1$ is also in $G^2$, and if all edges of $G^1$ satisfy the prior condition, returns \true. If even a single edge in $G^1$ is not in $G^2$, it would return \false.

If the condition of line 6 is fulfilled, we have to consider two different cases. The first one is when $\bm{S}$ consists of a single vertex ($|\bm{S}| = 1$) and hence we do not have a product of probabilities (the \class{Probability} object to return will not be recursive but instead very simple). We will obtain the conditional vertices in $\bm{v}_{G}^{(i-1)}$ by calling the \texttt{get\_previous\_order} method defined in Function \ref{fun:get_previous_order}.

\begin{table}[!ht]
\renewcommand\tablename{Function}
\centering
\begin{tabular}{p{0.13\textwidth}p{0.10\textwidth}p{0.675\textwidth}}
\hline
\multicolumn{3}{p{0.9\textwidth}}{\define{get\_previous\_order}{v, possible, ordering}}\\ \hline
\textbf{Description}                                        & \multicolumn{2}{p{0.8\textwidth}}{This function computes the vertices prior to a given one in a certain topological ordering. It excludes the given vertex and all other vertices not present in the possible set of vertices passed.}\\ \hline
\multirow{3}{*}[1.20em]{\textbf{Parameters}}    & \params{v}    & String, name of the initial vertex of the topological ordering.\\
                                                                                            & \params{possible} & Set of possible vertices that can be in the output set.\\
                                                                                            & \params{ordering} & List of vertices containing an ordering of the graph $G$.\\ \hline
\textbf{Returns}                                                & \multicolumn{2}{p{0.8\textwidth}}{A set of vertices strictly smaller than the given vertex in the given topological ordering, intersected with the possible vertices set.}\\ \hline
\end{tabular}
\vspace*{0.2cm}
\caption{Implemented \texttt{get\_previous\_order} function.}
\label{fun:get_previous_order}
\end{table}
Once we have the conditional variables of the output \class{Probability} object we must construct said object. If the probability distribution in the recursive call is simple and only consists of a single term, the output probability distribution will be also very simple. For example, if $P'(X, Y, Z) = P(X,Y,Z)$ and we want to calculate $P'(Y|X, Z)$, it is just $P(Y|X, Z)$. But if the given distribution is a little bit more complex and consists of a product or sums of probabilities the output distribution will also be complex. For instance, consider the previous example but with $P'(X, Y, Z) = P(X,Y)P(Z)$. Now we would have
\begin{equation*}
P'(Y|X, Z) = \frac{P'(X, Y, Z)}{P'(X, Z)} = \frac{P'(X, Y, Z)}{\sum_{Y}P'(X, Y, Z)} = \frac{P(X,Y)P(Z)}{\sum_{Y}P(X,Y)P(Z)} = \frac{P(X,Y)P(Z)}{P(X)P(Z)} = P(Y|X)\ .
\end{equation*}
To take this into account we have devised a function named \texttt{get\_new\_probability} (explained in Function \ref{fun:get_new_probability}).

\begin{table}[!ht]
\renewcommand\tablename{Function}
\centering
\begin{tabular}{p{0.13\textwidth}p{0.10\textwidth}p{0.675\textwidth}}
\hline
\multicolumn{3}{p{0.9\textwidth}}{\define{get\_new\_probability}{P, var, cond=\{\}}}\\ \hline
\textbf{Description}                                        & \multicolumn{2}{p{0.8\textwidth}}{Function that computes a conditional probability from a distribution \params{P} = $P'$ of the form $P'(\params{var}|\params{cond})$.}\\ \hline
\multirow{3}{*}[1.20em]{\textbf{Parameters}}    & \params{P}    & \class{Probability} object, with a probability distribution $P'$.\\
                                                                                            & \params{var}  & Set of variables of the new distribution.\\
                                                                                            & \params{cond} & Set of conditional variables of the new distribution.\\ \hline
\textbf{Returns}                                                & \multicolumn{2}{p{0.8\textwidth}}{A \class{Probability} object from the given distribution $P'$ encoding $P'(\params{var}|\params{cond})$.}\\ \hline
\end{tabular}
\vspace*{0.2cm}
\caption{Implemented \texttt{get\_new\_probability} function.}
\label{fun:get_new_probability}
\end{table}
The simplest case is when $\params{cond}=\varnothing$. In this scenario, if the given distribution is not recursive we only change the variables of the old distribution for \params{var}, while if it is a product of probabilities we add the variables $\texttt{V}\setminus\params{var}$ in the sum set (marginalizing). By contrast, if $\params{cond}\neq\varnothing$, we construct a new \class{Probability} object which is a fraction of the form $\frac{\sum_{\texttt{V}\setminus(\params{var}\cup\params{cond})}P'}{\sum_{\texttt{V}\setminus\params{cond}}P'}$, equal to $P'(\params{var}|\params{cond}) = \frac{P'(\params{var}\cup\params{cond})}{P'(\params{cond})}$. Before returning this probability we attempt to perform a simplification using \texttt{simplify} (Function \ref{fun:simplify}).

Once we have this new probability, this first case of line 6 can return that \class{Probability} object summed over $\bm{S}\setminus\bm{Y}$.

We have considered when $\bm{S}$ has only one element, and the remaining case when $\bm{S}$ has two or more elements is now simple. We construct a set of probabilities, and for each vertex in $\bm{S}$ we create a \class{Probability} object in the same manner as in the first scenario. Then we construct and return a recursive \class{Probability} object with this set of probabilities as children, and summed over $\bm{S}\setminus\bm{Y}$.

\noindent\textbf{\large Line 7}

\begin{figure}[!ht]
\vspace*{0.2cm}
\centering
\includegraphics{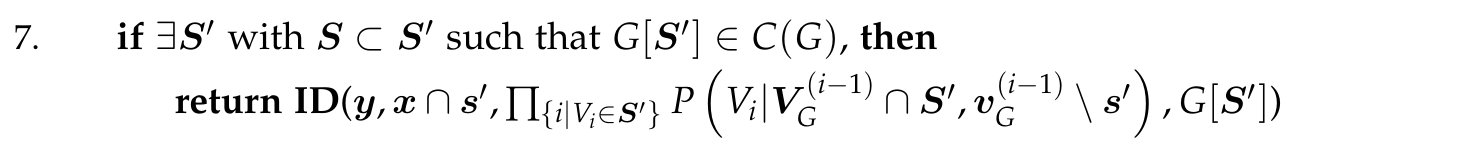}
\vspace*{-0.5cm}
\end{figure}
This line covers the last possible outcome. If the algorithm has reached this far it is clear that $C(G)$ has more than one C-component (otherwise line 5 would have been triggered), and that $G[\bm{S}]$ is not a maximal C-component of $G$ (or then line 6 would have been executed). Therefore there are bidirectional edges between $G[\bm{S}]$ and $G[\bm{X}]$ in $G$, and so there has to be a maximal C-component of $G$, $G[\bm{S}']$, such that $G[\bm{S}]\subset G[\bm{S}']$. The equivalence of intervened probabilities in line 7 relays on the following result.

\begin{lemma}{\bf ({\rm\cite[Lemma 8]{SP_2006a}})}
If the conditions of line 7 of \textbf{ID} are satisfied, $P_{\bm{x}}$ obtained from $P$ in $G$ is equal to $P_{\bm{x}\cap\bm{s}'}'$ obtained from $P' = \prod_{\{i|V_i\in\bm{S}'\}}P\left(V_{i}|\bm{V}_G^{(i-1)}\cap\bm{S}', \bm{v}_G^{(i-1)}\setminus\bm{s}'\right)$ in $G[\bm{S}']$.
\end{lemma}

The implementation we have devised is somewhat similar to the one performed in line 6. First it checks, for each component $G[\bm{S}_j']$ in $C(G)$, if $G[\bm{S}]\subset G[\bm{S}_j']$, by iterating over the components in $C(G)$ and using the \texttt{check\_subgraph} function earlier defined (see Function \ref{fun:check_subgraph}). Once it has found the maximal C-component of $G$ that satisfies $G[\bm{S}]\subset G[\bm{S}']\in C(G)$, it splits again the possible cases in $|\bm{S}'| = 1$ and $|\bm{S}'| > 1$.

When $|\bm{S}'| = 1$, we get the smaller vertices $\bm{V}_{G}^{(i-1)}$ using the \texttt{get\_previous\_order} function (Function \ref{fun:get_previous_order}), and perform the intersections, differences and unions necessary according to the conditional variables of the new probability stated by line 7. Next, using these conditional variables, we compute a new probability distribution as we did in line 6, by using function \texttt{get\_new\_probability} (Function \ref{fun:get_new_probability}). We finally return a recursive call to \texttt{ID\_rec} with $\bm{Y}$, ${\bm{X}\cap\bm{S}'}$, the new probability distribution and the induced subgraph $G[\bm{S}']$ (using \texttt{induced\_subgraph} from the \texttt{igraph} library).

When $|\bm{S}'| > 1$ we define a set of probabilities and for each vertex in $\bm{S}'$ we create a \class{Probability} object in the same fashion as in the case $|\bm{S}'| = 1$. We finally construct a recursive \class{Probability} object with this set of probabilities as children, which will be the distribution passed in the recursive call to \texttt{ID\_rec}. All the other parameters of this recursive call are the same as in the first scenario.

We have checked all lines of the \textbf{ID} algorithm, and since it is sound and complete, there are no more cases to explore. Our implementation of this algorithm, named \texttt{ID\_rec} in the developed package, consequently finishes here. Despite being very useful in the majority of cases, this algorithm only computes identifiable causal effects of the form $P_{\bm{x}}(\bm{y})$, but we could also think of conditional interventions. Shpitser and Pearl, authors of the recently explained identification algorithm, devised also an algorithm for such cases, as we explain in the following section.

\subsection{Identification of Conditional Interventional Distributions}

\begin{figure}[!ht]
\vspace*{-0.0cm}
\centering
\includegraphics{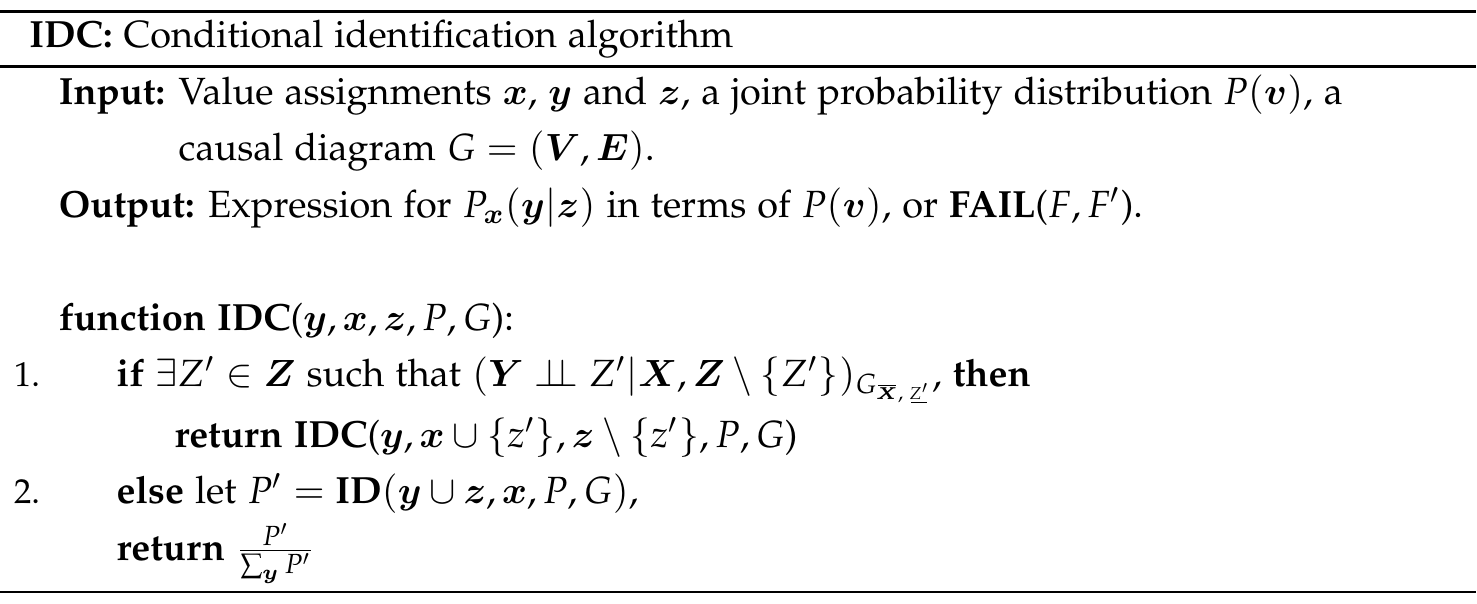}
\vspace*{-0.0cm}
\caption{Algorithm proposed by Shpitser and Pearl \cite{SP_2006b} to compute $P_{\bm{x}}(\bm{y}|\bm{z})$.}
\label{fig:IDC}
\vspace*{-0.0cm}
\end{figure}

We have seen an algorithm to compute the causal effect of the form $P_{\bm{x}}(\bm{y})$, but often we know that some variables hold, that is, the distribution is conditioned on a set of other variables. We now consider the problem of identifying distributions of the form $P_{\bm{x}}(\bm{y}|\bm{z})$, where $\bm{X}, \bm{Y}$ and $\bm{Z}$ are disjoint sets of variables. The basic idea is to reduce the problem to a solved case when $\bm{Z}=\varnothing$, so we can use the algorithm \textbf{ID} seen in the former section. A possible way to do so is to use rule 2 of $do$-calculus, that exchanges actions and observations, $P_{\bm{x}, \bm{z}}(\bm{y}|\bm{w}) = P_{\bm{x}}(\bm{y}|\bm{z}, \bm{w})$, under some conditions, and that is how the algorithm \textbf{IDC} is constructed. This conditional identification algorithm is presented in Figure \ref{fig:IDC}.

Similarly to the first identification algorithm, Shpitser and Pearl defined this algorithm in \cite{SP_2006b} and they also proved that it is sound and complete.

\begin{theorem}{\bf (Soundness and Completeness of IDC {\rm\cite[Theorem 7 and Theorem 8]{SP_2006b}})}
\textbf{IDC} always terminates, and whenever it returns an expression for $P_{\bm{x}}(\bm{y}|\bm{z})$, it is correct.
\end{theorem}

The first line of the \textbf{IDC} is a direct application of rule 2 of $do$-calculus where the exchange consists of a single variable $\{Z'\}$, because if ${(\bm{Y}\indep Z'|\bm{X}, \bm{Z}\setminus \{Z'\})_{G_{\mbox{\tiny\xoverline{\bm{X}}, \xunderline{Z'}}}}}$, then $P_{\bm{x}}(\bm{y}|\bm{z}) = P_{\bm{x}}(\bm{y}|\bm{z}\setminus\{z'\}, \{z'\}) = P_{\bm{x}\cup\{z'\}}(\bm{y}|\bm{z}\setminus\{z'\})$ follows.

The full construction and justification of the conditional identification algorithm can be found in \cite{SP_2006b}.

\subsubsection{Python Implementation of the \textbf{IDC} Algorithm}

Having seen how we represent graphs and probability distribution objects, and how the \textbf{ID} algorithm works (which is an essential requirement), we are ready to reveal our implementation of \textbf{IDC} in Figure \ref{fig:IDC}. In our package, algorithm \textbf{IDC} is called \texttt{IDC}, and it is defined in Function \ref{fun:IDC}.

\begin{table}[!ht]
\renewcommand\tablename{Function}
\centering
\begin{tabular}{p{0.13\textwidth}p{0.10\textwidth}p{0.675\textwidth}}
\hline
\multicolumn{3}{p{0.9\textwidth}}{\define{IDC}{Y, X, Z, P, G, ordering, verbose=\false, tab=0}}\\ \hline
\textbf{Description}                                        & \multicolumn{2}{p{0.8\textwidth}}{Recursive function that implements the conditional identification algorithm \textbf{IDC}, computing the causal effect $P_{\bm{x}}(\bm{y}|\bm{z})$ of a DAG $G$, by using the rule 2 of $do$-calculus and the \textbf{ID} algorithm.}\\ \hline
\multirow{8}{*}[4.20em]{\textbf{Parameters}}    & \params{Y}    & Set of strings containing the variables in $\bm{y}$.\\
                                                                                            & \params{X}    & Set of strings containing the intervened variables in $\bm{x}$.\\
                                                                                            & \params{Z}    & Set of strings containing the conditional variables in $\bm{z}$.\\
                                                                                            & \params{P}    & \class{Probability} object with the probability distribution $P$.\\
                                                                                            & \params{G}    & Graph object, encoding the DAG of the causal model $G$.\\
                                                                                            & \params{ordering} & List of strings containing a topological ordering of the nodes of $G$.\\
                                                                                            & \params{verbose}  & Boolean, prints some useful debugging information if enabled. \\
                                                                                            & \params{tab}          & Integer, keeps track of the depth of the recursion.\\ \hline
\textbf{Returns}                                                & \multicolumn{2}{p{0.8\textwidth}}{If the conditional causal effect is identifiable it returns a \class{Probability} object with de computed causal effect $P_{\bm{x}}(\bm{y}|\bm{z})$. If the algorithm encounters a hedge it raises an error, providing the two C-forests that form the hedge for $P_{\bm{x}}(\bm{y}|\bm{z})$ in $G$.}\\ \hline
\end{tabular}
\vspace*{0.2cm}
\caption{Implemented \texttt{IDC} function.}
\label{fun:IDC}
\end{table}
We are now going to explain in detail how we have implemented these two lines of algorithm \textbf{IDC} in our package.

\noindent\textbf{\large Line 1}

In the first line we have to see if there is a vertex $Z'$ in $\bm{Z}$ fulfilling the independence suggested. To do so, we must first obtain the graph in which this independence will be checked, and note that the graph is different for each vertex $Z'$ in $\bm{Z}$. Therefore, we first loop through the vertices in $\bm{Z}$, and then for each one we construct the required graph. To do so, we first use a newly defined function called \texttt{unobserved\_graph}, explained in Function \ref{fun:unobserved_graph}.

\begin{figure}[!ht]
\captionsetup[subfigure]{labelformat=empty}
\vspace*{-0.0cm}
\centering
\setlength{\mylength}{\textwidth}
\savebox{\largestimage}{\includegraphics[width=0.3\mylength]{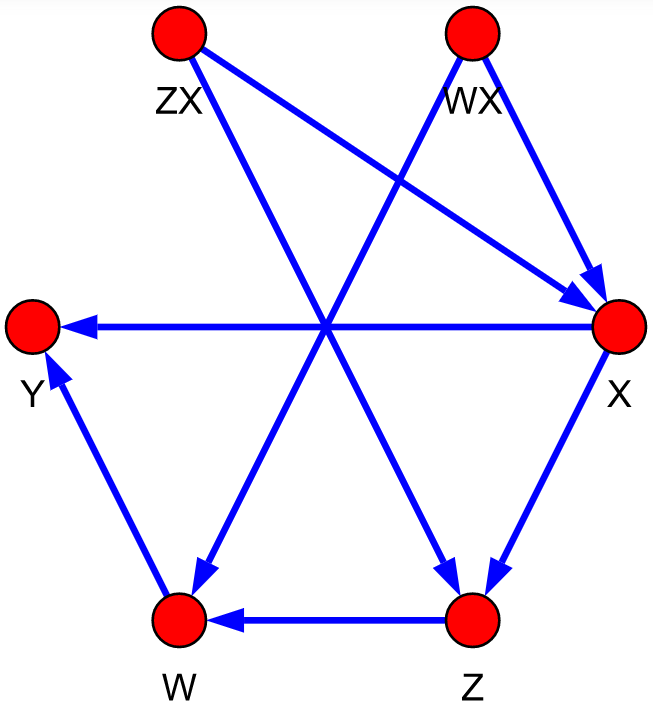}}%
\begin{subfigure}[t]{0.5\mylength}
        \centering
        \raisebox{\dimexpr.5\ht\largestimage-.5\height}{\includegraphics[width=0.3\mylength]{resources/code_1/graph.png}}
        \caption{\footnotesize (a) Graph with bidirected edges confounding variables.}
\end{subfigure}
\begin{subfigure}[t]{0.5\mylength}
        \centering
        \usebox{\largestimage}
        \caption{\footnotesize (b) Graph with explicit unmeasurable confounders.}
\end{subfigure}%
\vspace*{-0.0cm}
\caption{We always use the graph representation in (a), but for the purposes of checking $d$-separation we must explicitly state the direct arrows from confounders to variables as in (b).}
\label{fig:graph_explicit}
\vspace*{-0.0cm}
\end{figure}

\begin{table}[!ht]
\renewcommand\tablename{Function}
\centering
\begin{tabular}{p{0.13\textwidth}p{0.10\textwidth}p{0.675\textwidth}}
\hline
\multicolumn{3}{p{0.9\textwidth}}{\define{unobserved\_graph}{G}}\\ \hline
\textbf{Description}                                        & \multicolumn{2}{p{0.8\textwidth}}{Function that constructs a causal diagram where confounded variables have explicit unmeasurable nodes from a DAG of bidirected edges.}\\ \hline
\multirow{1}{*}[0.00em]{\textbf{Parameters}}    & \params{G}    & Graph object, encoding the DAG of the causal model $G$ with bidirected edges.\\ \hline
\textbf{Returns}                                                & \multicolumn{2}{p{0.8\textwidth}}{A graph object with no bidirected confounding edges, but instead explicit exogenous variables and direct edges to the the affected nodes.}\\ \hline
\end{tabular}
\vspace*{0.2cm}
\caption{Implemented \texttt{unobserved\_graph} function.}
\label{fun:unobserved_graph}
\end{table}
Function \ref{fun:unobserved_graph} changes the representation of a graph we have used up to this point to a more explicit representation, where exogenous variables confounding two nodes are now nodes in the graph object. This is necessary for the next steps (deleting arrows and computing $d$-separation of two sets of nodes), as we will see next, but let us first show how a graph object representation is changed when applying this function, by considering Figure \ref{fig:graph_explicit}.

In Figure \ref{fig:graph_explicit} (a) we see the plot of a graph object where the pairs of variables $(X,W)$ and $(X, Z)$ are confounded by hidden common causes. We see that, after applying the \texttt{unobserved\_graph} function, these unobserved common causes appear explicitly, named $WX$ for the first pair and $ZX$ for the second, in Figure \ref{fig:graph_explicit} (b). Note that this explicit representation will \textit{only} be used to compute $d$-separation of sets of measurable variables, and nowhere else.

\begin{figure}[!ht]
\captionsetup[subfigure]{labelformat=empty}
\vspace*{-0.0cm}
\centering
\setlength{\mylength}{\textwidth}
\savebox{\largestimage}{\includegraphics{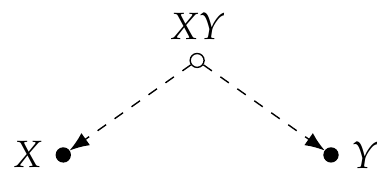}}%
\begin{subfigure}[t]{0.5\mylength}
        \centering
        \raisebox{\dimexpr.5\ht\largestimage-.5\height}{\includegraphics{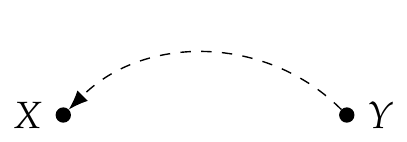}}
        \caption{\footnotesize (a) Bidirected representation.}
\end{subfigure}
\begin{subfigure}[t]{0.5\mylength}
        \centering
        \usebox{\largestimage}
        \caption{\footnotesize (b) Explicit representation.}
\end{subfigure}%
\vspace*{-0.0cm}
\caption{$X$ and $Y$ are confounded by an exogenous variable $XY$, and outgoing edges from $X$ have been deleted (incorrectly in (a)).}
\label{fig:graph_12}
\vspace*{-0.0cm}
\end{figure}

The following step is to remove incoming edges to $\bm{X}$ and outgoing edges from $Z'$. This is where the previously constructed graph comes into play: imagine we have to remove outgoing edges from a confounded vertex $X$. In the first representation we would delete a directed confounding edge from the bidirected pair (see Figure Figure \ref{fig:graph_12} (a)), while in reality a variable that is confounded only has incoming edges, not outgoing! To see this more clearly, consider Figure \ref{fig:graph_12}, where all outgoing edges of $X$ have been deleted.

In Figure \ref{fig:graph_12}, if we were to remove outgoing edges from $X$ in the bidirected representation we would be deleting an arrow that it really does not exist. In reality, $X$ has no outgoing edges, and we need the explicit representation to be aware of this.

In our implementation, we have deleted the appropriate incoming and outgoing edges from the explicit version of $G$ as stated in line 1, using the \texttt{delete\_edges} and \texttt{select} functions provided by \texttt{igraph}, as we did in line 3 of \textbf{ID}. Then we have checked if $\bm{Y}$ and $Z'$ are independent conditional to $\bm{X}$ and $\bm{Z}\setminus \{Z'\}$ in the modified graph, ${(\bm{Y}\indep Z'|\bm{X}, \bm{Z}\setminus \{Z'\})_{G_{\mbox{\tiny\xoverline{\bm{X}}, \xunderline{Z'}}}}}$,  by using Theorem \ref{thm:independence_d_separation} and seeing if they are $d$-separated. To do so, we have constructed Function \ref{fun:dSep}, named \texttt{dSep}.

\begin{table}[!ht]
\renewcommand\tablename{Function}
\centering
\begin{tabular}{p{0.13\textwidth}p{0.10\textwidth}p{0.675\textwidth}}
\hline
\multicolumn{3}{p{0.9\textwidth}}{\define{dSep}{G, Y, node, cond, verbose=\false}}\\ \hline
\textbf{Description}                                        & \multicolumn{2}{p{0.8\textwidth}}{Checks if the set \params{Y} and the node \params{node} are $d$-separated in $G$ conditional to a set of nodes \params{cond}.}\\ \hline
\multirow{5}{*}[2.40em]{\textbf{Parameters}}    & \params{G}                & Graph object, encoding the DAG of the causal model $G$.\\
                                                                                            & \params{Y}                & Set of strings containing some nodes in $G$. \\
                                                                                            & \params{node}         & String, node of the graph $G$.\\
                                                                                            & \params{cond}         & Set of strings containing some nodes in $G$. \\
                                                                                            & \params{verbose}  & Boolean, prints some useful debugging information if enabled. \\ \hline
\textbf{Returns}                                                & \multicolumn{2}{p{0.8\textwidth}}{Returns \true\ if all paths between \params{Y} and the node \params{node} conditional to \params{cond} in $G$ are $d$-separated, \false\ if there is a $d$-connected path between a node in \params{Y} and \params{node}.}\\ \hline
\end{tabular}
\vspace*{0.2cm}
\caption{Implemented \texttt{dSep} function.}
\label{fun:dSep}
\end{table}
Function \ref{fun:dSep} iterates through all possible paths between all nodes of \params{Y} and \params{node}, and for each path it computes if it is $d$-separated. To do so we have devised a function to verify if a given path in a graph is $d$-separated by a set of vertices, called \texttt{is\_path\_d\_separated} (see Function \ref{fun:is_path_d_separated}).

\begin{table}[!ht]
\renewcommand\tablename{Function}
\centering
\begin{tabular}{p{0.13\textwidth}p{0.10\textwidth}p{0.675\textwidth}}
\hline
\multicolumn{3}{p{0.9\textwidth}}{\define{is\_path\_d\_separated}{G, path, cond, verbose=\false}}\\ \hline
\textbf{Description}                                        & \multicolumn{2}{p{0.8\textwidth}}{Checks if a given path \params{path} is $d$-separated in $G$ conditional to a set of nodes \params{cond}.}\\ \hline
\multirow{4}{*}[1.80em]{\textbf{Parameters}}    & \params{G}                & Graph object, encoding the DAG of the causal model $G$.\\
                                                                                            & \params{path}         & List if indexes of vertices in $G$, consisting of a path between the first and last elements of the list.\\
                                                                                            & \params{cond}         & Set of strings containing some nodes in $G$. \\
                                                                                            & \params{verbose}  & Boolean, prints some useful debugging information if enabled. \\ \hline
\textbf{Returns}                                                & \multicolumn{2}{p{0.8\textwidth}}{Returns \true\ if the \params{path} conditional to \params{cond} in $G$ is $d$-separated, \false\ if it is $d$-connected.}\\ \hline
\end{tabular}
\vspace*{0.2cm}
\caption{Implemented \texttt{is\_path\_d\_separated} function.}
\label{fun:is_path_d_separated}
\end{table}
To certify that a path is $d$-separated in Function \ref{fun:is_path_d_separated} we have iterated through the vertices of the path, looking for the necessary structures that make a path $d$-separated (see Definition \ref{def:d_separation}). This is, we have looked for either a chain or a fork with a node of the conditional set in the middle ($I\to M \to J$ or $I\leftarrow M \to J$ with $M\in\params{cond}$ and $I,J\in\params{path}$), or for a collider with no descendants in the conditional set (i.e., $I \to M \leftarrow J$ with $\De(M)_{G}\cap\params{cond}=\varnothing$). Therefore, as we did with the ancestors, we have also created Function \ref{fun:get_descendants} to compute the descendants of a set of nodes, called \texttt{get\_descendants}.

\begin{table}[!ht]
\renewcommand\tablename{Function}
\centering
\begin{tabular}{p{0.13\textwidth}p{0.10\textwidth}p{0.675\textwidth}}
\hline
\multicolumn{3}{p{0.9\textwidth}}{\define{get\_descendants}{G, V}}\\ \hline
\textbf{Description}                                        & \multicolumn{2}{p{0.8\textwidth}}{Computes the set containing all descendants of a vertex or a set of vertices, including itself.}\\ \hline
\multirow{2}{*}[0.60em]{\textbf{Parameters}}    & \params{G}    & Graph object, encoding the direct portion of the causal diagram $G$ (which is a DAG).\\
                                                                                            & \params{V}    & String with the name of a vertex, or a set of strings containing the names of the vertices.\\ \hline
\textbf{Returns}                                                & \multicolumn{2}{p{0.8\textwidth}}{A set of strings containing the names of vertices in $G$ which are descendants of \params{V}.}\\ \hline
\end{tabular}
\vspace*{0.2cm}
\caption{Implemented \texttt{get\_descendants} function.}
\label{fun:get_descendants}
\end{table}
Analogously to the \texttt{get\_ancestors} function, if a single vertex is inputted, Function \ref{fun:get_descendants} computes its descendants by exploring the directed acyclic graph using BFS (breadth-first search). If, on the other hand, a set of vertices is given, it calls itself for every single vertex in the set and performs the union of the results.

Once we know, using the aforementioned functions, that there is a vertex $Z'$ such that ${(\bm{Y}\indep Z'|\bm{X}, \bm{Z}\setminus \{Z'\})_{G_{\mbox{\tiny\xoverline{\bm{X}}, \xunderline{Z'}}}}}$, we simply call recursively the function \texttt{IDC} with exactly the same parameters except for the fact that now $Z'$ is in the $do$ set instead of the conditional set (following line 1 of the algorithm \textbf{IDC}).

\noindent\textbf{\large Line 2}

This second line of \textbf{IDC} is really straightforward, as we do not have to check any condition. We just call the function of the first identification algorithm, \texttt{ID\_rec}, with the specified parameters, and then create a fractional \class{Probability} object with the numerator being the returned probability from \texttt{ID\_rec} and the denominator the same but summed over the variables in $\bm{y}$.

\subsection{Join Implementation of \textbf{ID} and \textbf{IDC} and Usage}

In our package we have decided to join the two identification algorithms in just one function, named \texttt{ID} (refer to Function \ref{fun:ID}).

\begin{table}[!ht]
\renewcommand\tablename{Function}
\centering
\begin{tabular}{p{0.13\textwidth}p{0.10\textwidth}p{0.675\textwidth}}
\hline
\multicolumn{3}{p{0.9\textwidth}}{\define{ID}{Y, X, G, cond=\{\}, verbose=\false}}\\ \hline
\textbf{Description}                                        & \multicolumn{2}{p{0.8\textwidth}}{Function that calls either the non-conditional (\textbf{ID}) or the conditional (\textbf{IDC}) identification algorithm, depending on the parameter \params{cond}.}\\ \hline
\multirow{5}{*}[2.40em]{\textbf{Parameters}}    & \params{Y}    & Set of strings containing the variables in $\bm{y}$.\\
                                                                                            & \params{X}    & Set of strings containing the intervened variables in $\bm{x}$.\\
                                                                                            & \params{G}    & Graph object, encoding the DAG of the causal model $G$.\\
                                                                                            & \params{cond} & Set of strings containing the conditional variables.\\
                                                                                            & \params{verbose}  & Boolean, prints some useful debugging information if enabled. \\ \hline
\textbf{Returns}                                                & \multicolumn{2}{p{0.8\textwidth}}{If the parameter \params{cond} is empty, this is, if it does not contain any element, the \texttt{ID\_rec} function is called. Otherwise, if it contains any element, then \texttt{IDC} is executed.}\\ \hline
\end{tabular}
\vspace*{0.2cm}
\caption{Implemented \texttt{ID} function.}
\label{fun:ID}
\end{table}
Function \ref{fun:ID} is the one in charge of checking if the the three sets of vertices inputted are pairwise disjoint, as required by both algorithms, and if the directed part of the graph is a DAG. Then it creates the first \class{Probability} object by setting all the nodes in $G$ as the variables of the distribution. It is also the one that has to compute a topological ordering of the vertices in $G$, which as we recall only has to be computed once. It does so by passing a graph object $G$ to a function named \texttt{get\_topological\_ordering}, which makes use of the provided function \texttt{topological\_sorting} of the \texttt{igraph} package over the directed part of $G$ (omitting bidirected edges).

\begin{figure}[!ht]
\vspace*{-0.0cm}
\centering
\includegraphics{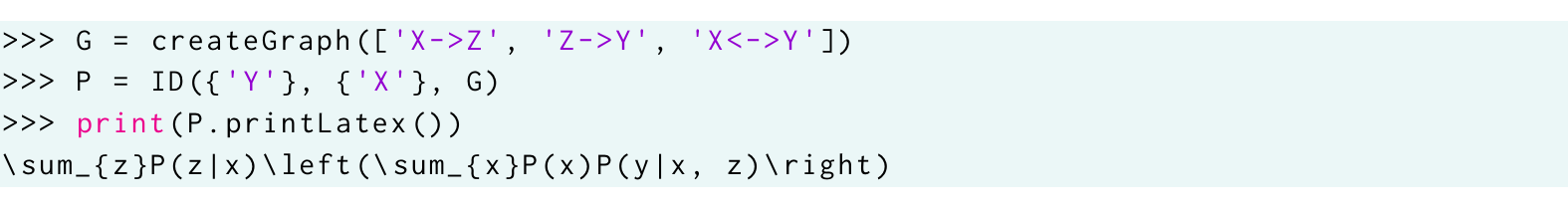}
\vspace*{-0.0cm}
\caption{How to create and identify a causal effect with the \texttt{caualeffect} package for Python.}
\label{fig:code_3}
\vspace*{-0.0cm}
\end{figure}

To use the developed identification algorithms in our \texttt{causaleffect} package, one first has to install \texttt{igraph} and \texttt{numpy}. If, additionally, one wants to draw plots of the causal diagrams, the package \texttt{pycairo} is also required. Once the \texttt{causaleffect} package has been imported, we can compute the causal effect in Example \ref{exp:algorithm}. As we can see from Figure \ref{fig:code_3}, we obtain the same result, printed in \LaTeX\ syntax.

We present a few more examples of the usage of the developed package.

\begin{example}
Consider the causal diagram shown in Figure \ref{fig:graph_13} (a), from \cite{SP_2006a}.
\begin{figure}[!ht]
\captionsetup[subfigure]{labelformat=empty}
\vspace*{-0.0cm}
\centering
\setlength{\mylength}{\textwidth}
\savebox{\largestimage}{\includegraphics{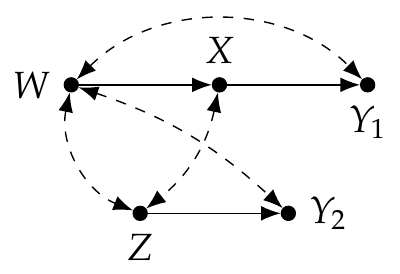}}%
\begin{subfigure}[t]{0.5\mylength}
        \centering
        \usebox{\largestimage}
        \caption{\footnotesize (a) Graph $G$.}
\end{subfigure}
\begin{subfigure}[t]{0.5\mylength}
        \centering
        \raisebox{\dimexpr.5\ht\largestimage-.5\height}{\includegraphics{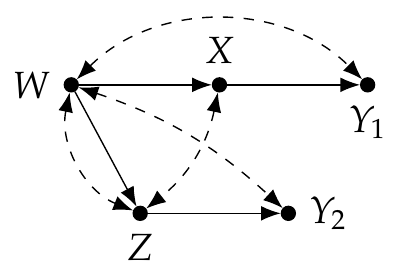}}
        \caption{\footnotesize (b) Graph $G'$.}
\end{subfigure}%
\vspace*{-0.0cm}
\caption{Causal diagrams.}
\label{fig:graph_13}
\vspace*{-0.0cm}
\end{figure}
This diagram could be the induced graph of a model for studying how the level of a certain toxin affects the survival rate of a pregnant mother and her child. In this model, $W$ and $Z$ are the afflictions of the mother and the unborn child, respectively. $X$ is the toxin produced in the mother's body due to the illness, and $Y_1$ and $Y_2$ are the survival rates of the mother and the child, respectively. Bidirected edges are reasonable confounding variables ($W\dashbidirectedarrow Y_1$ would be a common hidden cause affecting the affliction of the mother and her survival expectancy, for example, the existence of a substance affecting both the affliction and the chance of surviving). Suppose that we have a treatment that can artificially lower the amount of toxin $X$ in the mother's body, and we are interested in how this reduction in $X$ affects the survival rates of the mother and child. In causal theory this is equivalent to compute $P_{x}(y_1, y_2)$. To compute this causal effect in graph $G$ (Figure \ref{fig:graph_13} (a)), we just have to execute the code shown in Figure \ref{fig:code_4_1}.

\begin{figure}[!ht]
\vspace*{-0.0cm}
\centering
\includegraphics{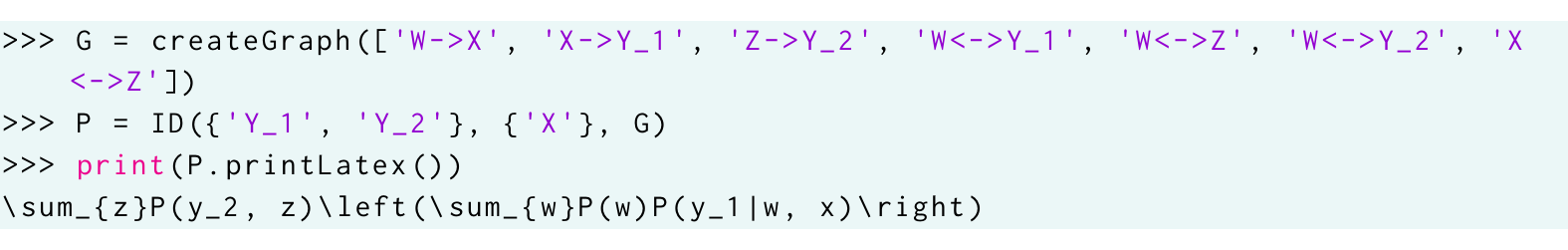}
\vspace*{-0.0cm}
\caption{Computing $P_{x}(y_1, y_2)$ from $G$.}
\label{fig:code_4_1}
\vspace*{-0.0cm}
\end{figure}
We see how the graph $G$ in Figure \ref{fig:graph_13} (a) is constructed in the first line, and how the \textbf{ID} algorithm is called next. In this case, as we see in Figure \ref{fig:code_4_1}, the effect is identifiable, and equal to
\begin{equation*}
P_{x}(y_1, y_2) = \sum_{z}P(y_2, z)\left(\sum_{w}P(w)P(y_1|w, x)\right) = P(y_2)\sum_{w}P(w)P(y_1|w, x)\ .
\end{equation*}
This means that we can actually know how this variation in $X$ would affect $Y_1$ and $Y_2$ without having to perform the potentially hazardous treatment in real patients.

Suppose now that our model is slightly changed due to recent studies that suggest that the affliction of the mother directly influences that of the child. Taking this new information into account in our causal diagram means adding one more edge, $W\to Z$ (see Figure \ref{fig:graph_13} (b)). If we now try to identify the same causal effect $P_{x}(y_1, y_2)$ in this new causal diagram $G'$ we will fail due to the existence of a hedge, as it can be seen in Figure \ref{fig:code_4_2}.

\begin{figure}[!ht]
\vspace*{-0.0cm}
\centering
\includegraphics{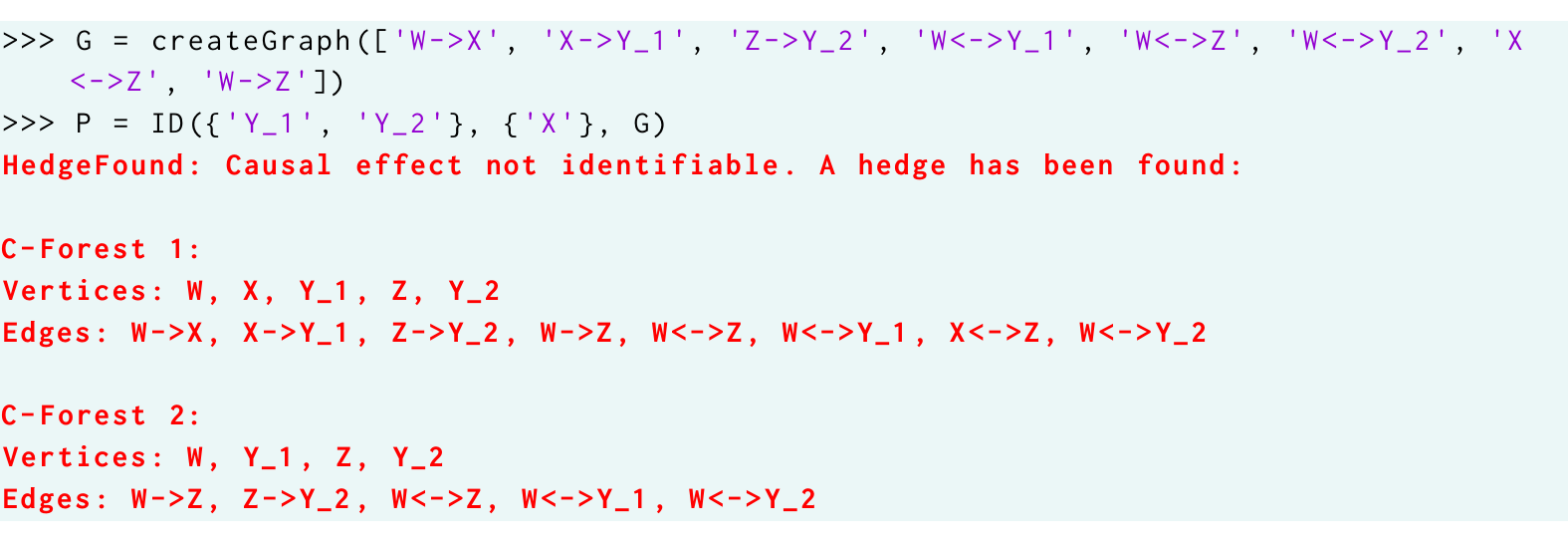}
\vspace*{-0.0cm}
\caption{Trying to identify $P_{x}(y_1, y_2)$ from $G'$.}
\label{fig:code_4_2}
\vspace*{-0.0cm}
\end{figure}
Indeed, we see that the first C-forest found by the algorithm is $G'$ itself, which has roots $\{Y_1, Y_2\}$. Moreover, we have ${\{Y_1, Y_2\}\subset\An(\{Y_1, Y_2\})_{G_{\xoverline{X}}}}$ and ${\{X\}\cap G' \neq\varnothing}$. The other $\{Y_1, Y_2\}$-rooted C-forest is $G'[\bm{V}\setminus X]$, which trivially fulfils ${\{X\}\cap G'[\bm{V}\setminus X] =\varnothing}$ and also ${G'[\bm{V}\setminus X]\subseteq G'}$. Hence $G'$ and $G'[\bm{V}\setminus X]$ form a hedge for $P_{x}(y_1, y_2)$ in $G'$, making this causal effect unidentifiable.
\end{example}

Consider now this following example of a conditional causal effect computation.

\begin{example}
Imagine we want to compute the causal effect $P_{x}(y|z)$ in the causal diagram $G$ presented in Figure \ref{fig:graph_14}.
\begin{figure}[!ht]
\vspace*{-0.0cm}
\centering
\includegraphics{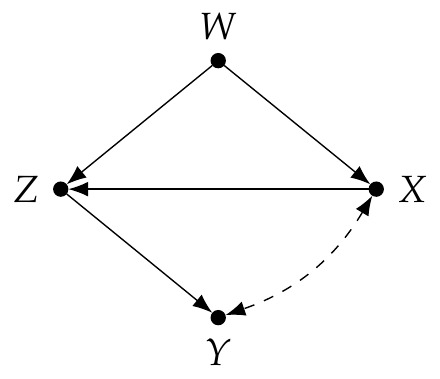}
\vspace*{-0.0cm}
\caption{Causal diagram $G$.}
\label{fig:graph_14}
\vspace*{-0.0cm}
\end{figure}
\begin{figure}[!ht]
\captionsetup[subfigure]{labelformat=empty}
\vspace*{-0.0cm}
\centering
\setlength{\mylength}{\textwidth}
\begin{subfigure}[t]{\mylength}
        \centering
        \includegraphics{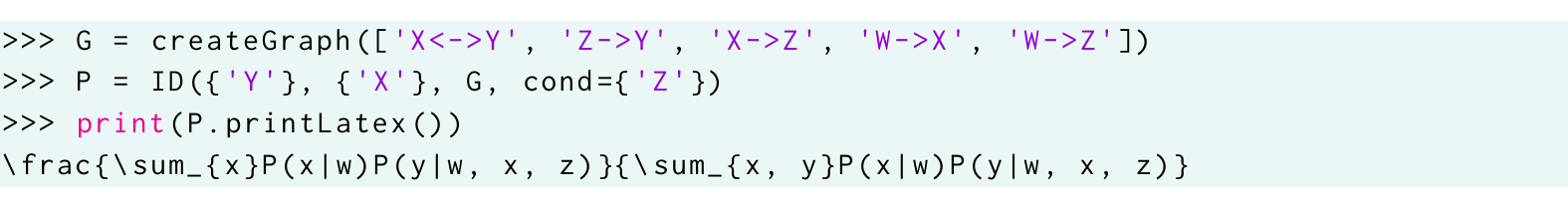}
        \caption{\footnotesize (a) Computing $P_{x}(y|z)$ from $G$ in Figure \ref{fig:graph_14}.}
\end{subfigure}
\vspace*{0.1cm}
\begin{subfigure}[t]{\mylength}
        \centering
        \includegraphics{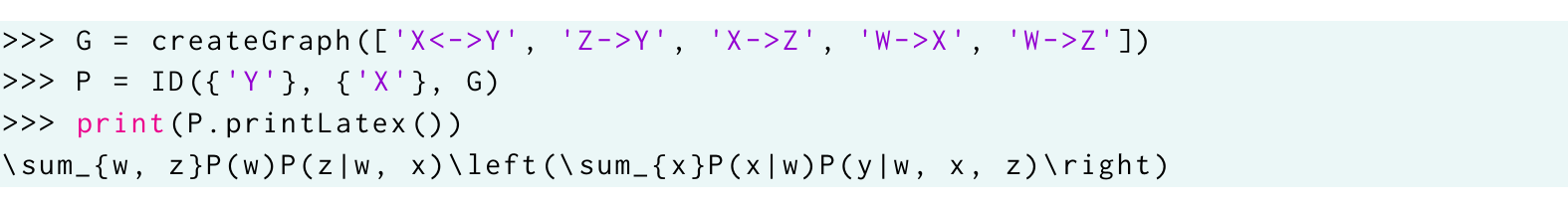}
        \caption{\footnotesize (b) Computing $P_{x}(y)$ from $G$ in Figure \ref{fig:graph_14}.}
\end{subfigure}%
\vspace*{-0.0cm}
\caption{Computing two causal effects with the implemented package.}
\label{fig:code_5}
\vspace*{-0.0cm}
\end{figure}
If we execute the code in Figure \ref{fig:code_5} (a), we retrieve the desired conditional causal effect, 
\begin{equation*}
P_{x}(y|z) = \frac{\sum_{x}P(x|w)P(y|w, x, z)}{\sum_{x, y}P(x|w)P(y|w, x, z)}\ .
\end{equation*}
Observe that, although our package is able to simplify a handful of expressions, the obtained result can be further reduced:
\begin{equation*}
P_{x}(y|z) = \frac{\sum_{x}P(x|w)P(y|w, x, z)}{\sum_{x, y}P(x|w)P(y|w, x, z)} = \frac{\sum_{x}P(x|w)P(y|w, x, z)}{\sum_{x}P(x|w)\sum_{y}P(y|w, x, z)} = \sum_{x}P(x|w)P(y|w, x, z)\ .
\end{equation*}
This expression is interesting, since we do not fix nor condition on the variable $W$, and yet it appears in the final expression as a \textit{free variable} (meaning it is not summed over all its possible values). In fact, despite appearing in the computed causal effect, said effect is independent of the value $w$, but in a practical setting some value within the domain of $W$ must be chosen for $w$. The independence can be easily seen from Figure \ref{fig:graph_14}, because when intervening $X$ and conditioning on $Z$ we $d$-separate $Y$ and $W$, thus making them independent.

If we now want to compute the causal effect $P_{x}(y)$ in the same causal diagram $G$, we just slightly change the call to the \texttt{ID} function, as we see in Figure \ref{fig:code_5} (b), and we obtain
\begin{equation*}
P_{x}(y) = \sum_{w, z}P(w)P(z|w, x)\left(\sum_{x}P(x|w)P(y|w, x, z)\right)\ ,
\end{equation*}
which cannot be further simplified.
\end{example}

Consider the following example, from the causal model and data in Example \ref{exp:bayesian}.
\begin{example}
In Example \ref{exp:bayesian} we had a causal diagram that modelled the probability of wearing sunscreen depending on the season and on the weather of that particular day. We want to see if a sunny day causes to wear sunscreen, and to do so we will compute $P(y)$ and $P_x(y)$ and compare them. First of all, we need to calculate an expression for the interventional distribution, and we do it with our \texttt{causaleffect} package, as in Figure \ref{fig:code_6}.

\begin{figure}[!ht]
\vspace*{-0.0cm}
\centering
\includegraphics{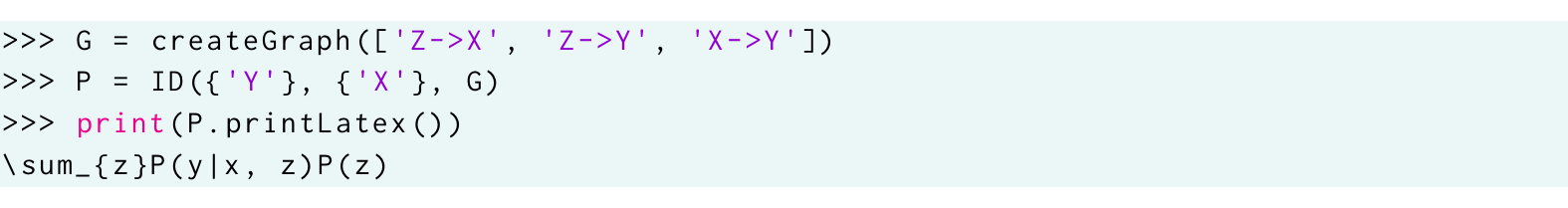}
\vspace*{-0.0cm}
\caption{Creation of graph $G$ from Example \ref{exp:bayesian} and computation of $P_x(y)$ from $G$.}
\label{fig:code_6}
\vspace*{-0.0cm}
\end{figure}
We have obtained $P_x(y) = \sum_{z}P(y|x, z)P(z)$, which is exactly the Back-Door Adjustment (note that $Z$ fulfils the back-door criterion relative to $(X, Y)$). Now we will compute these probabilities from data in Table \ref{tab:table1}, but to do so we need to remember that we did not have $P(X, Y, Z)$, but instead the conditional probabilities: $P(X, Y, Z) = P(Y|X, Z)P(X|Z)P(Z)$. We first compute what is the probability of wearing sunscreen.
\begin{align*}
P(Y=\text{T}) &= \sum_{x, z\in\{\text{T}, \text{F}\}}P(X=x, Y=\text{T}, Z=z)\\
&=\sum_{x, z\in\{\text{T}, \text{F}\}}P(Y=\text{T}|X=x, Z=z)P(X=x| Z=z)P(Z=z)\\
&= 0.99\cdot0.90\cdot0.25 + 0.20\cdot0.70\cdot0.75 + 0.60\cdot0.10\cdot0.25 + 0.05\cdot0.30\cdot0.75\\
&= 0.354\ .
\end{align*}
Then, we compute the probability of wearing sunscreen imposing that it is sunny. This intervention cannot be done in real life, and that is the beauty of it!
\begin{align*}
P(Y=\text{T}|do(X=\text{T})) &= \sum_{z\in\{\text{T}, \text{F}\}}P(Y=\text{T}|X=\text{T}, Z=z)P(Z=z)\\
&= 0.99\cdot0.25 + 0.20\cdot0.75\\
&= 0.3975\ .
\end{align*}
To compare even more results, we also compute the probability of wearing sunscreen given that we see it is sunny.

\begin{align*}
P(Y=\text{T}|X=\text{T}) &= \frac{\sum_{z\in\{\text{T}, \text{F}\}}P(X=\text{T}, Y=\text{T}, Z=z)}{\sum_{y, z\in\{\text{T}, \text{F}\}}P(X=\text{T}, Y=y, Z=z)}\\
&= \frac{\sum_{z\in\{\text{T}, \text{F}\}}P(Y=\text{T}|X=\text{T}, Z=z)P(X=\text{T}| Z=z)P(Z=z)}{\sum_{y, z\in\{\text{T}, \text{F}\}}P(Y=y|X=\text{T}, Z=z)P(X=\text{T}| Z=z)P(Z=z)}\\
&= \frac{\sum_{z\in\{\text{T}, \text{F}\}}P(Y=\text{T}|X=\text{T}, Z=z)P(X=\text{T}| Z=z)P(Z=z)}{\sum_{z\in\{\text{T}, \text{F}\}}P(X=\text{T}| Z=z)P(Z=z)}\\
&= \frac{0.99\cdot0.90\cdot0.25 + 0.20\cdot0.70\cdot0.75}{0.90\cdot0.25 + 0.70\cdot0.75} = \frac{0.32775}{0.75}\\
&= 0.437\ .
\end{align*}
We see that $P(Y=\text{T}|do(X=\text{T})) > P(Y=\text{T})$, and this rise in probability reveals that the sun causes to wear sunscreen. Additionally, we also see that the conditional probability is bigger than the interventional, ${P(Y=\text{T}|X=\text{T})}>{P(Y=\text{T}|do(X=\text{T}))}$. We believe that the probability of wearing sunscreen increases more when we see it is sunny than when we force it to be sunny because when it is sunny it is also more likely to be summer, and this also affects the probability of wearing sunscreen. On the other hand, when we intervene to be sunny it does not care if it is summer or not.
\end{example}

\subsection{From Causal Effects to Counterfactual Queries}

We have already seen how, using causal diagrams and the $do$-operator, we can answer causal queries from the second level of the Ladder of Causation when they are identifiable. The next natural step would be to study and formalize the concept of counterfactuals, and try to build a technique to answer counterfactual questions from observational data. This journey was undertaken by none other than Shpitser and Pearl \cite{SP_2007}, authors of the previous algorithms for computing causal effects. In this section, we will briefly explain an idea of their work regarding the identifiability of counterfactual queries and some algorithms made to evaluate them, although we have not implemented those in our package as they were far too complex and out of the scope of this project.

To be able to answer counterfactual queries, the authors introduced some notation. The reader can refer to \cite{pearl_causality} for an extensive discussion on counterfactuals and the notation used. A variable $Y$ affected by an intervention $do(x)$ is changed into a \textit{counterfactual variable}, and it is denoted by $Y_{x}$. Then, questions such as ``\textit{what if $\bm{X}$ were $\bm{x}$}'' would be represented as $P(Y_{\bm{x}}|e)$, where $e$ are observations that induce the probability distribution. There is an intrinsic problem with these types of queries and it is that actions $\bm{x}$ and evidence $e$ can stand in logical contradiction, and no experimental setup exists which would emulate both the evidence and actions. For instance, there is no experiment that allows us to know the percentage of deaths that could be avoided among people who received a given treatment, had they not taken the treatment. So it is unclear if counterfactual expressions like $P(Y_{\bm{x}}|e)$, with $e$ and $\bm{x}$ incompatible, can be estimated consistently.

\begin{figure}[!ht]
\captionsetup[subfigure]{labelformat=empty}
\vspace*{-0.3cm}
\centering
\setlength{\mylength}{\textwidth}
\savebox{\largestimage}{\includegraphics{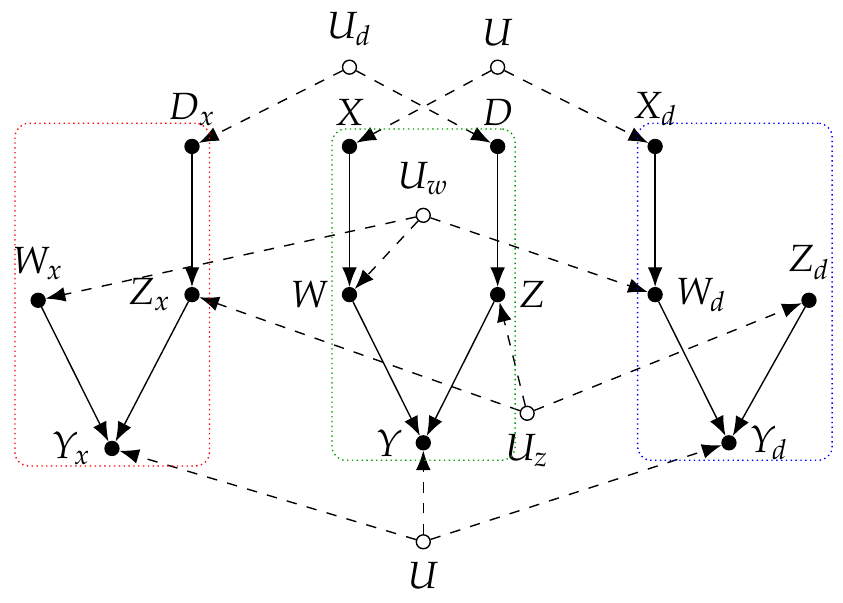}}%
\begin{subfigure}[t]{0.3\mylength}
        \centering
        \raisebox{\dimexpr.5\ht\largestimage-.5\height}{\includegraphics{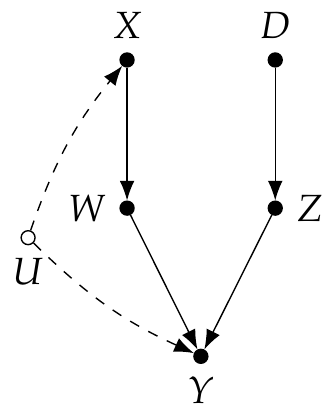}}
        \caption{\footnotesize (a)}
\end{subfigure}
\begin{subfigure}[t]{0.7\mylength}
        \centering
        \usebox{\largestimage}
        \caption{\footnotesize (b)}
\end{subfigure}%
\vspace*{-0.0cm}
\caption{(a) Causal graph $G$ of model $M$. (b) Parallel world graph of $G$ for $P(y_x|x', z_d, d)$, formed by three worlds: the original model $M$ (centre, green) and the submodels $M_x$ (left, red) and $M_d$ (right, blue).}
\label{fig:graph_15}
\vspace*{-0.0cm}
\end{figure}

The authors designed two algorithms \cite{SP_2007} to identify counterfactuals and evaluate them from interventional probabilities when identifiable. To do so, they introduce the concept of \textit{parallel world graphs}, which can be thought as multiple causal diagrams (each one of them representing a possible world, with a precise intervention) sharing the same exogenous variables. An example of a parallel world graph taken from \cite{SP_2007} is shown in Figure \ref{fig:graph_15}.

Figure \ref{fig:graph_15} (a) is the induced causal graph $G$ of the causal model $M$. $M$ models how drugs $X$ and $D$ affect intermediate symptoms $W$ and $Z$, which in turn influence another symptom $Y$. Suppose that we want to know how likely the patient would be to have a symptom $Y$ given a certain dose $x$ of drug $X$, assuming we know that the patient has taken dose $x'$ of drug $X$, dose $d$ of drug $D$, and we also know how the intermediate symptom $Z$ responds to treatment $d$. This causal query can be expressed as $P(y_x|x', z_d, d)$, and the parallel world graph would be formed by three worlds: the original model $M$ and also the submodels $M_x$ and $M_d$, as seen in Figure \ref{fig:graph_15} (b). Note that we have not included nodes fixed by actions in the parallel world graph since we already know their values, which are constant.

Parallel world graphs can have duplicate nodes (for instance, in Figure \ref{fig:graph_15} (b) we have $Z = Z_x$ since $Z$ is not a descendant of $X$), and this could arise some errors when computing $d$-separation. So what they do is they merge duplicate nodes following rigorous criteria and create what they call \textit{counterfactual graphs}, which is equivalent to counterfactual queries as a causal graph is to causal queries.

They present two algorithms similar to the ones for causal queries, called \textbf{ID$^{*}$} (for unconditional queries) and \textbf{IDC$^{*}$} (for conditional queries), that given a causal graph and a counterfactual query return either an error or an expression for computing the counterfactual query from interventional probabilities. Analogously to the causal case, they are recursive, and \textbf{IDC$^{*}$} calls \textbf{ID$^{*}$} as a subroutine. They have to work with counterfactual graphs, so they construct them from the inputted causal graphs using another algorithm also defined in the paper, named \textbf{make-cg}. Note that both \textbf{ID$^{*}$} and \textbf{IDC$^{*}$} return interventional probabilities, so to finally obtain results from observational data one has to use the causal effect algorithms already defined in this section.

The implementation of these counterfactual identifying algorithms is highly non-trivial and would require the construction of additional classes and functions.

\section{Conclusions}

Causal theory is a field in statistics that until recently had not been studied much. The influence of powerful statisticians, like Karl Pearson, discouraged the use of causal graphs to compute causal queries in the first half of the twentieth century, but in the last thirty years, thanks to Pearl and many other scientists, a formalized causal theory has been built.

This mathematization of causal questions reached a peak with the design of deterministic algorithms that identify and, when possible, compute causal effects from observational data. Although known by many scientists, these results are unknown to many others, and to bring them to a wider audience the author has implemented a new Python library that computes causal effects. This package is very practical since in only two lines of code one can construct a causal graph and query a causal effect.

This is not the first developed package with these results. To the best of the author's knowledge, the only implementation of these identification algorithms before the one presented in this work is the R package called \texttt{causaleffect}, by Tikka and Karvanen \cite{causaleffect_R}. In this project, the author has also studied in utmost detail this package and its associated paper and has found a subtle bug in the implementation of \textbf{IDC}. This error was reported to the authors, and it has already been fixed in version 1.3.13 (June 14$^{\text{th}}$, 2021).

The implementation and analysis of the identification algorithms presented in this work have required the author to go through the necessary background mathematical results on the theories involved. I believe this has been vital to present the results in this work in a more organized and clearer way, and I expect this to be a gateway for more scientists to discover Pearl's remarkable results.

As the author has stated in this work, a logical extension of this project would involve the implementation of counterfactual identification algorithms. To do so, one would have to sail through the obscure notation of counterfactual formalization. If successful, this would finally enable the computation of counterfactual queries, questions of the third level of the Ladder of Causation.

\section[Source Code]{Source Code}

The source code of the developed Python library, \texttt{causaleffect}, can be found in the following GitHub repository:
\begin{equation*}
\text{\url{https://github.com/pedemonte96/causaleffect}}
\end{equation*}
Every figure containing code in this work has its equivalent Python script in the provided repository.

\bibliographystyle{unsrtnat}

\end{document}